\documentclass[final,12pt]{elsarticle}
\pdfoutput=1
\usepackage{amssymb}

\biboptions{sort&compress} 

\usepackage{longtable}
\usepackage{algorithm,algpseudocode}
\usepackage{listings}
\usepackage{amsmath,amsfonts}
\usepackage{color}
\usepackage{hyperref}
\renewcommand{\vec}[1]{\mbox{\boldmath $#1$}}
\usepackage[detect-none]{siunitx}
\usepackage{ifthen}
\renewcommand{\j}[1]{\textcolor{black}{#1}}
\newcommand{\jj}[1]{\textcolor{black}{#1}}
\newcommand{\jjj}[1]{\textcolor{black}{#1}}
\newcommand{\jh}[1]{\textcolor{black}{#1}}
\usepackage[normalem]{ulem} 

\newcommand{\jl}[1]{\textcolor{black}{#1}}

\newcounter{bla}

\journal{Journal of \LaTeX\ Templates}

\begin{document}

\begin{frontmatter}



\title{
\texttt{gVOF}: An open-source package for unsplit geometric volume of fluid methods on arbitrary grids \\
}
\author[Cartagena]{Joaquín López\corref{author}} 
\author[Madrid]{Julio Hernández} 
\address[Cartagena]{Dept. de Ingenier\'{\i}a Mec\'anica, Materiales y
Fabricaci\'{o}n, ETSII, Universidad Polit\'{e}cnica de Cartagena,
E-30202 Cartagena, Spain}
\address[Madrid]{Dept. de Mec\'{a}nica, ETSII, UNED, E-28040
Madrid, Spain} 
\cortext[author] {Corresponding author.\\\textit{E-mail address:} joaquin.lopez@upct.es}
\begin{abstract}
The \texttt{gVOF} package includes \jh{a complete and self-contained set of} routines for volume of fluid initialization,  interface reconstruction and fluid advection, which are used to implement several accurate and efficient geometric volume of fluid (VOF) \jh{methods on} arbitrary grids, either structured or unstructured with convex or non-convex cells, based on  multidimensional unsplit advection and  piecewise linear interface calculation (PLIC) schemes. The package uses the non-convex analytical and geometrical tools developed by López et al. [J. Comput. Phys. 392 (2019) 666-693] and the isosurface extraction procedure proposed by López et al. [\jj{J. Comput. Phys. 444 (2021) 110579}]. 
The operations
required \jh{during} the advection \jh{step} of  \jh{unsplit} geometric VOF  methods \jh{in general} involve \jh{handling of}
high-complex non-convex flux polyhedra, \jh{even with} self-intersecting
faces, which are  robustly and efficiently treated in this work without the need of costly techniques based on convex decomposition.  
Results for the accuracy, computational efficiency, and volume
\jh{(local and global)} conservation properties  of different combinations of the implemented
advection and reconstruction methods are presented for several
numerical tests on structured and unstructured grids. 
An extensive comparison with  results  obtained by other authors using advanced geometric VOF methods shows the \jh{outstanding performance of the} \texttt{gVOF} package in terms of efficiency and accuracy. 
To demonstrate the performance of the package \jh{in solving} complex two-phase flow problems, the implemented methods are combined with an existing
in-house code to simulate the impact of a water \jl{drop} on a free surface.
\jh{The package}
can be used in
\texttt{FORTRAN} or \texttt{C} languages and will be publicly available
for download.
The \texttt{OpenMP} application programming interface is also used to \jl{improve computational} efficiency. 
The goal of this work is to 
facilitate and extend the use of advanced 
unsplit geometric VOF methods in new or existing computational fluid dynamics  codes. 
\end{abstract}
\end{frontmatter}
{\bf Program summary}

\begin{small}
\noindent
{\em Program Title:}  \texttt{gVOF}                                        \\
{\em Licensing provisions:} GPLv3                                   \\
{\em Programming language:}  \texttt{FORTRAN} and \texttt{C}, with \texttt{C} interfaces     \\
{\em Nature of problem:} The software package includes efficient and accurate routines for volume of fluid initialization,  reconstruction of interfaces and fluid advection on arbitrary grids, either structured or unstructured with convex or non-convex cells, which are used to implement advanced \jh{unsplit geometric} VOF methods. In particular, \jh{the package includes} a fluid volume fraction initialization method, six PLIC reconstruction methods and three multidimensional unsplit advection methods. Routines to visualize interfaces \jj{and compute reconstruction errors}; test programs to assess the accuracy, computational efficiency, and volume conservation of the implemented methods \jl{for the} reconstruction and advection of complex interfaces on arbitrary grids; and a user manual have also been included in the supplied package \jl{(temporal link to download the software: \url{https://data.mendeley.com/datasets/k6556xngbp/draft?a=1ee7cb83-8c31-453a-a17f-488cbe7d3685})}. \\
{\em Solution method:} Basically, the implemented methods, which can be used on grids with polyhedral cells of arbitrary geometry, have the following general characteristics.  
The implemented fluid volume fraction initialization \jj{and reconstruction error computation methods are} based on the refinement procedure proposed in \cite{01,02}. 
The implemented interface reconstruction methods are the following:
a least-squares gradient interface reconstruction (LSGIR) method;  \jj{improved versions} of the local (LLCIR), extended (ELCIR) and conservative (CLCIR) \jh{isosurface-based} interface reconstruction  methods proposed in \cite{3}; an extension to 3D of the iterative Swartz interface reconstruction (SWIR) method proposed in \cite{3b}; and a version of the least-squares fit interface reconstruction (LSFIR) method proposed in \cite{3c,3d}. The three implemented advection methods \jj{are} \jh{the following}: an extension to 3D arbitrary grids of the edge-matched flux polygon advection (EMFPA) method proposed in \cite{1}; a new version of the  face-matched flux polyhedron advection (FMFPA) method proposed in \cite{2}; and the non-matched flux polyhedron advection (NMFPA) method, which can be considered as an extension to 3D arbitrary grids of the unsplit advection method proposed in \cite{2b}.
The supplied software package uses the \texttt{VOFTools} \cite{4} and \texttt{isoap} \cite{5} libraries.
All the implemented methods can be combined by the user to solve different tests on arbitrary grids. The implemented routines can be used in \texttt{FORTRAN} or \texttt{C}. \jh{The} \texttt{OpenMP} application programming interface is used to improve the computational efficiency. \\
   \\

\end{small}

\section{Introduction}

There are many methods for interface capturing. Among them (see, for example, Reference \cite{tryggvason11} for a complete review), the volume of fluid (VOF) method is one of the most popular. This method uses an auxiliary function, referred to in this work as $f$, to implicitly capture the interface. Mass conservation is among its main advantages and the recent improvements in the computation of geometric characteristics such as interface curvature \jj{(e.g., \cite{renardy02,cummins05,afkhami07,afkhami08,popinet09,lopez09,lopez10,ivey15,jibben19})} or orientation (e.g., \cite{scardovelli03,pilliod04,liovic05,aulisa07,lopez04,lopez05,lopez08,ivey15}) make this method even more competitive compared to others. The VOF methods could be classified in two main groups: (1) algebraic and (2) geometric methods. 
In algebraic type VOF methods, the auxiliary function $f$ is usually represented algebraically by a polynomial or trigonometric function while in geometric type VOF methods, a discretized version of $f$ is represented geometrically  by the region delimited by the considered cell and the interface, which is generally defined by a line in two dimensions (2D) or a plane in three dimensions (3D). The geometric VOF methods, although they require a relatively high complex implementation, are more accurate in the computation of fluxes through cell faces and may be more computational efficient since only the closest cells to the interface are involved in the computation (algebraic VOF methods involve 
larger grid cell stencils). However, probably due to the above geometric implementation complexity, algebraic VOF methods are still widely used in many commercial and non-commercial codes (for example STAR-CCM$+$
or ANSYS Fluent,
 among others).
A recent review of interface-capturing methods for two-phase flows \cite{mirjalili17} concludes that geometric VOF methods are among the most promising interface-capturing methods for future investment. An overview of different geometric VOF methods, mainly focused on unstructured grids and three dimensions, is recently presented in \cite{maric20}.

Geometric VOF methods basically consist of two steps: interface reconstruction and fluid advection.
At each time interval, these two steps can be computed in only one stage using unsplit schemes or in multiple stages  using split schemes that perform one dimensional advection for each spatial dimension (2 stages in 2D or 3 stages in 3D).  
The most advanced geometric VOF methods used today generally involve unsplit schemes with PLIC (piecewise-linear interface calculation) reconstruction\jl{, although competitive geometric VOF methods based on split schemes can also be found, e.g., in \cite{aulisa03,aulisa07,baraldi14}}. PLIC  reconstruction is being used since the first implemented geometric VOF method \cite{debar74} and prevails over the less accurate piecewise-constant SLIC (simple line interface calculation) reconstruction \cite{hirt81}. Unsplit schemes, although \jl{they} require higher complex geometric operations than split schemes, especially in 3D, are \jl{computationally} efficient because only one interface reconstruction per time interval is required and have the potential to be readily implemented on unstructured grids.  \jj{This work is focused on these schemes.}

Among others, \citet{kothe96}, \citet{rider98}, \citet{harvie00,harvie01}, \citet{lopez04,lopez05} and \citet{pilliod04},  for 2D, and \citet{rider98}, \citet{miller02}, \citet{liovic05}, \citet{hernandez08}, \citet{lopez08}, and more recently \citet{owkes14} or \j{\citet{maric18}} for 3D, developed   geometric-unsplit VOF methods, but only results on structured square or cubic grids were provided.  
Some of the first results on 2D unstructured grids can be found, for example, in the works by \citet{mosso96,mosso96b}.
Successful implementations of \jl{unsplit geometric} VOF \jl{methods} coupled with a level set method were also developed by \citet{ningegowda14} and \citet{cao18} on 2D structured and unstructured grids, respectively.  
Due to the highly complex geometric operations involved in these methods, very scarce results can be found in the literature on 3D grids with non-cubic cells. Thanks to the appearance of the \texttt{VOFTools} routines, which have also been used by many of the authors referenced in this section, 
the first results on  grids with non-cubic cells were obtained by \citet{lopez08b}. Later, \citet{ivey12} and \citet{jofre14}  implemented
methods that use, like in the edge-matched flux polygon advection (EMFPA) method proposed by \citet{lopez04} and following the suggestion made by \citet{hernandez08}, the velocities in cell vertices to construct flux regions in cell faces on 3D unstructured grids,  minimizing  overlaps between them and reducing bounding errors in the fluid volume fraction distribution.
Very recently, \citet{ngo21} developed a similar geometric-unsplit VOF method coupled with a level set method and provided results on unstructured triangular and tetrahedral grids.
\citet{ivey17} proposed the non-intersecting flux polyhedron advection (NIFPA) method that uses an iterative procedure to satisfy conservation and  boundedness of the liquid volume fraction irrespective of the underlying flux polyhedron geometry on structured and unstructured grids.
A recent  advanced geometric VOF method based on isosurface constructions for general grids with arbitrary polyhedral cells was proposed by \citet{roenby16} and improved later in \cite{scheufler19} by using reconstructed distances to PLIC interfaces.

The main objective of this work is to 
implement a publicly available software 
which includes several accurate and efficient  PLIC reconstruction  and unsplit advection methods with the purpose to spread their use to simulate complex interface dynamics in arbitrary grids, with convex or non-convex cells.
In particular, the following seven methods, described in Section~\ref{sec:methods}, are implemented: 
\begin{enumerate}[(i)]
\item Six PLIC-based reconstruction methods:
\begin{itemize}
  \item an interface reconstruction method based on a least-squares gradient technique proposed by \citet{barth90} (Section~\ref{sec:lsgir}), 
\item three new versions of the isosurface based interface reconstruction methods proposed by \citet{lopez08} (Section~\ref{sec:llcir}), 
\item a 3D  version of the iterative Swartz \cite{swartz89} interface reconstruction \jl{method} (Section~\ref{sec:swir}), and
\item a 3D version of the least-squares fit interface reconstruction methods of \citet{scardovelli03} and \citet{aulisa07}. 
\end{itemize}
 \item Three unsplit advection methods: 
\begin{itemize}
\item an extension to 3D arbitrary grids of the edge-matched flux polygon advection, also referred to hereafter as EMFPA (edge-matched flux polyhedron advection), method proposed by \citet{lopez04} (Section~\ref{sec:emfpa}), 
\item a new version of the face-matched flux polyhedron advection (FMFPA) method proposed by \citet{hernandez08} (Section~\ref{sec:fmfpa}), and
\item an extension to 3D arbitrary grids of the unsplit advection method proposed by \cite{rider98}, which will be referred \jl{to} as non-matched flux polyhedron advection (NMFPA) method (Section~\ref{sec:nmfpa}).
\end{itemize}
\end{enumerate}
Also, a fluid volume fraction initialization method based on the refinement procedure proposed in \cite{lopez09,lopez19} and routines for interface visualization are implemented.
A brief description of the routines included in the \texttt{gVOF} \jjj{package}  is presented in Section~\ref{sec:routines} and further details can be found in the user manual included in the supplied software.
In Section~\ref{sec:testprograms}, an analysis of    
accuracy, computational efficiency, and volume conservation properties
of different combinations of the implemented advection and reconstruction methods is carried out for several numerical tests using structured and unstructured grids with convex and non-convex polyhedral cells. Also, 
an exhaustive comparison with results obtained with previous advanced geometric-unsplit VOF methods is included.
The implemented \jjj{package} has been combined with an existing code developed by our group \cite{lopez09,hernandez08,gomez05,lopez05,gomez19} to simulate  complex phenomena involved in the impact of a water \jl{drop} onto a free surface, and its results are presented in Section~\ref{sec:drop-impact}. Finally, the parallel performance of the \texttt{gVOF} \jjj{package} is assessed in Section~\ref{sec:parallel}.
To avoid doing an excessively long work, the detailed analysis of the high number of parameters involved in the implemented algorithms will be published elsewhere.

\section{Problem statement}
\label{sec:methods}

The \texttt{gVOF} \jjj{package} solves the time-evolution equation
\begin{equation}
  \label{vof}
  \frac{\partial f}{\partial t} + \vec{\nabla} \cdot (\vec{u}f) - f \vec{\nabla} \cdot \vec{u}= 0,
\end{equation}
where  the function $f$ is equal to 1 in the fluid  and 0 otherwise, and $\vec{u}$ is the velocity field. 
This equation is integrated over a time interval from $t^n$ to $t^{n+1}$ and  a given {cell}, $\Omega$, of volume
$V_\Omega$, to obtain, at each time step,
\begin{equation}
  \label{advec}
  F^{n+1}=F^{n} -  \frac{1}{V_\Omega}
    \int_{t^{n}}^{t^{{n}+1}}\int_\Omega
     \vec{\nabla} \cdot (\vec{u}f) \,  \mbox{d} \Omega\,
    \mbox{d} t + \frac{F^{n+1}+F^{n}}{2V_\Omega}
    \int_{t^{n}}^{t^{{n}+1}}\int_\Omega
     \vec{\nabla} \cdot (\vec{u}) \,  \mbox{d} \Omega\,
    \mbox{d} t,
\end{equation}
where $F$ is a discretized version of the function $f$, whose value in each cell of the computational grid is the fraction of the cell occupied by the fluid. $F$ will be denoted hereafter as fluid volume fraction.
 As suggested by \citet{rider98}, the application of the last term in Eq.~(\ref{advec}), even for incompressible flows, can help to improve the local and global volume conservation. Note that this term must be null if the velocity vector field $\vec{u}$ is discretely solenoidal.
The first integral in Eq.~(\ref{advec})
represents the net volume of fluid advected out of the cell. This volume is computed geometrically using an unsplit advection method, for which the fluid interface must be previously reconstructed using a PLIC method. Both  reconstruction and advection represent the most complex and time-consuming steps in any advanced geometric VOF code. Below, the methods implemented in the \texttt{gVOF} \jjj{package} for these two steps are briefly described.

\subsection{Interface reconstruction}
\label{sec:reconstruction}

Based on the value of $F$, the grid cells are classified in two types:
\begin{itemize}
\item uniform, if $F<\epsilon$ or $F>1-\epsilon$, and
\item interfacial, otherwise,
\end{itemize}
where $\epsilon$ is a small value close to zero (in the order of $10^{-10}$). For all the uniform cells, the volume fraction is cut off before reconstructing as 
\begin{equation}
F=\left\{
\begin{split}
1&, \,\,\,\mbox{if}\,\,\, F>1-\epsilon \\
0&, \,\,\,\mbox{if}\,\,\, F<\epsilon. 
\end{split}
\right.
\end{equation}
For each interfacial cell, the interface is represented by a planar interface defined  as
\begin{equation}
\vec{n}\cdot \vec{x}+C=0,
\label{plic-eq}
\end{equation}
where the unit vector $\vec{n}$ normal to the interface and pointing into the fluid is firstly determined  from any of the methods amenable to arbitrary grids described below, and then, the constant $C$ is computed  so that the interface splits the cell $\Omega$ 
of volume $V_\Omega$ into two sub-cells of volumes $FV_\Omega$ and
$(1-F)V_\Omega$. 
In this work, the CIBRAVE (coupled interpolation-bracketed analytical volume enforcement) method of \citet{lopez16b} is generally used to compute $C$, except when using grids with rectangular parallelepiped cells, for which the efficient analytical method of \citet{scardovelli00} is used.
The implementation of these two volume conservation enforcement methods is included in the \texttt{VOFTools} package \cite{lopez17,lopez19b,lopez20}.

This work implements into the \texttt{gVOF} \jjj{package}: a weighted least-squares gradient technique (Section~\ref{sec:lsgir}), also known as Youngs' method, which shows very good performance in terms of efficiency and accuracy in regions where the grid resolution is not enough; three methods based on isosurface extractions (Section~\ref{sec:isorec}), which generally show superior accuracy in regions with enough grid resolution; and two interface reconstruction methods  which can be applied iteratively  (Sections~\ref{sec:swir} and \ref{sec:lsfir}).

\subsubsection{Least-squares gradient interface reconstruction, LSGIR}
\label{sec:lsgir}

The unit  vector normal to the interface, $\vec{n}$, is obtained for every interfacial cell from the gradient of the fluid volume fraction distribution $\vec{\nabla} F$ as
\begin{equation}
\vec{n} = \frac{\vec{\nabla} F}{|\vec{\nabla} F|}
  \end{equation}
using a weighted least-squares technique \cite{barth90}, amenable to any grid. This technique, which can be considered as an extension of the Youngs' finite difference approximations for $\vec{\nabla} F$ \cite{youngs84} to arbitrary grids \cite{kothe96,rider98}, has been implemented as follows.

A stencil involving neighbor cells \jl{of every interfacial cell is considered, in which each neighbor cell $k$, with geometric centers given by $\vec{x}_k\equiv (x_k,y_k,z_k)$,} shares at least one vertex with the interfacial \jl{cell,} with geometric center given by $\vec{x}\equiv (x,y,z)$.
Then,  the sum $\sum\limits_{k=1}^n \left(\widetilde{F_k}-F_k \right)^2$\jl{, over} all the $n$ neighbor \jl{cells,} of the quadratic differences between the Taylor series expanded $\widetilde{F_k}$ and $F_k$ values,
 is minimized using a weighted least-squares fit that yields the following equation for the volume fraction \jl{gradient:}
\begin{equation}
  \vec{\nabla} F = \left(A^T A \right)^{-1} A^T \vec{b},
\end{equation}
where
\begin{equation}
  A= 
  \begin{pmatrix}
    w_1(x_1-x)  &   w_1(y_1-y) &   w_1(z_1-z) \\
    w_2(x_2-x)  &  w_2(y_2-y)  &   w_2(z_2-z) \\
    \vdots &  \vdots & \vdots \\
    w_n(x_n-x) & w_n(y_n-y) & w_n(z_n-z) \\
  \end{pmatrix},
  \label{eq:lsgir1}
  \end{equation}
\begin{equation}
  \vec{b}= 
  \begin{pmatrix}
    w_1(F_1-F) \\
    w_2(F_2-F) \\
    \vdots    \\
    w_n(F_n-F)\\
  \end{pmatrix},
  \label{eq:lsgir2}
  \end{equation}
and the weight
\begin{equation}
w_k=\frac{1}{|\vec{x}-\vec{x}_k|^\beta}
\label{lsgir-beta}
\end{equation} 
is used to soften the influence of remote stencil points on the results of the least squares approximation. A detailed analysis of the effect of the $\beta$ parameter on the accuracy of the LSGIR method will be published elsewhere.

\subsubsection{Isosurface based interface reconstruction}
\label{sec:isorec}

\begin{figure}[htbp]
\begin{center}
\includegraphics{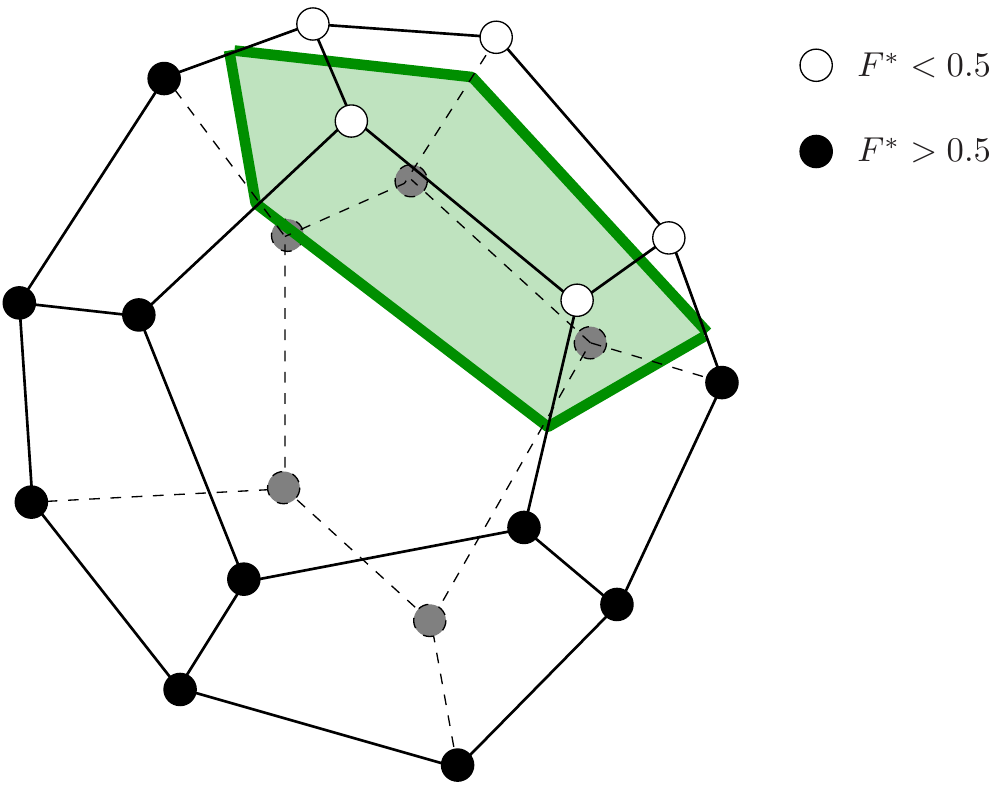}
\caption{Isosurface extracted on an irregular polyhedral cell from the volume fraction values interpolated at its vertices.}
\label{isosurface-rec}
\end{center}
\end{figure}

The  vector  $\vec{n}$ is obtained with the aid of the  isosurface extracted from the volume fraction distribution interpolated at cell vertices\jl{, $F^*$}.
For every interfacial cell whose \jl{maximum  and minimum  interpolated $F^*$} values 
satisfy the condition 
\begin{equation}
F_{min}^*<0.5<F_{max}^*,
\label{isocondition}
\end{equation}
the isosurface corresponding to the isovalue
$0.5$ is constructed using the procedure proposed by \citet{lopez20b} (see the example of Fig.~\ref{isosurface-rec}). 

The interfacial cells that do not satisfy the condition of Eq.~(\ref{isocondition}) or produce more than one isosurface, situations that frequently occur in regions of low grid resolutions, are reconstructed using the LSGIR method, which, as  mentioned, performs well in such cases.

Below, three different variants of the interface reconstruction based on isosurface extraction will be briefly described. Considerations for the domain boundaries can be found in \cite{lopez08} and further details will be published elsewhere.

\paragraph{Local level contour-based interface reconstruction, LLCIR}
\label{sec:llcir}

\begin{figure}[htbp]
\begin{center}
\includegraphics{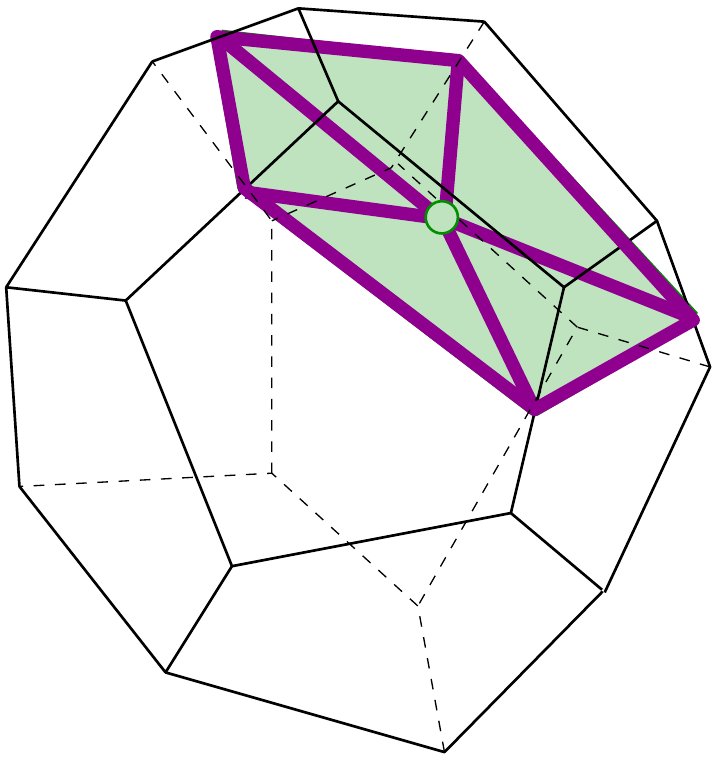}
\caption{Triangulation of the isosurface of Fig.~\ref{isosurface-rec} used \jl{in} the LLCIR method.}
\label{isosurface-rec-local}
\end{center}
\end{figure}
\begin{figure}
\begin{center}
  \includegraphics{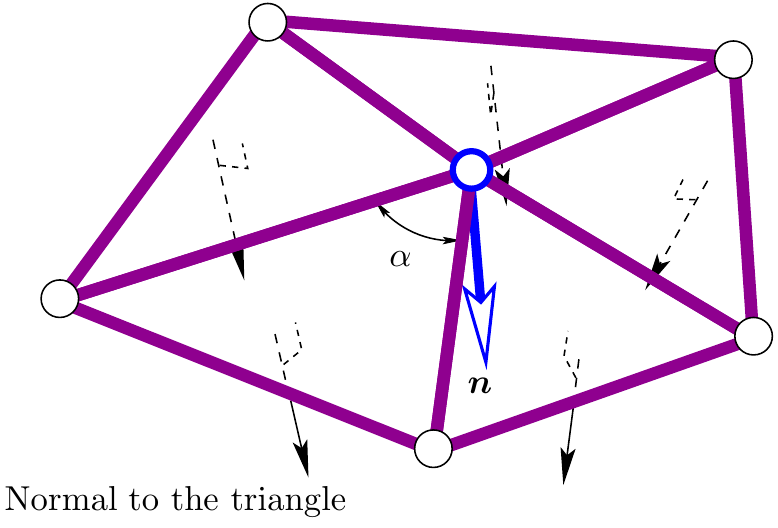}
\end{center}
\caption{Interface orientation  vector\jl{, $\vec{n}$,} obtained by \jl{averaging  the} unit vectors normal to the triangles. The angle $\alpha$, \jl{along with other parameters of each triangle, are} used for some of the weight factors included in the \texttt{gVOF} \jjj{package} that can be applied to obtain the average interface normal.}
\label{average-normal2}
\end{figure}

The extracted isosurface, which generally is non-planar, is triangulated using its geometric center\jj{, which is obtained by simple averaging the position vectors of the isosurface vertices,} as it is sketched in the example of Fig.~\ref{isosurface-rec-local}, and the interface unit vector $\vec{n}$ in the considered interfacial cell is obtained from a weighted average of the unit vector normals to the isosurface triangles (see the example in Fig.~\ref{average-normal2}). 
Different weighted parameters, based on angles, edge lengths or areas of the isosurface triangles, are considered in the implemented \jjj{package} and a detailed analysis of the performance of each one will be carried out elsewhere.

\paragraph{Extended level contour-based interface reconstruction, ELCIR}
\label{sec:elcir}

\begin{figure}[htbp]
\begin{center}
\includegraphics{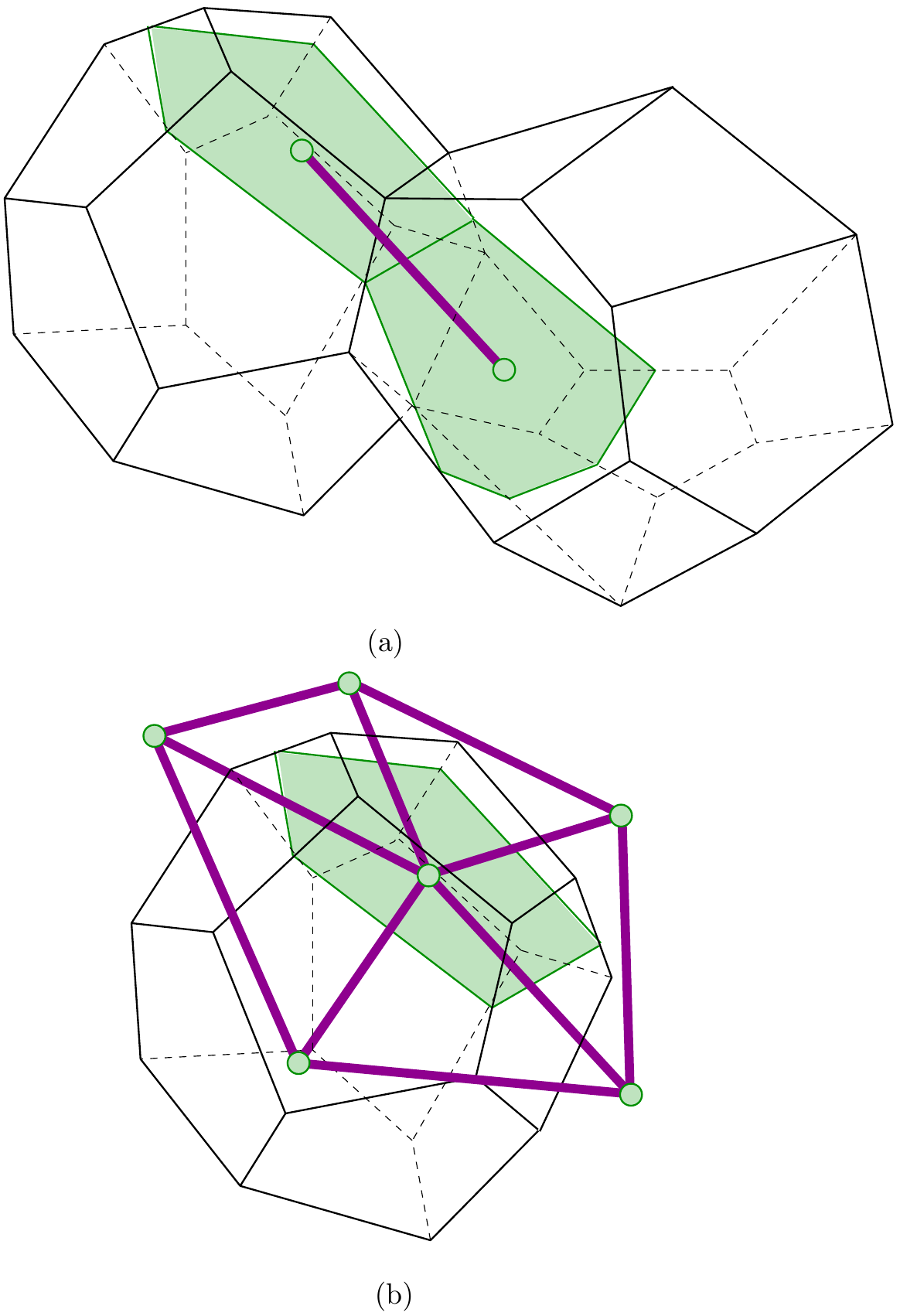}
\caption{Sketch of the ELCIR method. (a) Extension to an adjacent cell by connecting the geometric centers \jj{of the set of vertices} of the extracted isosurfaces. (b) Triangulated surface extended to the geometric centers \jj{of the set of vertices} of the isosurfaces extracted from adjacent cells.}
\label{isosurface-rec-ext}
\end{center}
\end{figure}
The vector $\vec{n}$ is obtained from an extended triangulated isosurface involving adjacent cells. 
The extension to adjacent cells is made by connecting the geometric centers \jj{of the set of vertices} of the extracted isosurfaces, as sketched in the example of Fig.~\ref{isosurface-rec-ext}(a), resulting in a new triangulated surface (see the extended triangulated surface in the example of Fig.~\ref{isosurface-rec-ext}(b)) from which $\vec{n}$ is finally obtained by averaging  as mentioned above (Fig.~\ref{average-normal2}).
If the extended triangulated surface \jl{yields} an orientation that differs by more than 1.2 rad with respect to that \jl{provided} by the LLCIR method, the vector $\vec{n}$ is obtained from the previous local triangulated isosurface.

\paragraph{Conservative level contour-based interface reconstruction, CLCIR}
\label{sec:clcir}

\begin{figure}[htbp]
\begin{center}
\includegraphics{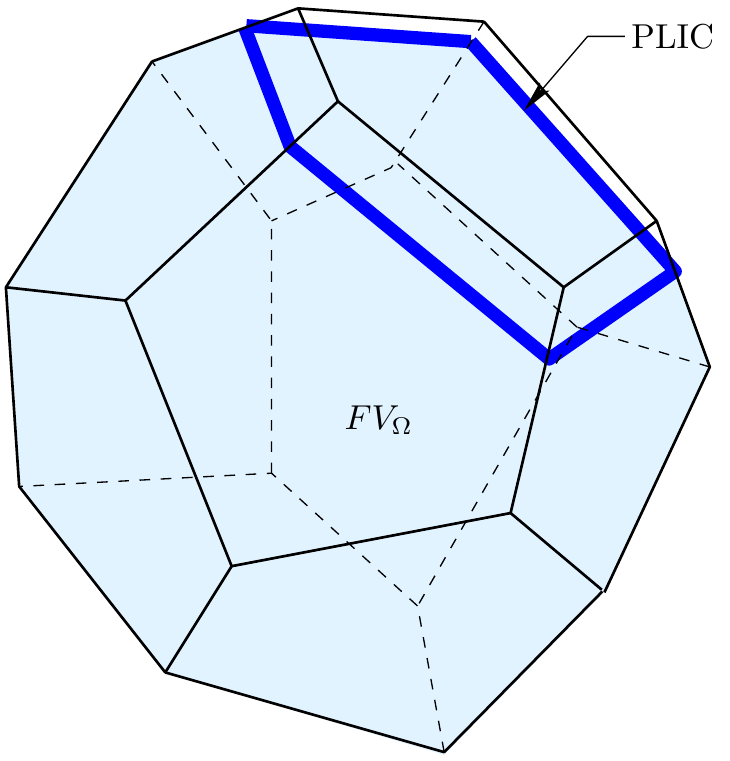}
\caption{PLIC interface that encloses the fluid volume $FV_\Omega$ in the cell.}
\label{isosurface-plic}
\end{center}
\end{figure}

\begin{figure}[htbp]
\begin{center}
\includegraphics{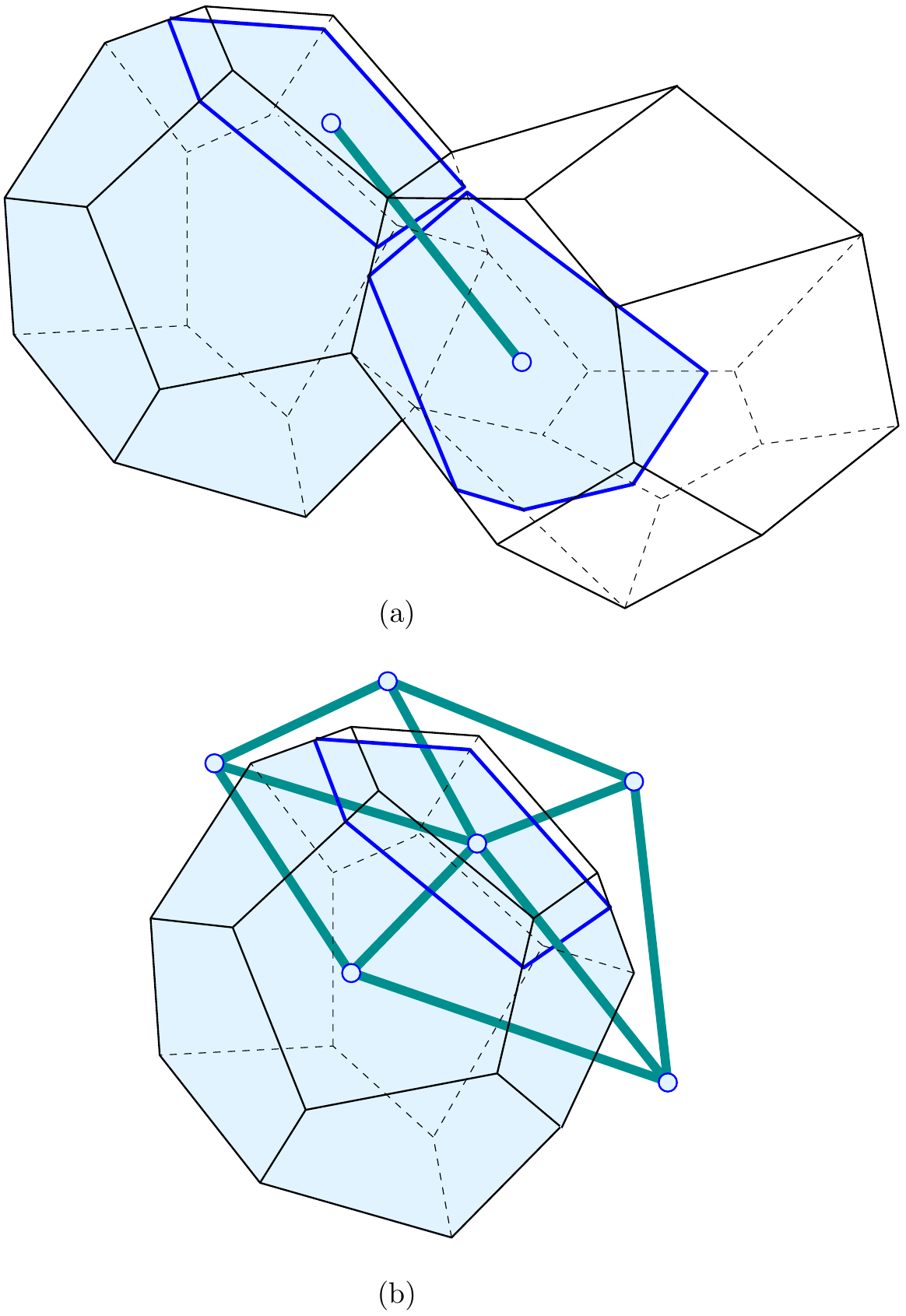}
\caption{Sketch of the CLCIR method. (a) Connection to an adjacent cell by joining the geometric centers \jj{of the set of vertices} of the corresponding PLIC interfaces. (b) Resulting triangulated surface involving adjacent interfacial cells.}
\label{isosurface-plic2}
\end{center}
\end{figure}

After applying the ELCIR method and  the volume conservation procedure on each interfacial cell to obtain the position of the PLIC interface (Fig.~\ref{isosurface-plic} shows the reconstructed PLIC interface that encloses  the fluid volume corresponding to the value of $F$ in the cell),
the vertices of the extended triangulated surface are translated to the geometric centers \jj{of the set of vertices} of the corresponding PLIC interfaces (see the example in Fig.~\ref{isosurface-plic2}(a)), resulting in a conservative extended triangulated surface (Fig.~\ref{isosurface-plic2}(b)) from which \vec{n} is again obtained by averaging. A filter  like that used in the above section for orientation differences higher than 1.2 rad is also applied here. 

\subsubsection{Swartz interface reconstruction, SWIR}
\label{sec:swir}

The unit normal vector  $\vec{n}$ is obtained  using a procedure based  on the iterative second-order interface  reconstruction method of \citet{swartz89}, which uses the fact that for a pair of certain neighbor interfacial cells there exists a``common orientation" that can be obtained from the fluid volume fractions in the two cells. Neighbor cells are those sharing at least one vertex. 
 For a given interfacial cell, $\vec{n}$ is computed by averaging its common orientations with some of the neighbor interfacial cells, as described below.
The method implemented in this work is similar, although with some considerations, to the variants described by \citet{dyadechko05} and \citet{garimella05}, which are somewhat different from the previous description given  by \citet{mosso96}. 

For a given interfacial cell, two types of iterations (inner and outer) are performed as follows. The inner iteration is applied over every valid neighbor interfacial cell 
until the common orientation reaches a prescribed tolerance, while the outer iteration involves all its valid neighbor interfacial cells as follows.

\paragraph{Inner iteration}

A pair of neighbor interfacial cells are valid to perform the inner iteration if their interface orientations obtained in the previous outer iteration 
differ by less than $45^\circ$. Every valid pair must be iterated over as follows until the difference of the \jl{common orientation} is below $10^{-6}$ rad.
\begin{enumerate}
\item Connect the geometric centers \jj{of the set of vertices} of the paired PLIC interfaces.
\item Compute the perpendicular to the common segment joining the geometric centers.
\item Use this perpendicular to estimate the common orientation  and locate the PLIC interfaces to conserve the associated fluid volume fractions.
\end{enumerate}

\paragraph{Outer iteration}

The interface orientation computed in the previous outer iteration  is updated by the new orientation 
if  it differs by less than $30^\circ/I_\mathrm{out}$, where $I_\mathrm{out}$ is the number of repetitions performed in the  outer iteration,
and the PLIC interface is located to match the corresponding fluid volume fraction.
It should be mentioned that the update of an interface orientation is made only after the new orientation is computed
over all the interfacial cells.
 The outer iteration is repeated until the angle between the previous ($I_\mathrm{out}-1$) and current ($I_\mathrm{out}$) computed interface orientation reaches a value lower than a prescribed tolerance or a maximum $n_\mathrm{out}$ of  repetitions
(the prescribed tolerance and the maximum number of repetitions of the outer iteration can be set by the \texttt{gVOF} user). 
The previous interface orientation values for the first outer iteration ($I_\mathrm{out}=1$) are obtained using the LSGIR method from Section~\ref{sec:lsgir}.

\subsubsection{Least squares fit interface reconstruction, LSFIR}
\label{sec:lsfir}

The unit normal vector  $\vec{n}$ is obtained using a version applied to arbitrary polyhedral grids of the least-squares fit interface reconstruction methods presented by \citet{scardovelli03} and \citet{aulisa07}.

The interface is first reconstructed using the LSGIR method of
Section~\ref{sec:lsgir} and the geometric center \jj{of the set of vertices} of every
reconstructed PLIC interface is computed\jj{. Using the centroid of the polygonal PLIC interface, like in \cite{aulisa07}, instead of the above geometric center provides similar results at the cost of a higher CPU time.
} The interface orientation of a given interfacial cell is updated from the orientation of the plane passing through its PLIC geometric center $\vec{x}$ that minimizes the distances $\delta_k$ to every PLIC geometric center $\vec{x}_k$ of neighbor interfacial cell $k$. The solution is obtained using a least-squares procedure that minimizes the functional $H$ defined by
\begin{equation}
H=\sum\limits_k^{n_\mathrm{valid}} w_k \delta_k^2,
\end{equation}
where $n_\mathrm{valid}$ is the total number of valid neighbor interfacial cells and the weight $w_k=1/|\vec{x}_k-\vec{x}|^{2.5}$ (the exponent 2.5 is roughly obtained by trial-and-error analysis). A neighbor interfacial cell is valid if it satisfies the same condition imposed for the inner iteration in the SWIR method.
The LSFIR method can be applied using the outer iteration described in  Section~\ref{sec:swir}.

\subsection{Fluid advected through cell faces}
\label{sec:advection}

\begin{figure}[htbp]
\begin{center}
\includegraphics{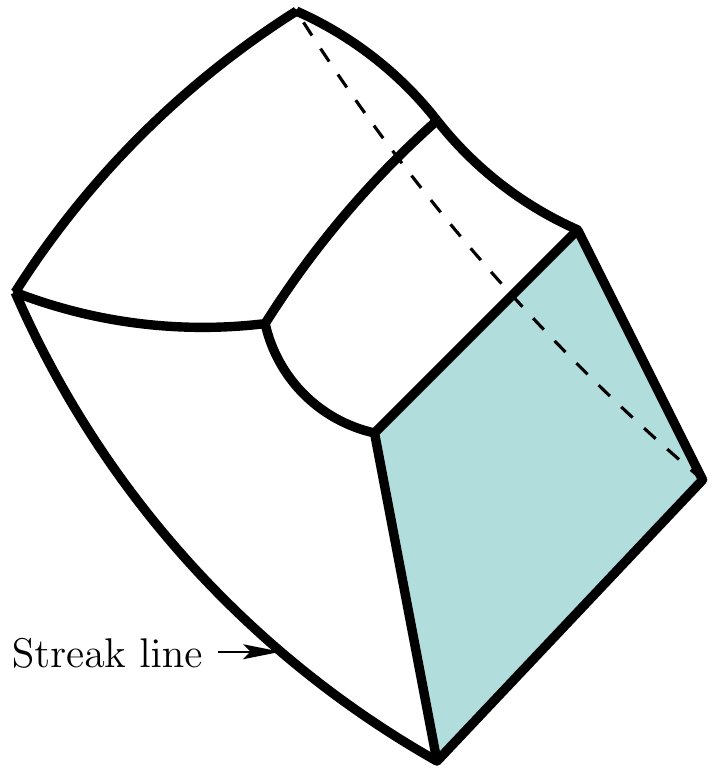}
\caption{Example of a flux region constructed on a cell face (shaded region).}
\label{newflux-pol}
\end{center}
\end{figure}

The following procedure is applied
to determine the net volume of fluid advected out of the cell (first integral in Eq.~(\ref{advec})).
On each polygonal cell face $j$ of $I$ vertices, the flux region of volume
  \begin{equation}
    V_{d_j} = \left( \vec{u}_j^{n+\frac{1}{2}} \cdot \vec{n}_{j} \right ) A_{j} \Delta t,
    \label{a-donacion}
  \end{equation}
where $\vec{n}_{j}$ is the unit  vector normal to it
pointing out of the cell, $\vec{u}_j^{n+\frac{1}{2}}$ is the velocity vector at its center and intermediate time $t^{n+\frac{1}{2}}= \frac{1}{2} \left( t^n + t^{n+1} \right)$, and $A_{j}$ is its area, is delimited by 
$I+2$ flux region faces (see the example of Fig.~\ref{newflux-pol} corresponding to a quadrilateral cell face (shaded region)), one of which is planar and coincides with the considered cell face (first flux region face) and the other $I+1$ flux region faces  may
be non-planar, self-intersected  and have curved edges: $I$ faces that have a common edge
with the first flux region face (each one of these faces can be
considered as the streak surfaces emanating from the corresponding
common edge) and the last one is opposite to the first one.
To make this complex flux 
region computationally affordable, its edges and faces will be approximated by 
straight lines and planar surfaces, respectively. Hereafter
this discrete flux region will be denoted as flux polyhedron
$\Omega_j^p$ and may be non-convex and even have self-intersecting
faces.  

Three different schemes, the edge and face matched flux polyhedron    advection schemes described respectively in Sections~\ref{sec:emfpa} and \ref{sec:fmfpa}, and the non-matched flux polyhedron advection scheme described in Section~\ref{sec:nmfpa}, are used to construct the flux polyhedron $\Omega^p_j$.
Hereafter, the unsplit advection procedure implemented in 3D in combination with  the construction of edge-matched, face-matched, or non-matched flux polyhedra will be referred to as EMFPA, FMFPA or NMFPA methods, respectively. 

\begin{figure}[htbp]
\begin{center}
\includegraphics{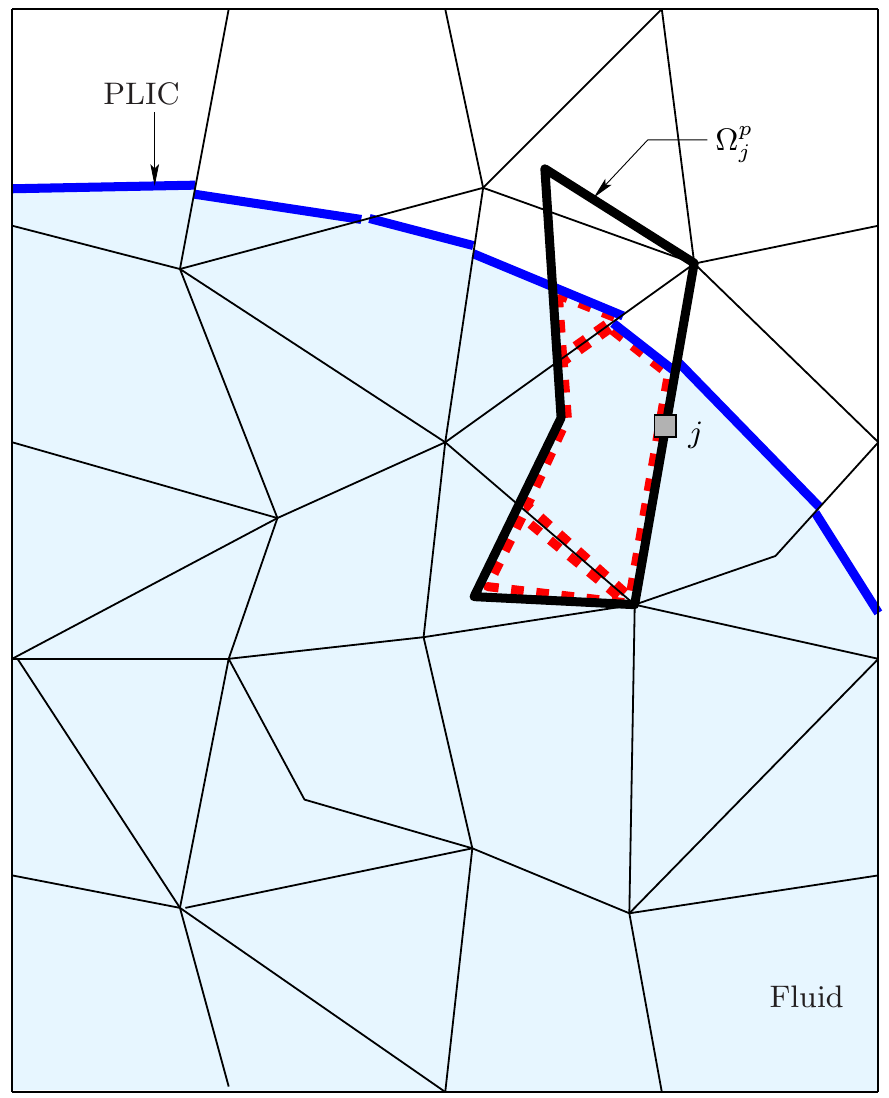}
\caption{2D example for the computation of $V_{F_j}$. The truncated fluid regions are highlighted with red dashed lines.}
\label{donating-flux-vfj}
\end{center}
\end{figure}

The volume of fluid $V_{F_j}$ advected through the cell face $j$,
 which is taken to be positive when the fluid leaves the cell and negative otherwise, will depend on the shapes of $\Omega^p_j$ and the fluid regions determined by  the reconstructed PLIC interfaces in cells intersected by $\Omega^p_j$. Note that
 donating regions from several cells need to be considered. 
This is the most expensive task, in which recursive intersections between half-spaces and the generally non-convex $\Omega_j^p$ are required to compute $V_{F_j}$ (the resulting truncated fluid regions are highlighted with red dashed lines in the 2D example of Fig.~\ref{donating-flux-vfj}). The use of the non-convex tools proposed in \cite{lopez19} allows to perform  these recursive intersections efficiently without the need to additionally decompose $\Omega_j^p$, which will have a very positive impact on the computational efficiency of the whole advection scheme. 
The details of the procedure used to perform the recursive intersection operations will be published elsewhere.

Finally, the new volume of fluid fraction at
$t^{n+1}$  is obtained from Eq.~(\ref{advec}) as 
\begin{equation}
  \label{fnext}
  F^{{n}+1}=\left[F^{n}\left(1+\frac{V_{d_T}}{2V_\Omega} \right) - \frac{V_{F_T}}{V_\Omega}\right]\left( 1-\frac{V_{d_T}}{2V_\Omega} \right)^{-1},
\end{equation}
where 
\begin{equation}
V_{d_T}=\sum_{j} V_{d_j}
\end{equation}
is the total net flux volume in the cell (second integral in Eq.~(\ref{advec})),  and 
\begin{equation}
V_{F_T}=\sum_{j}  V_{F_j}
\label{eq:vft}
\end{equation} 
is the total net volume of fluid that
leaves (or enters) the cell (first integral in Eq.~(\ref{advec})), and  the summations extend over all the faces of the considered grid cell.

The \texttt{gVOF} \jjj{package} does not use additional algorithms to redistribute the very small liquid volumes which are out of the limits given by the volume fraction. Instead, the updated $F^{n+1}$ value is finally adjusted by simply making
\begin{equation}
F^{n+1}=\mbox{max}\left[\mbox{min}(F^{n+1},1.0),0.0\right].
\end{equation}

\subsubsection{Edge-matched flux polyhedron}
\label{sec:emfpa}

\begin{figure}[htbp]
\begin{center}
\includegraphics{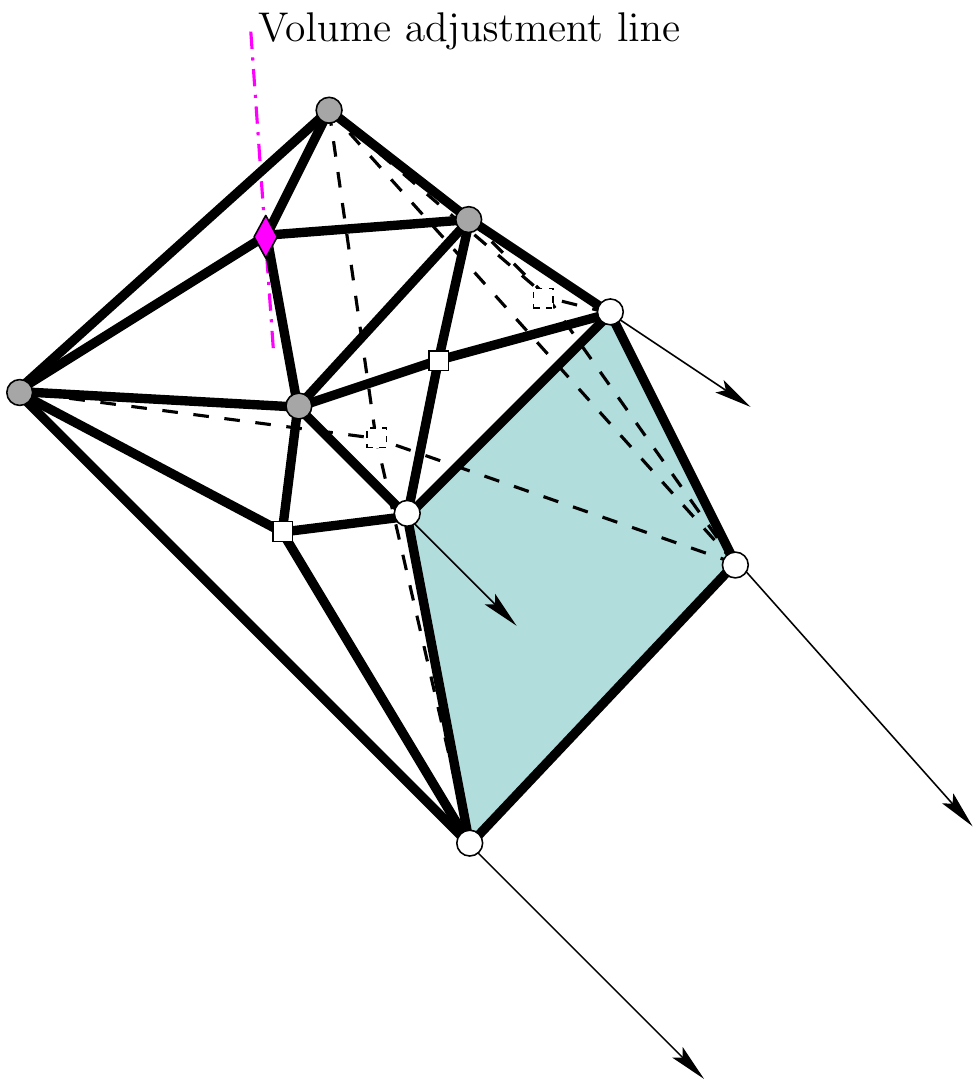}
\caption{Example of a face-matched flux polyhedron construction. The arrows represent the velocity vectors on the cell face vertices.}
\label{newflux-pol-emfpa}
\end{center}
\end{figure}

The procedure used to
construct an edge-matched flux polyhedron  
is analogous to that of the 2D version
proposed by \citet{lopez04}, in which the flux
polygons on cell edges that have a vertex in common are constructed by having an edge with a common orientation, thus
avoiding over/underlaps between flux polygons, provided that the time interval is low enough. 
In 3D, the
proposed procedure also avoids over/underlaps between flux polyhedra on cell faces with a common vertex.
The flux region corresponding to a given cell face of $I$ vertices is delimited by a polyhedron $\Omega^p_j$ with $5I+1$ faces and $3I+1$ vertices as described below (see the example of Fig.~\ref{newflux-pol-emfpa}).

The first face of $\Omega^p_j$ is matched to the considered cell face. It will be assumed that all the computational cells are composed by planar faces.  Each vertex $i$ (white circles in Fig.~\ref{newflux-pol-emfpa}) of the first face is transported back in time along its streak line by solving 
\begin{equation}
\frac{\mathrm{d}\vec{x}}{\mathrm{d}t} = \vec{u}.
\label{streak-line}
\end{equation}
The following approximation is used in this work to integrate Eq.~(\ref{streak-line}) from $t=t^{n+1}$ to $t^{n}$ with $\vec{x}(t)=\vec{x}_i$ at $t=t^{n+1}$
\begin{equation}
\vec{x}_{I+i}=\vec{x}_i-\Delta t \vec{u}_i^{n+\frac{1}{2}},
\label{streak-line2}
\end{equation}
where $\vec{x}_{I+i}$ is the position vector of the transported face
vertex $i$ at time $t^n$ (gray circles in Fig.~\ref{newflux-pol-emfpa}). 
All the results presented in this work were obtained with $\vec{u}_i^{n+\frac{1}{2}}$ defined at the intermediate time
$t^{n+\frac{1}{2}} = \frac{1}{2}(t^n+t^{n+1})$.

Each edge of the first face  and the approximated streak lines corresponding to the back-traced of its vertices delimit a flux face which is generally non planar. The next step is to replace this generally non-planar face by
four triangles using its geometric center \jj{obtained by simple averaging of vertex positions} (white square in Fig.~\ref{newflux-pol-emfpa}), as shown in the sketch of Fig.~\ref{newflux-pol-emfpa}.

The remaining flux region face is delimited by  the $I$ back-traced vertices, producing a generally non planar face which is also triangulated to obtain a surface of $I$ triangles. Instead of directly using the geometric center of the $I$ back-traced vertices in the triangulation, which would produce a flux region without the required volume, this geometric center is translated in the direction normal to the face (volume adjustment line indicated in Fig.~\ref{newflux-pol-emfpa}) so that the volume of the flux polyhedron coincides with $V_{d_j}$ (the final point is depicted with a rhombus symbol in Fig.~\ref{newflux-pol-emfpa}). This approach is like that used in \cite{owkes14}.

\subsubsection{Face-matched flux polyhedron}
\label{sec:fmfpa}

\begin{figure}[htbp]
\begin{center}
\includegraphics{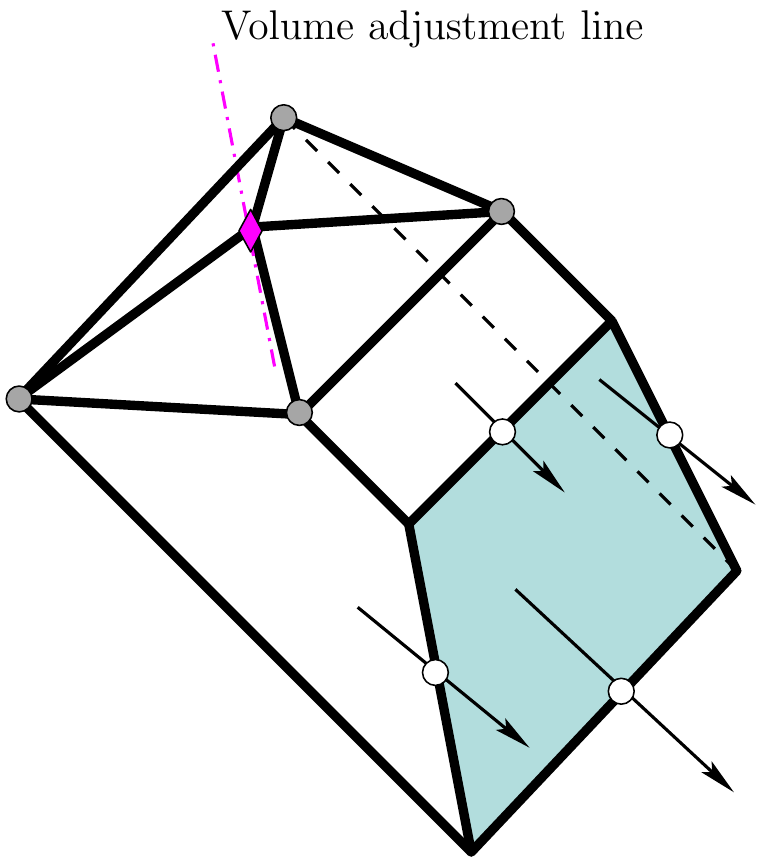}
\caption{Example of a face-matched flux polyhedron construction. The arrows represent the velocity vectors on the centers of the cell face edges.}
\label{newflux-pol-fmfpa}
\end{center}
\end{figure}
A face-matched flux polyhedron $\Omega^p_j$ is constructed using a new version of the procedure proposed in \cite{hernandez08}.
This new procedure  produces flux polyhedra defined by $2I+1$ faces and vertices (see the example of Fig.~\ref{newflux-pol-fmfpa}), which obviously is  computationally less expensive compared to the
flux polyhedra produced by the EMFPA method  (this can be clearly seen by comparing the examples in Figs.~\ref{newflux-pol-emfpa} and \ref{newflux-pol-fmfpa}). 
Following the work in \cite{hernandez08}, the basic idea is to ensure that the flux polyhedra constructed on cell faces with a common edge have a face with a common orientation. The details of this face-matched flux polyhedron construction are given below.

As in Section~\ref{sec:emfpa}, the first face of $\Omega^p_j$ coincides
with the considered cell face. Each face of $\Omega^p_j$ that has a common edge with the first face is made parallel to the velocity vector at the center of the common edge (white circles in Fig.~\ref{newflux-pol-fmfpa}). 
The vertices of the remaining face of the flux region (gray circles in Fig.~\ref{newflux-pol-fmfpa}), which are initially obtained by back tracing along the lines of intersection of the planes that contain the  faces  contiguous to the cell face, determine a  generally non-planar flux region face which is adjusted by triangulation using the same volume conservation enforcement mentioned in the above section. This approach is found to be more robust than that used by \citet{hernandez08} to impose the volume constraint of $\Omega^p_j$, although also slightly more complex geometrically (in \cite{hernandez08}, $\Omega^p_j$ is defined by only $I+2$ faces and $2I$ vertices).
As it was mentioned, 
 the FMFPA method, either the original or the new version, avoids over/underlaps between flux polyhedra constructed on cell faces with a common edge, although over/underlaps may still occur between flux polyhedra constructed on cell faces with only a common vertex.

\subsubsection{Non-matched flux polyhedron}
\label{sec:nmfpa}

\begin{figure}[htbp]
\begin{center}
\includegraphics{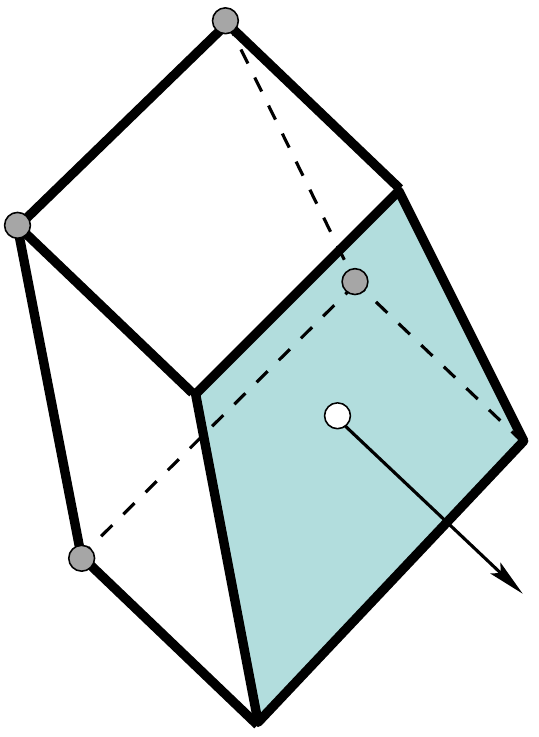}
\caption{Example of a non-matched flux polyhedron construction. The arrow represents the velocity vector at the face center.}
\label{newflux-pol-nmfpa}
\end{center}
\end{figure}

A non-matched flux polyhedron $\Omega^p_j$ is constructed using solely the velocity vector defined at the center of face $j$. The procedure described below, which could be considered as an extension to 3D arbitrary grids of the procedure described by \citet{rider98}, produces a flux polyhedron defined by $I+2$ faces and $2I$ vertices (see the example of Fig.~\ref{newflux-pol-nmfpa}), which further reduces the geometric complexity of the methods described in Sections~\ref{sec:fmfpa} and \ref{sec:emfpa}. 
\citet{liovic05} developed a similar procedure which was applied to cubic grids.  
Note that with this procedure the absence of over/underlaps between flux polyhedra constructed on cell faces with some vertex (just one or both) in common is not warranted. 

As in Sections~\ref{sec:emfpa} and \ref{sec:fmfpa}, the first face of $\Omega^p_j$ is made coincident with the considered cell face.
Each vertex  of the first face is transported back in time from an expression like that of Eq.~(\ref{streak-line2}), but using a unique velocity vector  \jl{(the face center velocity vector)} for all the other face vertices. 
Because of \jl{this}, (1) each edge of the first face  and the approximated streak lines corresponding to the back-traced of its vertices delimit a flux face which is planar, (2) the $I$ back-traced vertices also delimit  a planar flux region face, and (3) the resulting volume of the flux polyhedron must coincide with the volume $V_{d_j}$. Therefore, the triangulation of non-planar faces and  the additional step used in the EMFPA and FMFPA methods to impose the volume constraint given by  $V_{\Omega_j^p}=V_{d_j}$ are not required.

\section{Description of the routines}
\label{sec:routines}

Below, the main routines implemented to solve Eq.~(\ref{vof})  are presented (a brief description of each one is summarized in Table~\ref{tab:routines}). 
\begin{table}[htp]
\caption{\jl{Brief} description of the main routines included in the \texttt{gVOF} \jjj{package}.}
\begin{center}
\scalebox{0.95}{\footnotesize \begin{tabular}{lp{10.5cm}}
\hline
Name & Description \\
\hline
\texttt{defgrid} & constructs the grid \\
\texttt{neigbcell} & obtains the list of cells that share at least one node with a given cell \\
\texttt{compgrid} & computes different geometric parameters related with the cells and faces of the grid \\
\texttt{printgrid} & writes the geometry of the grid to a file in \texttt{VTK} format \\
\texttt{initfgrid} & initializes the fluid volume fraction in the grid cells \\
\texttt{taggrid} & interpolates the fluid volume fraction to the grid nodes and tags the nodes, faces and cells of the grid \\
\texttt{clcir} & reconstructs the PLIC interfaces using the isosurface-based CLCIR method \\
\texttt{elcir} & reconstructs the PLIC interfaces using the isosurface-based ELCIR method \\
\texttt{llcir} & reconstructs the PLIC interfaces using the isosurface-based LLCIR method \\
\texttt{lsgir} & reconstructs the PLIC interfaces using the gradient of the fluid volume fraction distribution \\
\texttt{swir} & reconstructs the PLIC interfaces using the iterative SWIR method \\
\texttt{lsfir} & reconstructs the PLIC interfaces using the LSFIR method \\
\texttt{recerr} & computes the interface reconstruction errors \\
\texttt{printplic} & writes the PLIC interfaces to a \texttt{file} in \texttt{VTK} format \\
\texttt{printvoxel} & remeshes the main grid onto a Cartesian grid (voxel grid) preserving the fluid volume and writes the computed fluid volume fraction at the voxel grid nodes to a file in \texttt{VTK} format \\
\texttt{faceflux} & computes the volume of fluid advected during a time interval through the grid faces \\
\texttt{emfp} & constructs a flux polyhedron using the EMFPA method \\
\texttt{fmfp} & constructs a flux polyhedron using the FMFPA method \\
\texttt{nmfp} & constructs a flux polyhedron using the NMFPA method \\
\texttt{vofadv} & advances the volume fraction distribution to the next time step \\ 
\hline
\end{tabular}}
\end{center}
\label{tab:routines}
\end{table}%
It must be mentioned that \texttt{gVOF} requires the  \texttt{VOFTools} v5  \cite{lopez19,lopez20} and  \texttt{isoap}  \cite{lopez20b,mendeleylink} external libraries.
 For the sake of brevity, the calling convection and 
the list of input and output parameters for each routine described hereafter\jj{, as well as others auxiliary routines not mentioned here,} are left to be provided in the user manual included in the supplied software package.

\subsection{Grid construction}

\begin{figure}
\begin{center}
\includegraphics{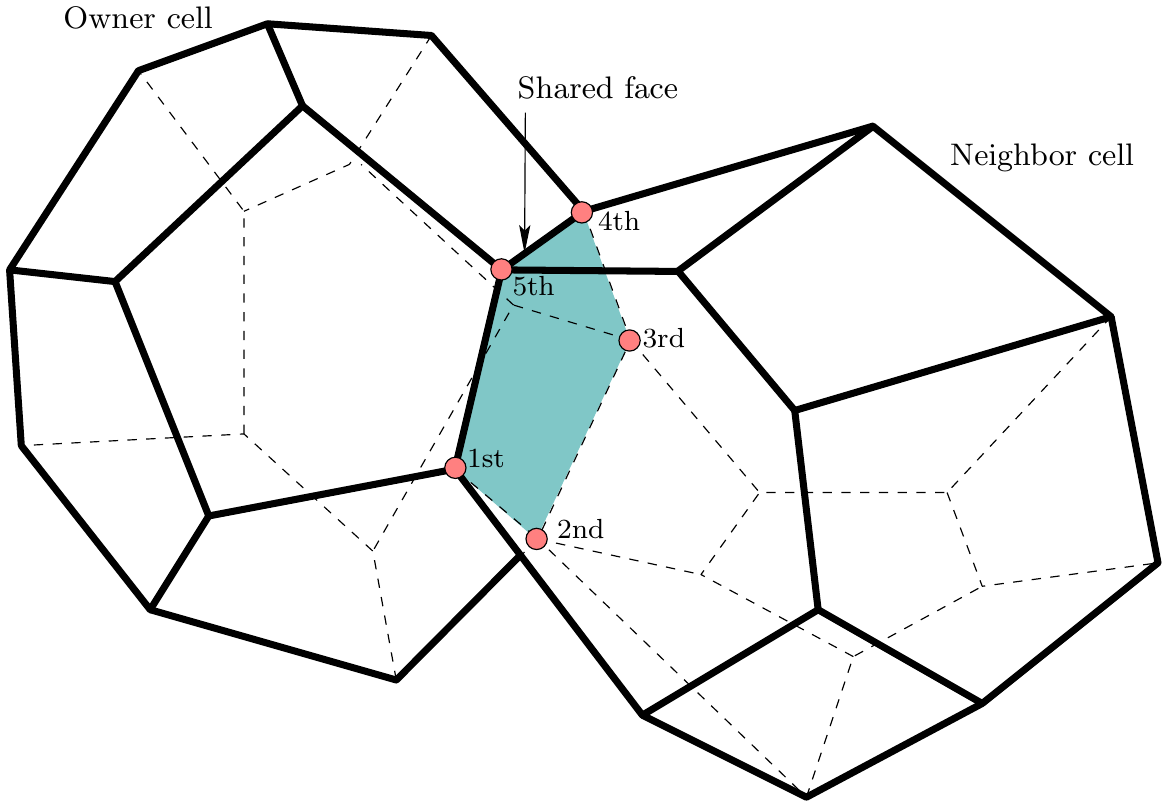}
\caption{Owner and neighbor cells that share a common face, with indication of the sequential order of  nodes that define the shared face.}
\label{owner-face-neighbour}
\end{center}
\end{figure}

The  \texttt{defgrid} routine has been implemented to  construct the computational grid.
The  grid is defined by a  structure like that used by other codes, such as \texttt{OpenFOAM} \cite{openfoam}. The computational domain is divided into discrete non-overlapping polyhedral cells to form the grid. Each polyhedral cell is defined by a set of nodes and is bounded by planar polygonal faces. These faces may be internal or belong to a grid boundary.
An internal face is shared by two cells, which are referred to \jl{in} the example of Fig.~\ref{owner-face-neighbour} as owner and neighbor cells, while  a face  located in the domain boundary belongs to only one cell. 
The nodes of each face are arranged  sequentially  so that the vector joining two consecutive nodes leaves the face to the left (counterclockwise order) when viewed from outside the owner cell (see the node order illustrated in the example of Fig.~\ref{owner-face-neighbour}).

The \texttt{neigbcell} routine has been implemented to obtain the list of cells that share at least one node with a given cell and the \texttt{compgrid} routine has been implemented to compute different geometric parameters related with the cells and faces of the computational grid, such as sizes, areas, volumes, \jj{geometric centers}, or orientations, among others.

The \texttt{printgrid} routine has been implemented to write the geometry of the constructed grid to a file in \texttt{VTK} (visualization toolkit \cite{vtk}) format which can be visualized by using, for example, the \texttt{ParaView} program \cite{paraview}.
\begin{figure}
\begin{center}
\includegraphics{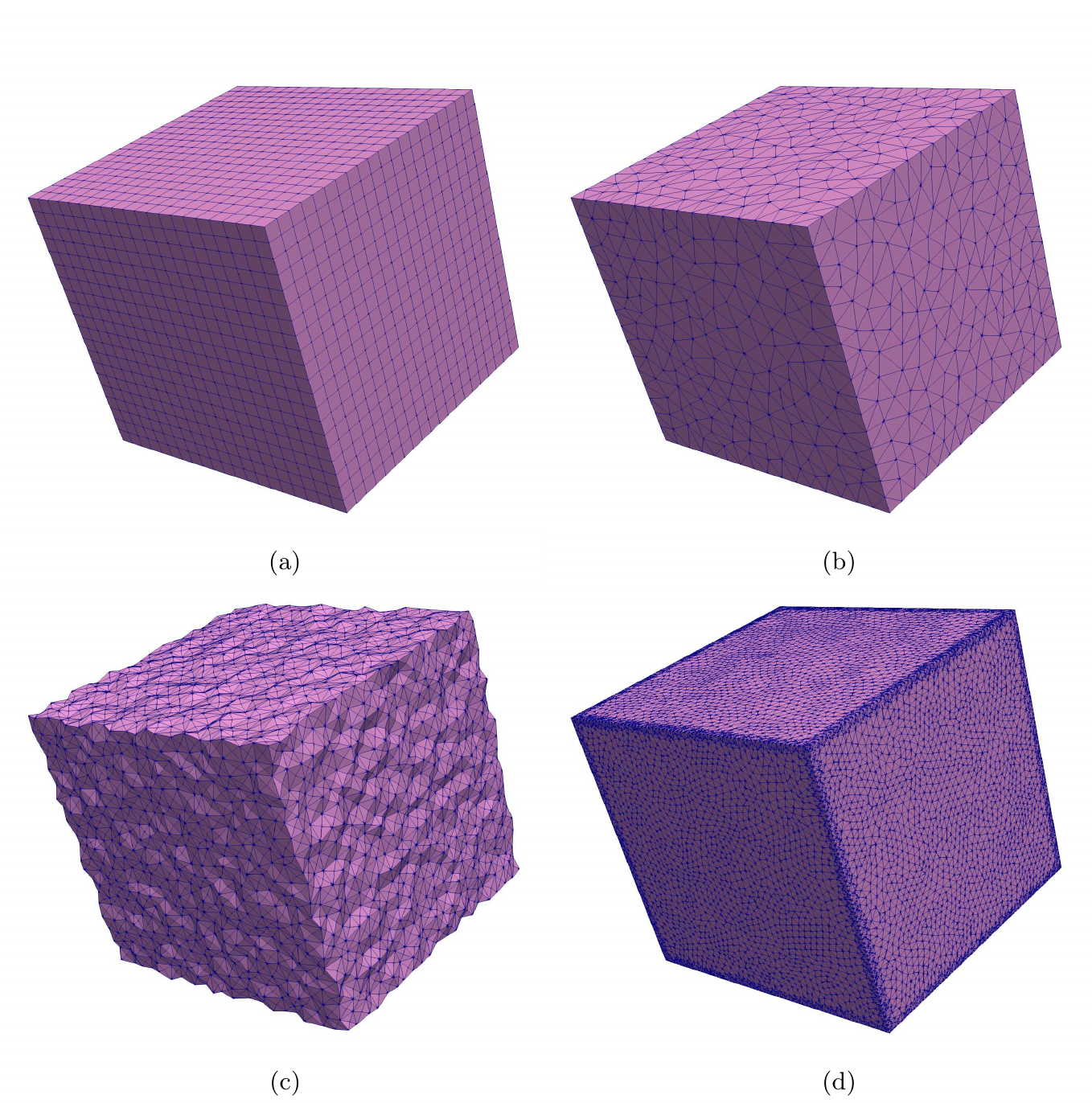}
\caption{Grids with around $20^3$ cells in a unit domain (convex cells on the top and non-convex cells on the bottom): (a) uniform grid with rectangular parallelepipedal cells and non-uniform grids with (b) convex  and non-convex  cells ((c) distorted cubic and (d) irregular polyhedrical cells).}
\label{fig:grids_type-paper}
\end{center}
\end{figure}
As an example, Fig.~\ref{fig:grids_type-paper} shows  four  different types of grids included in the supplied software with around $20^3$ cells in a unit domain.

\subsection{Fluid volume fraction initialization}

The \texttt{initfgrid} routine has been implemented to initialize the fluid volume in the grid cells using an accurate refinement procedure \cite{lopez09,lopez19,lopez20}, which allows to obtain the volume fraction of the fluid contained in a convex or non-convex polyhedral cell. The shape of the fluid interface is defined by implicit external functions.  
The degree of refinement (given by the number of divisions along each coordinate axis of the  Cartesian cell superimposed to a given cell  in which the fluid volume fraction is computed) can be increased by the users to improve the accuracy of the initialization. 
Detailed assessments of the accuracy of this initialization procedure can be found in  \cite{lopez09,lopez16b,lopez19}.

\subsection{Grid tagging}

\begin{figure}
\begin{center}
\includegraphics{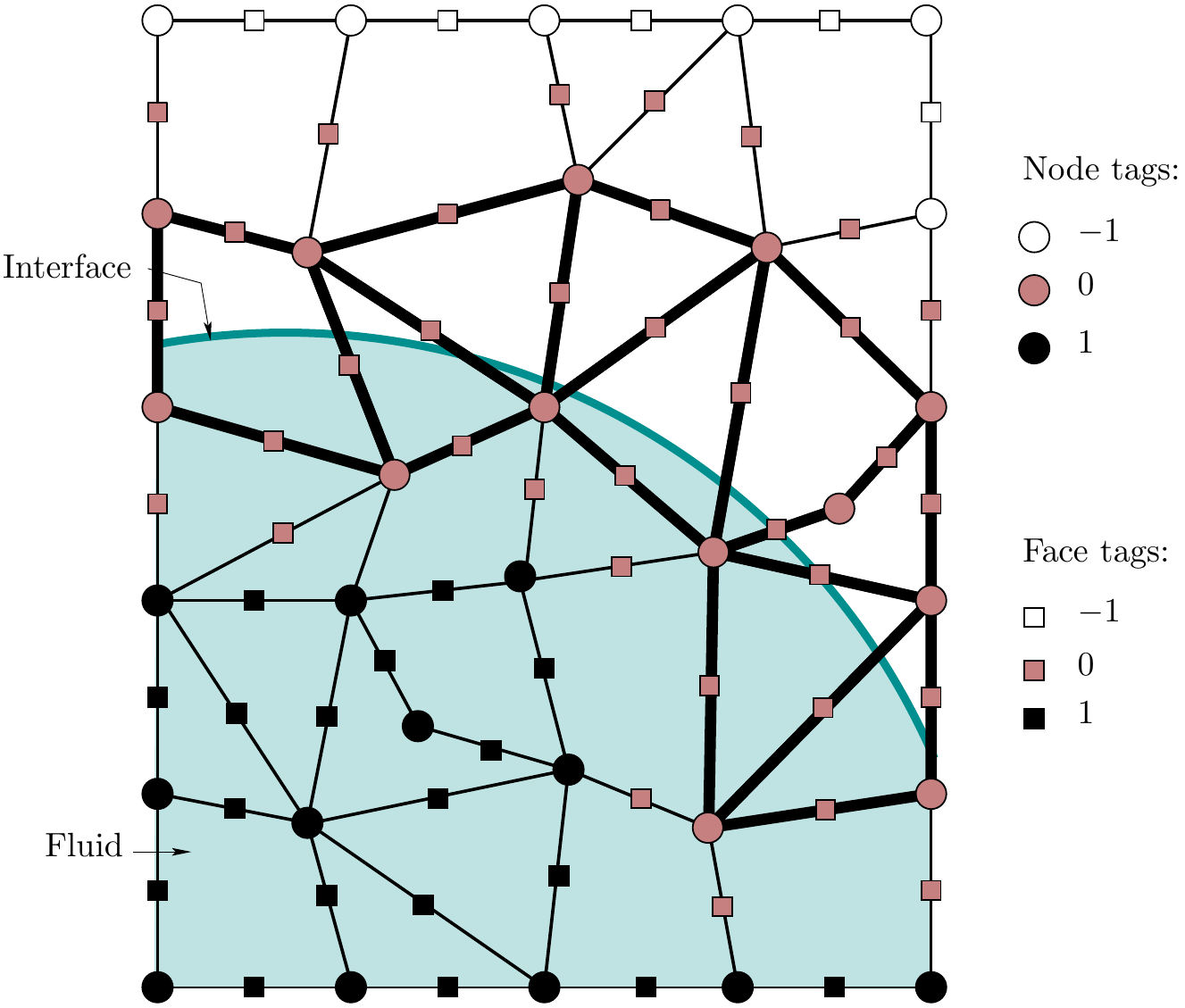}
\caption{2D example of node (circle symbols) and face (square symbols) tagging. The interfacial cells are highlighted with thick lines.}
\label{grid-tags-1}
\end{center}
\end{figure}

\begin{figure}
\begin{center}
\includegraphics{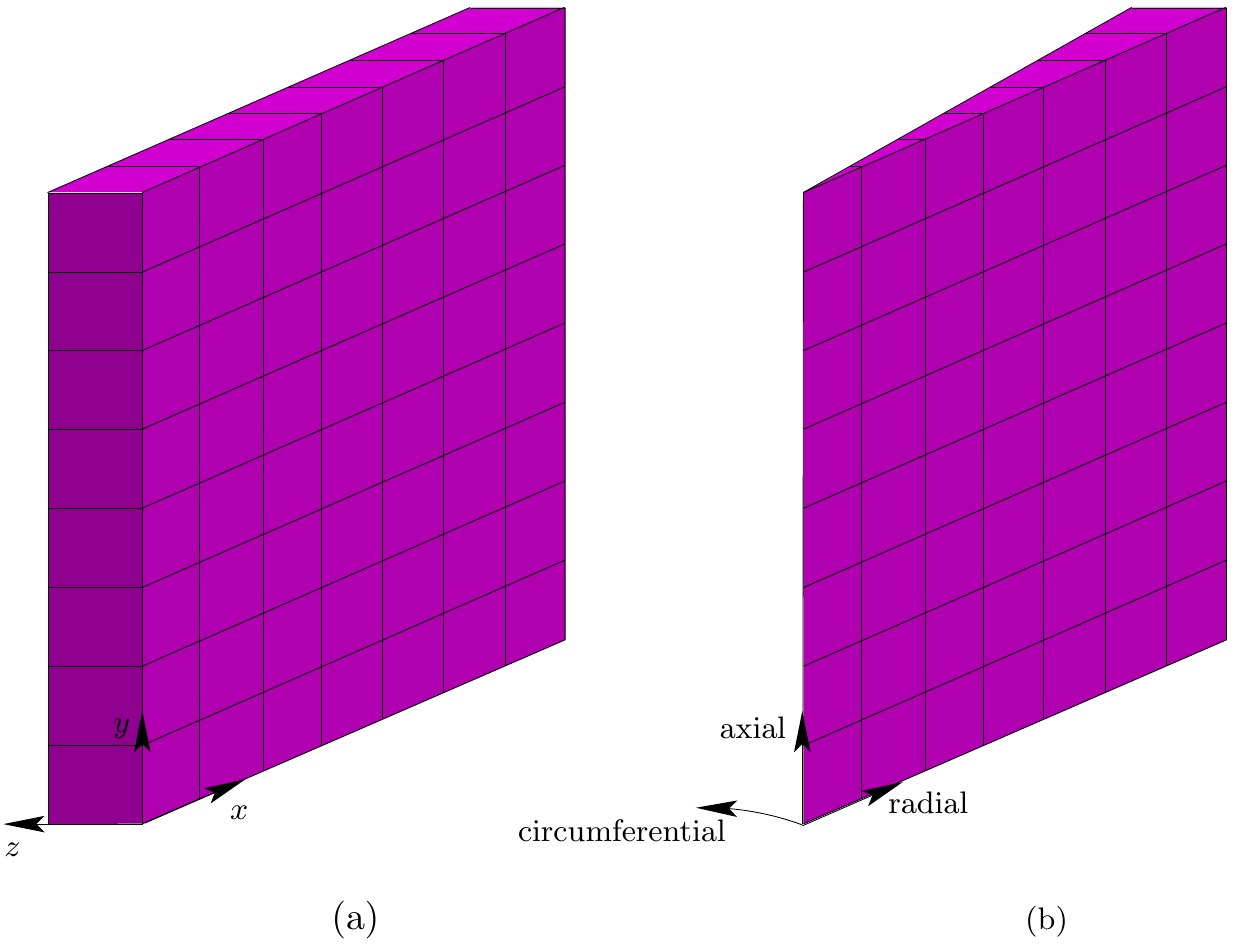}
\caption{Examples of 3D grids used to solve (a) 2D and (b) axisymmetric problems.}
\label{2d-axisymmetric-grids}
\end{center}
\end{figure}

The nodes, faces and cells of the grid are tagged as described below to improve the computational efficiency of the implemented interface reconstruction and fluid advection algorithms (note that these algorithms only need to be applied in a region close to the interface).
The \texttt{taggrid} routine first  computes the volume of fluid fraction at every grid node ($F^*$) using an inverse distance weighting interpolation from the volume fractions  in the cells that share it. 
To illustrate the tagging procedure used in \texttt{gVOF}, the 2D example of Fig.~\ref{grid-tags-1} is used for clarity, although the \jjj{package} works on 3D. 
As it will be seen in Section~\ref{sec:2ddef}, the users can simulate 2D (see the example in Fig.~\ref{2d-axisymmetric-grids}(a)) or axisymmetric (Fig.~\ref{2d-axisymmetric-grids}(b)) problems by using 3D grids with only one cell along one of the coordinates directions (circumferential direction for axisymmetric cases). 
The tags are assigned to nodes, faces and cells as, respectively,
\begin{equation*}
\mbox{Node tag} = \left\{
\begin{split}
-1&\,\,\,\,\,\,\mbox{if}\,\,\,F^*<\epsilon, \\
1&\,\,\,\,\,\,\mbox{if}\,\,\,F^*>(1-\epsilon) \,\,\,\mbox{or} \\
0&\,\,\,\,\,\,\mbox{otherwise}, \\
\end{split}
\right.
\end{equation*}
\begin{equation*}
\mbox{Face tag} = \left\{
\begin{split}
-1&\,\,\,\,\,\,\mbox{if all its nodes are tagged  with $-1$}, \\
1&\,\,\,\,\,\,\mbox{if all its nodes are tagged with 1} \,\,\,\mbox{or} \\
0&\,\,\,\,\,\,\mbox{otherwise}, \\
\end{split}
\right.
\end{equation*}
and
\begin{equation*}
\mbox{Cell tag} = \left\{
\begin{split}
-1&\,\,\,\,\,\,\mbox{if}\,\,\,F<\epsilon, \\
1&\,\,\,\,\,\,\mbox{if}\,\,\,F>(1-\epsilon) \,\,\,\mbox{or} \\
0&\,\,\,\,\,\,\mbox{otherwise}. \\
\end{split}
\right.
\end{equation*}

\jl{Provided} that the CFL number is limited by a maximum value equal to 1, the following considerations about the above face tagging can be made. If a face is tagged with $-1$, the volumetric flux of fluid must be zero, and therefore, no geometric operation will be carried out to compute the flux of fluid. Note that any flux polyhedron constructed on a face tagged with $-1$  ($\Box$ symbols in Fig.~\ref{grid-tags-1}) will be far enough away from the fluid region. If a face is tagged with 1, the volumetric flux of fluid is approximately obtained in the \texttt{gVOF} \jjj{package} without geometrical operations from the fluid velocity defined at the considered face. Note that any flux polyhedron constructed on a face tagged with 1  ($\blacksquare$ symbols in Fig.~\ref{grid-tags-1}) will be completely inside the fluid region. Otherwise, a flux polyhedron  must be constructed using the EMFPA or FMFPA method (Sections~\ref{sec:emfpa} or \ref{sec:fmfpa}, respectively) at the considered face and a series of recursive truncation operations are needed to obtain the corresponding volumetric flux of fluid.
Also note that only the cells tagged with 0  are interfacial (thick lines in Fig.~\ref{grid-tags-1}) and any of the interface reconstruction routines described in the next section must be applied to all of them to obtain the corresponding PLIC interfaces.

\subsection{Interface reconstruction}

The \texttt{lsgir}, \texttt{llcir}, \texttt{elcir}, \texttt{clcir}, \texttt{swir} and \texttt{lsfir} routines have been implemented to reconstruct the PLIC interfaces using, respectively, the least-squares gradient interface reconstruction of Section~\ref{sec:lsgir}, the three isosurface based interface reconstruction of  Section~\ref{sec:llcir},  the iterative Swartz interface reconstruction of Section~\ref{sec:swir} and the least squares fit interface reconstruction of Section~\ref{sec:lsfir}.

\begin{figure}
\begin{center}
\includegraphics{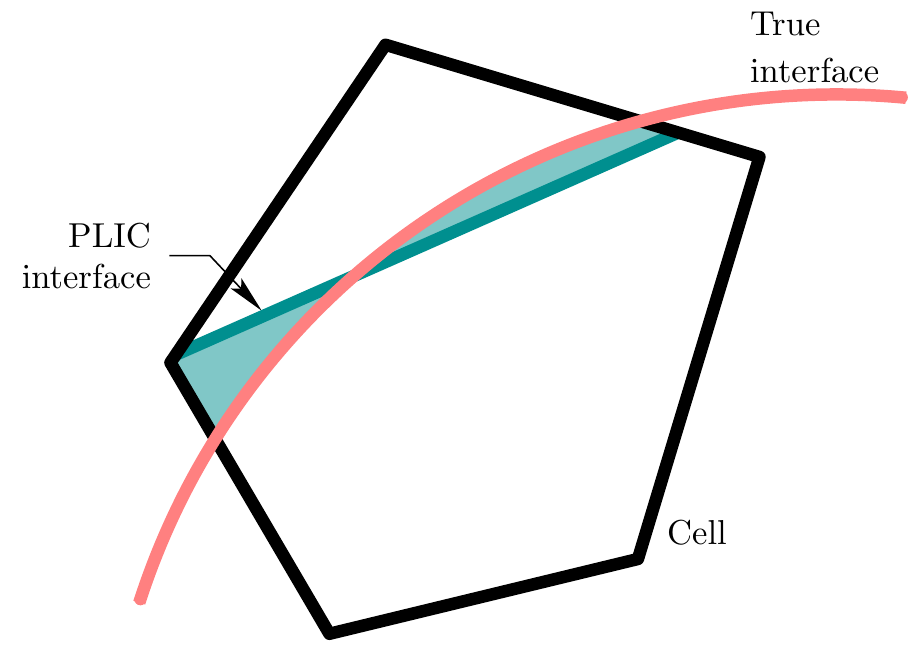}
\caption{2D example with a sketch of the reconstruction error definition (shaded region).}
\label{rec-errors-usermanual}
\end{center}
\end{figure}

The \texttt{recerr} routine has been implemented to compute the errors of the PLIC reconstruction of an interface given by an implicitly defined function.
The reconstruction error is defined within a given cell as the volume enclosed by the true interface and the PLIC interface (shaded region in the 2D example in Fig.~\ref{rec-errors-usermanual}).
To compute this volume, a procedure like that used in the \texttt{initfgrid} routine to compute the initial fluid volume in a cell is used.
Additionally, this routine returns the fluid volume initialization error which is computed as the relative absolute difference between the exact volume of the fluid body and total initialized fluid volume.

\begin{figure}
\begin{center}
\includegraphics{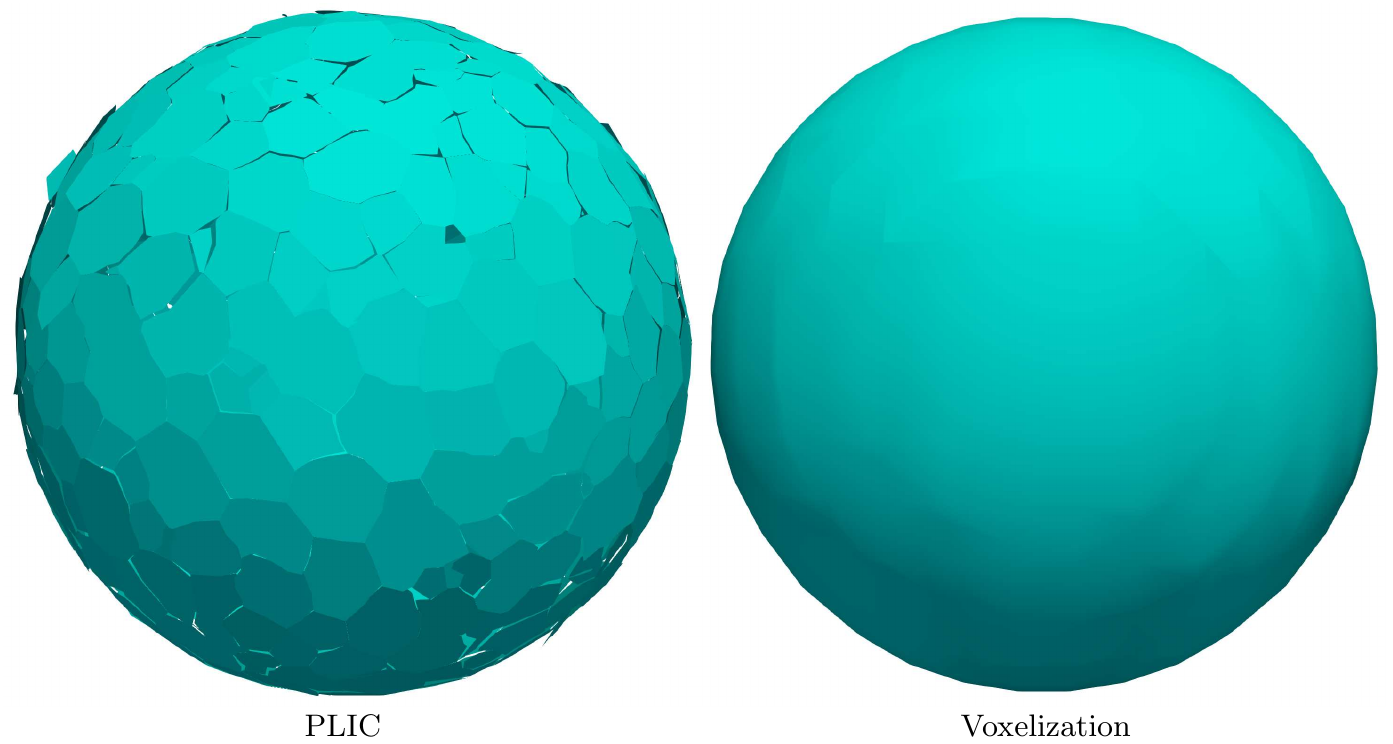}
\caption{Comparison between the spherical interface resulting from PLIC using the \texttt{printplic} routine (left picture) and voxelization using the \texttt{printvoxel} routine (right picture)  in a non-convex irregular polyhedral grid.}
\label{plic-voxel-comparison}
\end{center}
\end{figure}

To visualize the reconstructed interface, the following two routines have also been implemented. 
The  \texttt{printplic} routine writes the reconstructed PLIC interface \jl{into} a \texttt{VTK}-format file. Due to the discontinuities of the PLIC reconstructed in adjacent cells, this interface representation may be unpleasant for visualization purposes. 
To get a smoother visualization, the  \texttt{printvoxel} routine remeshes the main grid onto a Cartesian grid (voxel grid) where a new volume fraction distribution is computed by truncation from the PLIC interface reconstructed on the main grid to exactly preserve the total fluid volume, and writes the computed fluid 
volume fraction at the voxel grid nodes to a \texttt{VTK}-format file, which can be  visualized using programs like \texttt{ParaView} \cite{paraview} (see the comparison in Fig.~\ref{plic-voxel-comparison} between the spherical interface resulting from PLIC using the \texttt{printplic} routine (left picture) and from voxelization using the \texttt{printvoxel} routine (right picture) in a non-convex irregular polyhedral grid). 
The details and assessment of the voxelization procedure implemented in this routine will be published elsewhere.  It must be mentioned that the \texttt{isovtkgrid} routine, which is part of the \texttt{isoap} library \cite{lopez20b,mendeleylink}, is also used in this work to write isosurfaces extracted on arbitrary grids.

\subsection{Fluid advection}

The \texttt{faceflux} routine  has been implemented to compute the volume of fluid advected during a time interval through the grid faces using the EMFPA (Section~\ref{sec:emfpa}),  FMFPA (Section~\ref{sec:fmfpa}) or NMFPA (Section~\ref{sec:nmfpa}) method by calling, respectively, the \texttt{emfp}, \texttt{fmfp} or \texttt{nmfp} subroutine also included in the \texttt{gVOF} \jjj{package}. This routine also returns the volumes of the flux polyhedra constructed at the grid faces.

The \texttt{vofadv} routine has been implemented to advance the volume fraction distribution to the next time step by solving the Eq.~(\ref{fnext}) using the volumetric fluxes computed from the above routines.

\section{Test programs and performance analysis}
\label{sec:testprograms}

Two test programs are included in the supplied software package to assess the performance of the implemented algorithms to solve  different static (Section~\ref{sec:staticresults}) and dynamic (Section~\ref{sec:dynamicresults}) problems with prescribed velocity fields using arbitrary grids. 
The following  grids with convex and non-convex cells are considered (further details of the generation of grids used in this work can be found in \cite{lopez20b} and some of these non-uniform grids are also included in the supplied software package):
\begin{itemize}
\item Uniform grids of rectangular parallelepiped cells.
\item Unstructured grids with tetrahedral cells  obtained using the program \texttt{TetGen} \cite{tetgen}. 
\item Structured grids with non-convex cells  obtained by distorting cubic cells.  Each of the eight corner vertices of every initial uniform cubic cell of size $h$ is randomly moved to the surface
of a sphere with radius $0.25 h$ and centered in the corresponding initial
vertex position. The faces of the distorted cell, which are generally non-planar,
are triangulated by joining its center with two consecutive vertices of each face, which results in a non-convex polyhedron of 14 vertices and 24 triangular faces. 
\item Unstructured grids with non-convex irregular polyhedral cells  obtained with the aid of \texttt{TetGen}  \cite{tetgen} and two \texttt{OpenFOAM}’s programs   \cite{openfoam} (\texttt{tetgenToFoam} and \texttt{polyDualMesh}). 
The faces of the cells generated in this way are generally non-planar and must be triangulated, as for the distorted cubic grids, by joining its center with two consecutive vertices of each face. 
  \item Unstructured grids of different types available at \cite{roenby16b} which were used to compare our results with those presented in \cite{roenby16}.
\end{itemize}

The grid size $\mathcal{N}$ is obtained as
\begin{equation}
  \mathcal{N}=\widetilde{N}_\mathrm{CELL}^{1/\mathcal{D}}
\end{equation}
where $\widetilde{N}_\mathrm{CELL}$ is the number of grid cells in an equivalent unit domain (note that for the cases presented below with a domain different to a unit one, $\widetilde{N}_\mathrm{CELL}$ will not coincide with the true number of grid cells ${N}_\mathrm{CELL}$) and $\mathcal{D}$ is the dimensions number of the problem (2 and 3 for 2D and  3D problems, respectively). Note that 
in this work the 2D simulations are performed using 3D grids with only one cell along one of the coordinate directions. 

In this and the next sections, 
the CPU-times were estimated using the \texttt{gfortran} compiler with the \texttt{-O3}  optimization and the \texttt{OpenMP} application programming interface \cite{deal65} on an iMac Pro 2017 with an 18-core Intel Xeon W 2.3GHz CPU. \j{The parallel performance of the \texttt{gVOF} \jjj{package} is assessed in Section~\ref{sec:parallel}.}
For the LSGIR method of Section~\ref{sec:lsgir}, a value $\beta=1.5$  in the weight of Eq.~(\ref{lsgir-beta}) was used, which provides good results, especially for the cubic grids. For the isosurface based interface reconstruction methods of Section~\ref{sec:isorec}, the weight factor introduced by \citet{max99} 
was used for cubic grids, the weight factor given by the angle $\alpha$ of each triangular face of the constructed isosurface (see the sketch of Fig.~\ref{average-normal2}) if $\alpha\le \pi/2$ or $\pi-\alpha$ otherwise, was used for the tetrahedral and distorted cubic grids and the weight factor, also introduced by \citet{max99}, given by the area of each triangular facet was used  for the non-convex irregular polyhedral grids.
For the SWIR method of Section~\ref{sec:swir} and the   LSFIR method of Section~\ref{sec:lsfir}, a tolerance of $ 10^{-3}$ rad and $n_\mathrm{out}=4$ for the outer iteration are used (different values of the tolerance and maximum number of repetitions of the outer iteration can be set by the \texttt{gVOF}).
A detailed analysis of the different weightings  and  iteration parameters available in the \texttt{gVOF} \jjj{package} 
will be published elsewhere. A volume fraction tolerance $\epsilon$ of $10^{-12}$ is used for all the results, except for the  2D deformation test of Section~\ref{sec:2ddef} and a particular case mentioned later, for which $\epsilon=10^{-10}$ is used. Ten divisions along each coordinate direction are used to initialize the fluid volume fraction  and compute the reconstruction error on each interfacial cell.

\subsection{Static test cases}
\label{sec:staticresults}

 \begin{algorithm}
{ \caption{Interface reconstruction test}
\label{alg:rec}
 \begin{algorithmic}[1]
  \State Define the test case and allocate arrays for grid construction  
\State Call \texttt{defgrid} and allocate arrays for reconstruction and assessment
\State Call \texttt{printgrid} 
\State Call \texttt{neigbcell}
\State Call \texttt{compgrid}
\State Call \texttt{initfgrid}
\State Call \texttt{taggrid}
\State Call \texttt{clcir}, \texttt{elcir}, \texttt{llcir}, \texttt{lsgir}, \texttt{swir} or \texttt{lsfir} to reconstruct the interface
\State Call \texttt{printplic}, \texttt{isovtkgrid} or \texttt{printvoxel}  to print the interface shape
\State Call \texttt{recerr} to obtain the reconstruction error $E_\mathrm{rec}^{L_1}$
\State Deallocate arrays
\end{algorithmic}}
\end{algorithm}
The test program used to assess the performance of the different interface reconstruction methods implemented in the \texttt{gVOF} \jjj{package} is briefly presented in Algorithm~\ref{alg:rec}. Two static interface reconstruction test cases are considered:
\begin{enumerate}
\item Sphere of radius 0.325 centered at $(0.525,0.464,0.516)$. 
\item Torus centered at $(0.525,0.464,0.516)$ with minor and major radius of 0.1 and 0.2. 
\end{enumerate}

\begin{figure}[htbp]
\begin{center}
\includegraphics{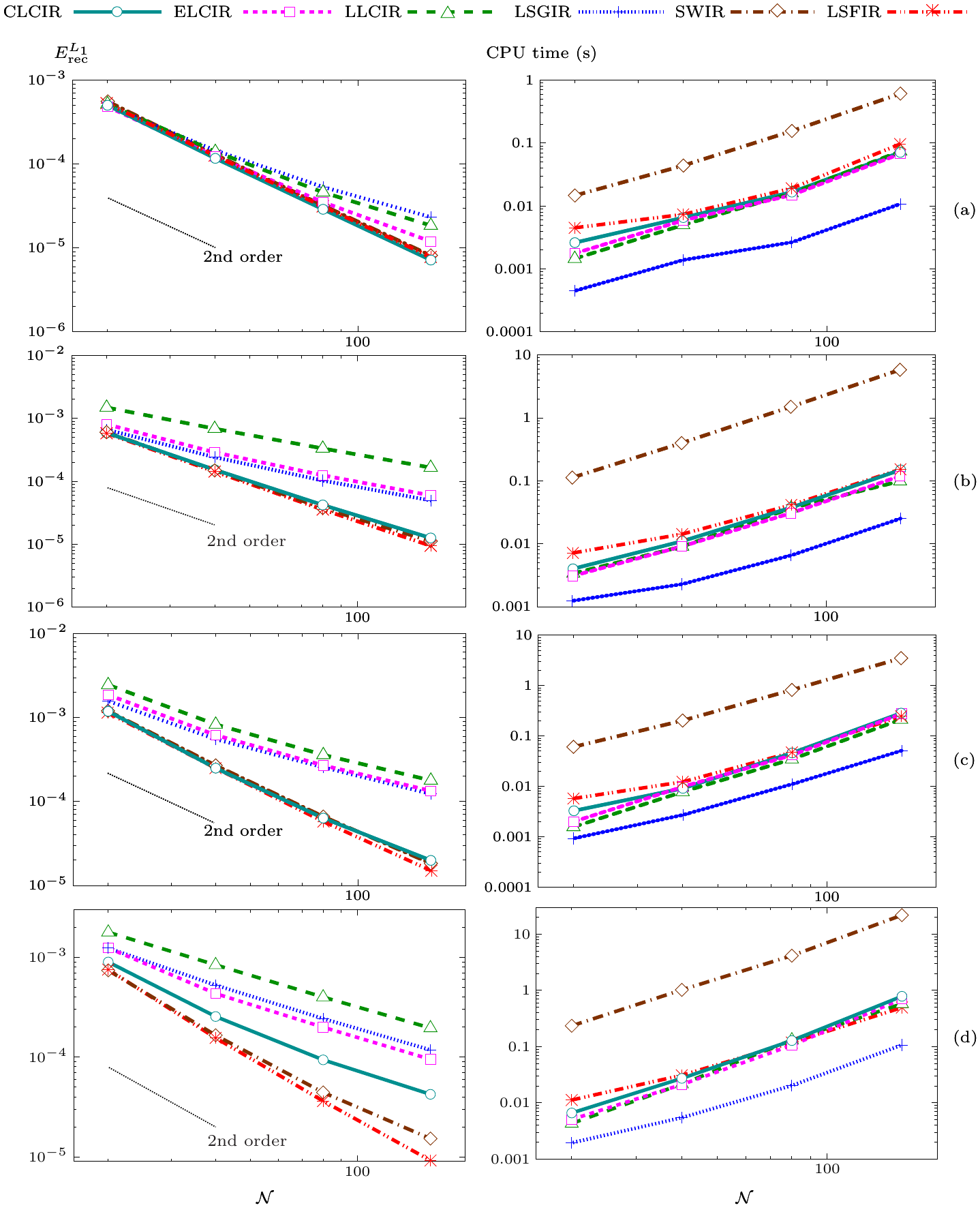}
\caption{Error $E_\mathrm{rec}^{L_1}$ (left pictures) and CPU time (right pictures) for the reconstruction of the spherical fluid body using (a) cubic, (b) distorted cubic, (c) tetrahedral and (d) non-convex irregular polyhedral grids.}
\label{sphere-error-cputime}
\end{center}
\end{figure}

\begin{figure}[htbp]
\begin{center}
\includegraphics{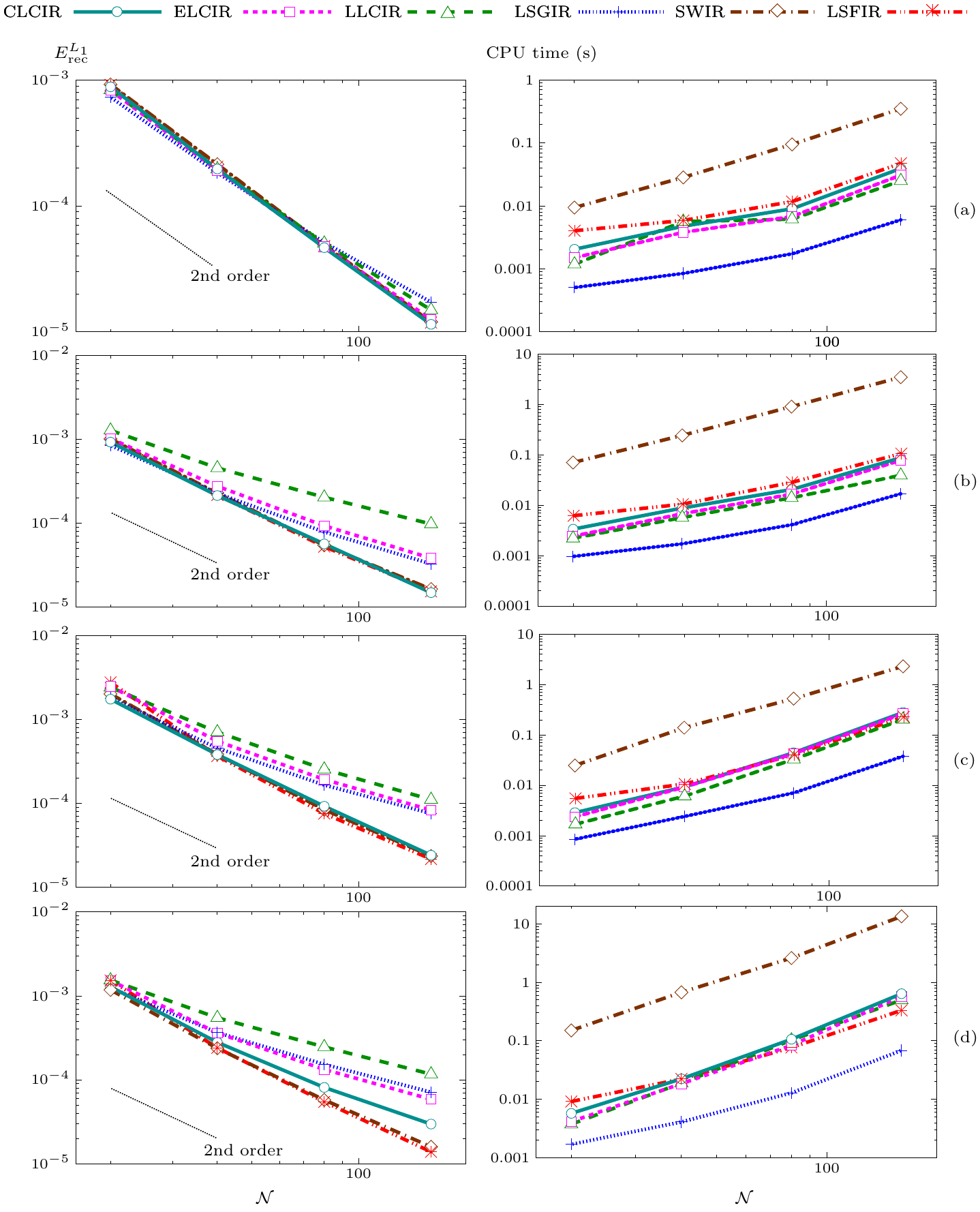}
\caption{Same results as in Fig.~\ref{sphere-error-cputime}, but for the toroidal fluid body.}
\label{torus-error-cputime}
\end{center}
\end{figure}

\begin{figure}[htbp]
\begin{center}
\includegraphics{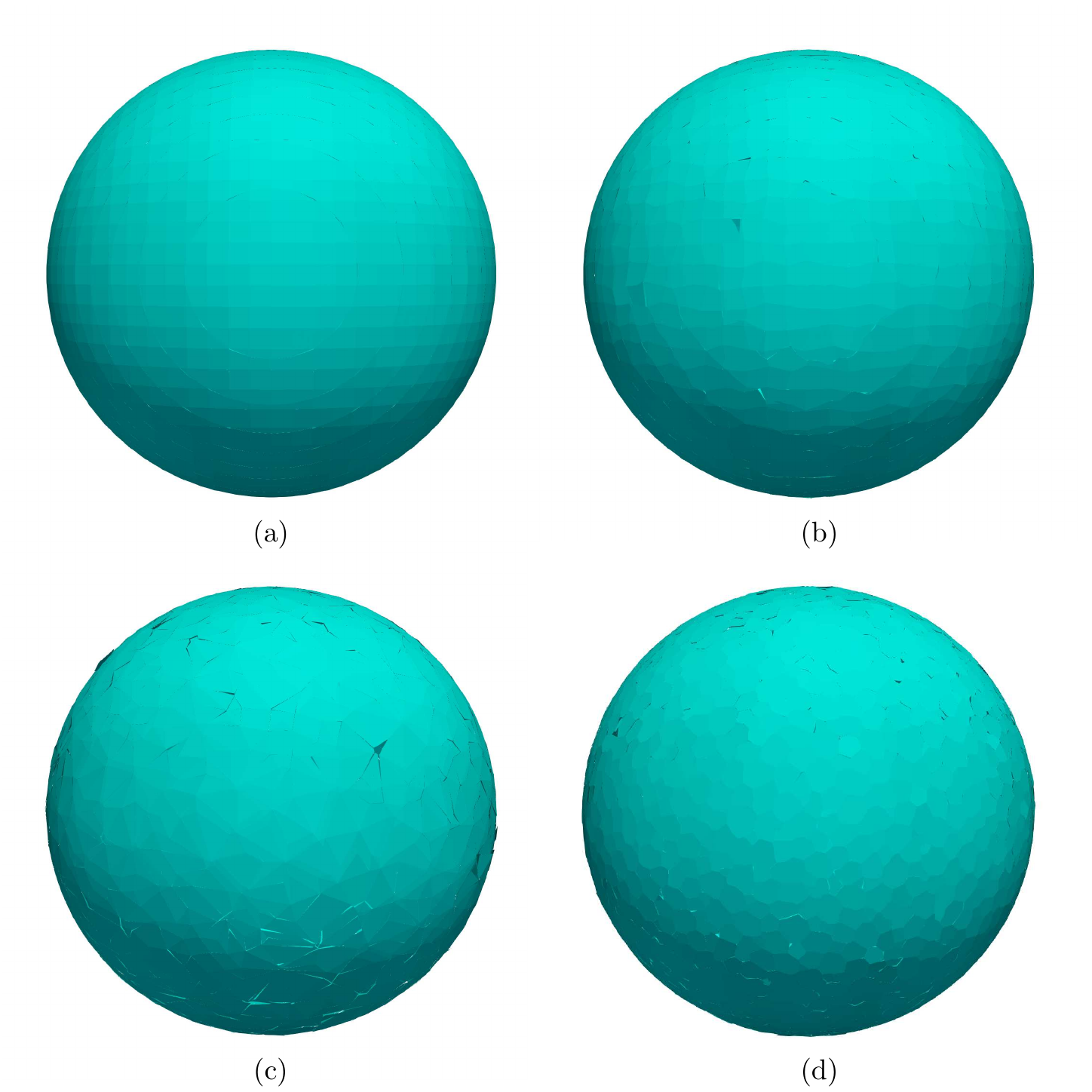}
\caption{PLIC interfaces for the spherical fluid body reconstructed using the CLCIR method and (a) cubic, (b) distorted cubic, (c) tetrahedral and (d) non-convex irregular polyhedral grids with $\mathcal{N}=40$.}
\label{sphere-plic-grids40}
\end{center}
\end{figure}

\begin{figure}[htbp]
\begin{center}
\includegraphics{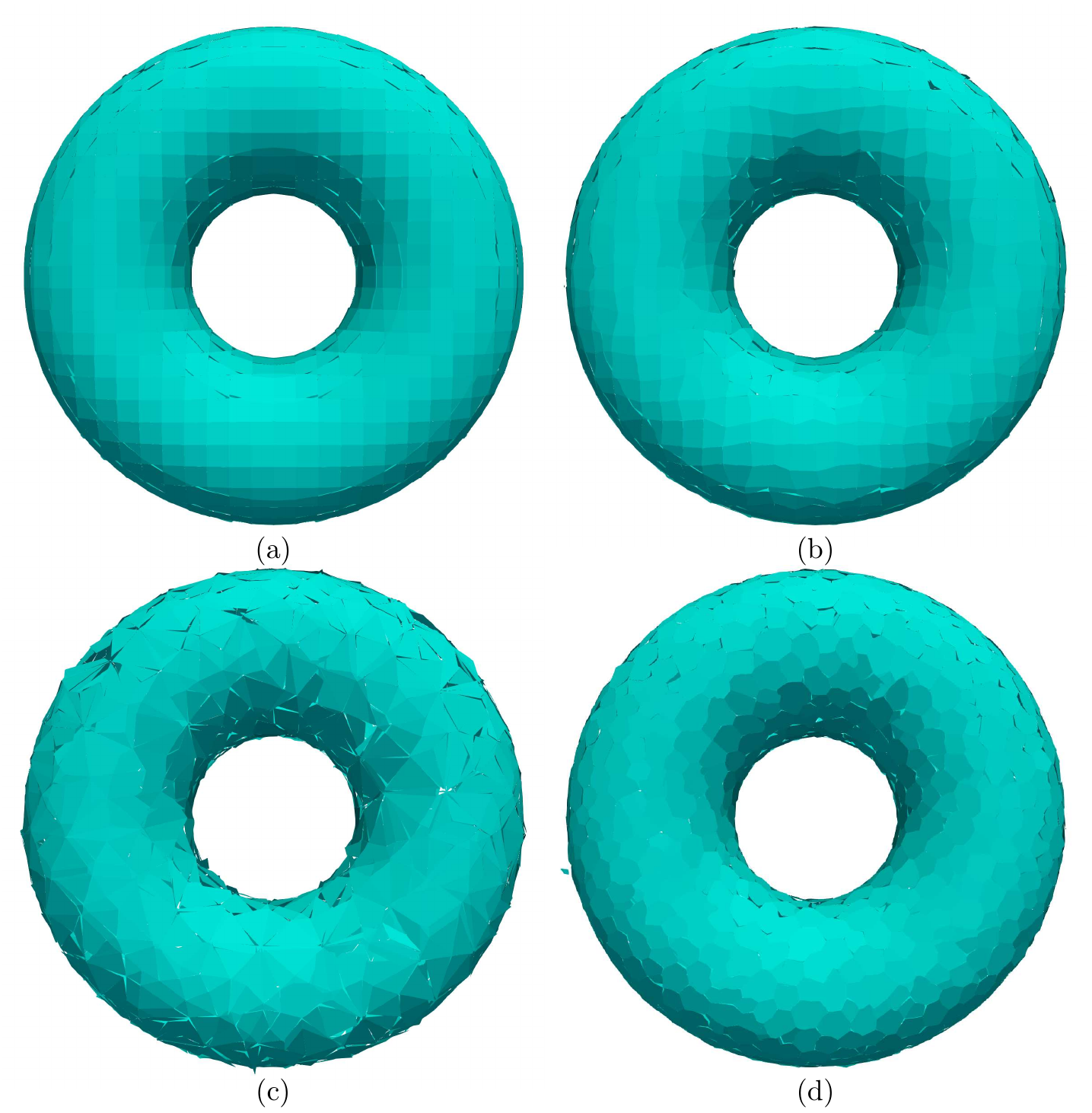}
\caption{Same results as in Fig.~\ref{sphere-plic-grids40}, but for the toroidal fluid body.}
\label{torus-plic-grids40}
\end{center}
\end{figure}

Figs.~\ref{sphere-error-cputime} and \ref{torus-error-cputime} show the error $E_\mathrm{rec}^{L_1}$ (left pictures), which is obtained by summing the error volumes computed in all the interfacial cells,  and execution time (right pictures)  for the reconstruction of the spherical and toroidal fluid bodies using different grids. It should be mentioned that changes in the relative differences in CPU time between methods could be observed depending on the architecture or compilation options used in the comparison.
For the grids with cubic, non-convex distorted cubic and tetrahedral cells, the CLCIR, SWIR and LSFIR methods are the most accurate providing similar accuracy with second-order convergence. However, the iterative SWIR method is clearly less computationally efficient being, on average, 20 and 100 times more time consuming than the CLCIR and LSGIR methods, respectively. Note that the computational efficiency of the CLCIR and LSFIR methods is relatively similar. For the grids with non-convex irregular polyhedral cells, the LSFIR method is the most accurate exhibiting second-order convergence. 
The LSGIR method is the least time consuming, showing very good performance at low grid resolutions. 
As an example, Figs.~\ref{sphere-plic-grids40} and \ref{torus-plic-grids40} show  the reconstructed PLIC interfaces for the spherical and toroidal fluid bodies using the CLCIR method and different grids with $\mathcal{N}=40$.

\subsection{Dynamic test cases with prescribed velocity field}
\label{sec:dynamicresults}

Algorithm~\ref{alg:adv} presents the
test program used for the assessment of the coupling between the different interface reconstruction and advection methods implemented in the \texttt{gVOF} \jjj{package} to solve dynamic problems. Extensive comparisons with some results available in the literature and obtained using other advanced geometric VOF methods are included in this section.
Others results, which can also be found in the literature, are left out of this comparison for several reasons, among  which, apart from brevity, are: 
the use of unstructured grids which are not provided and cannot be easily reproduced to be able to compare or the lack of information about parameters required for a rigorous comparison. 
 \begin{algorithm}
{\caption{Fluid advection test}
\label{alg:adv}
 \begin{algorithmic}[1]
  \State Define the test case and allocate arrays for grid construction 
\State Call \texttt{defgrid} and allocate arrays for reconstruction, advection, and \jl{a\-ssess\-ment}
\State Call \texttt{printgrid} 
\State Call \texttt{neigbcell}
\State Call \texttt{compgrid}
\State Call \texttt{initfgrid}
\State Call \texttt{taggrid}
\State Call \texttt{clcir}, \texttt{elcir}, \texttt{llcir}, \texttt{lsgir}, \texttt{swir} or \texttt{lsfir} and print the interface
\State $t_0=0$ and $t=0$
\State Set the velocity field at the initial instant $t_0$
\State Set the initial time step  $\Delta t$
\While{$t+\Delta t<t_\mathrm{end}$}
\State $t_0=t$ and $t=t_0+\Delta t$
\State Set the velocity field at the instant $\frac{1}{2}\left(t_0+t\right)$ 
\State Call \texttt{faceflux}
\State Call \texttt{vofadv}
\State Call \texttt{taggrid}
\State Call \texttt{clcir}, \texttt{elcir}, \texttt{llcir}, \texttt{lsgir}, \texttt{swir} or \texttt{lsfir}  and print, if required, the interface
\State Set the time step $\Delta t$
\EndWhile
\State Errors computation
\State Deallocate arrays
\end{algorithmic}}
\end{algorithm}

The time step $\Delta t$ used to solve Eq.~(\ref{vof}) is determined at each instant $t$ from the following expression
\begin{equation}
\Delta t =  \mathrm{min} \left[ \frac{  \min\limits_{i=1,\cdots,N_\textrm{CELL}} (h_{x_i})}{\max\limits_{j=1,\cdots,N_\textrm{FACE}}(u_{j})}, \frac{\min\limits_{i=1,\cdots,N_\textrm{CELL}} (h_{y_i})}{\max\limits_{j=1,\cdots,N_\textrm{FACE}} (v_{j})}, \frac{\min\limits_{i=1,\cdots,N_\textrm{CELL}} (h_{z_i})}{\max\limits_{j=1,\cdots,N_\textrm{FACE}} (w_{j})} \right] \mathrm{CFL},
\label{eq:cfl}
\end{equation}
where CFL is the Courant-Friedrich-Levy number, $u_{j}$, $v_{j}$ and $w_{j}$ are the components of the velocity vectors at the  centers of every grid face $j$, $N_\mathrm{FACE}$ is the number of faces in the grid, and $h_{x_i}$, $h_{y_i}$ and $h_{z_i}$ are the sizes along the corresponding coordinate axis of the 
minimum-size rectangular parallelepiped that encloses every cell $i$.
\j{Other strategies can be found in the literature to fix the time step from the CFL number.  For example, \citet{roenby16}, and  \citet{scheufler19}, adjusted the time step by using a CFL number that only concerns cells near the interface, or 
\citet{ivey17} fixed the time step by choosing a sufficiently low CFL number  to maintain the bounding  volume errors close to the machine precision. Some authors are not very clear on this respect, therefore the comparisons should be seen with certain reservations given the great influence of the chosen time step on the final accuracy.
}

To quantify the accuracy of the interface shape, the following error norm, used for example by \citet{lopez04,lopez08} or \citet{owkes14}, among others, is considered:
\begin{equation}
  E_{\textrm{shape}}^{L_1}(t)=\sum\limits_{i=1}^{N_{\textrm{CELL}}} V_{\Omega_i} \left| F_i^\mathrm{e}(t) - F_i(t)\right|,
\end{equation}
where  $F_i^\mathrm{e}(t)$ and $F_i(t)$ are, respectively, the exact and computed fluid volume fraction of cell $i$ at instant $t$. Other authors, such as \citet{roenby16}, used the same error norm but relative to the exact total fluid volume:
\begin{equation}
  E_{\textrm{shape}^*}^{L_1}(t)=\frac{\sum\limits_{i=1}^{N_{\textrm{CELL}}} V_{\Omega_i} \left| F_i^\mathrm{e}(t) - F_i(t)\right|}{\sum\limits_{i=1}^{N_{\textrm{CELL}}} V_{\Omega_i} F_i^\mathrm{e}(t)}.
\end{equation}

To quantify the change in the total fluid volume, the following error norm is used
\begin{equation}
  E_{\textrm{vol}}^{L_1}(t)=\left|\sum\limits_{i=1}^{N_{\textrm{CELL}}}  V_{\Omega_i} F_i^\mathrm{e}(t) - \sum\limits_{i=1}^{N_{\textrm{CELL}}} V_{\Omega_i} F_i(t)\right|.
\end{equation}
  
To quantify the unboundedness of the volume of fluid fractions (values lower and higher than 0.0 and 1.0, respectively), the error norm used by \citet{owkes14} is also considered in this work
\begin{equation}
  E_\textrm{bound}^{L_\infty} (t) = \textrm{max} \left[-\mathop{\textrm{min}}_{i=1,\cdots,N_\textrm{CELL}} V_{\Omega_i} F_i(t), \mathop{\textrm{max}}_{i=1,\cdots,N_\textrm{CELL}} V_{\Omega_i} \left( F_i(t) -1 \right) \right].
\end{equation}
The corresponding maximum and average values produced during the simulation are obtained, respectively, as
\begin{equation}
E_\textrm{bound}^{L_\infty} = \mathop{\textrm{max}}_{i=1,\cdots,N_\mathrm{STEP}} E_\textrm{bound}^{L_\infty} (t_i)
\end{equation}
\begin{equation}
E_\textrm{bound}^{L_1} = \frac{\sum\limits_{i=1}^{N_\mathrm{STEP}} E_\textrm{bound}^{L_\infty} (t_i)}{N_\mathrm{STEP}},
\end{equation}
where $N_\mathrm{STEP}$ is the number of time steps required to complete the numerical test.
It must be emphasized that unlike other VOF methods referenced to compare, the \texttt{gVOF} \jjj{package}, as mentioned, does not use special algorithms to redistribute the very small liquid volumes out of the bounds 0 and 1 for the volume fraction $F$. 

Results for the total execution time $t_\mathrm{cpu}=t_\mathrm{rec}+t_\mathrm{adv}$, where $t_\mathrm{rec}$ and $t_\mathrm{adv}$ are the total CPU times consumed by the reconstruction and advection steps, respectively,  
and its average value per time step $\widetilde{t}_\mathrm{cpu}$ are also presented below.

All the results presented in this section were obtained using the exact velocity field at both the face centers and cell vertices.
It has been checked that when the velocities at the cell vertices are interpolated from the prescribed velocities at the face centers using a simple average, the accuracy values of the interface shape estimated using any of the error norms considered in this work 
are very close to those presented below.
For the complex 3D deformation test of Section~\ref{sec:3ddef}, the observed variations in the estimated accuracy by using prescribed or interpolated velocities at the cell vertices are 
in the order of \jl{only 1\%}, being negligible for the finest grids.

\subsubsection{Translation test}

This test is used to compare with the results obtained by \citet{roenby16} using unstructured grids available in \cite{roenby16b}.
A sphere of fluid with radius 0.25, which is initially centered at (0.5, 0.5, 0.5) in a domain $[0,1]\times [0,1] \times [0,5]$, is translated by a velocity field given by $(0,0,1)$ from time 0 to $t_\mathrm{end}=4$.
Table~\ref{table:translation} compares the \jl{$E_\mathrm{shape^*}^{L_1}(t=4)$} error values obtained using \texttt{gVOF}, combining the FMFPA method with the CLCIR, SWIR and LSFIR methods, and those obtained using  two VOF methods: the algebraic VOF MULES (multidimensional universal limiter with explicit solution) \cite{deshpande12} and the geometric VOF isoAdvector \cite{roenby16},  both implemented in the \texttt{OpenFOAM} software \cite{openfoam}. 
All these results correspond to $\mathrm{CFL}=0.5$.
\begin{table}[htp]
\caption{Results for the translation test using tetrahedral grids and $\mathrm{CFL}=0.5$. Comparison with results from \cite{roenby16}. The CPU times reported in \cite{roenby16} were obtained using a single core of an Intel Xeon 3.1 GHz CPU (E5-2687W).}
\begin{center}
\scalebox{1}{\begin{tabular}{ccccc}
\hline
Grid size, $\mathcal{N}$ & \jl{$E_\mathrm{shape^*}^{L_1}(t=4)$}  & \jl{$E_\mathrm{vol}^{L_1}(t=4)$} & $E_\mathrm{bound}^{L_1}$  & $t_\mathrm{cpu}$ (s)  \\
\hline
\multicolumn{5}{l}{\texttt{gVOF} (FMFPA, CLCIR)} \\
41 & $3.1\times10^{-2}$ & $5.7\times 10^{-10}$ & $7.7\times 10^{-12}$  & 22 \\
71 & $1.3\times10^{-2}$ & $4.6\times 10^{-9}$ & $1.1\times 10^{-11}$  & 162 \\
\multicolumn{5}{l}{\texttt{gVOF} (FMFPA, SWIR)} \\
41 & $3.1\times10^{-2}$ & $3.7\times 10^{-10}$ & $1.1\times 10^{-12}$  & 104 \\
71 & $1.3\times10^{-2}$ & $1.5\times 10^{-9}$ & $2.4\times 10^{-12}$  & 601 \\
\multicolumn{5}{l}{\texttt{gVOF} (FMFPA, LSFIR)} \\
41 & $3.9\times10^{-2}$ & $3.4\times 10^{-10}$ & $1.0\times 10^{-11}$  & 22 \\
71 & $1.6\times10^{-2}$ & $2.5\times 10^{-9}$ & $1.0\times 10^{-11}$  & 145 \\
\multicolumn{5}{l}{isoAdvector \cite{roenby16}} \\
41 & $4.6\times10^{-2}$ & $5.1\times 10^{-13}$ & $--$  & 157 \\
71 & $2.1\times10^{-2}$ & $3.9\times 10^{-12}$ & $--$  & 1411 \\
\multicolumn{5}{l}{MULES \cite{roenby16}} \\
71 & $4.2\times10^{-1}$ & $2.9\times 10^{-14}$ & $--$  & 7306 \\
\hline
\end{tabular}}
\end{center}
\label{table:translation}
\end{table}%
Note that for this test, the final shape errors are reduced on average by 35\% compared to the results obtained using the isoAdvector method \jl{and  97\%} compared to the MULES method, which represents a significant improvement in accuracy. The computational efficiency of the FMFPA and CLCIR methods, which produce errors like those obtained using the SWIR method but with lower consumed CPU times, can also be observed in the results presented in the table.  It must be considered that the execution times reported in \cite{roenby16} were obtained using a  processor different to that used in this work and, therefore, those results should be viewed as a qualitative reference rather than for  direct comparison. It must be mentioned that the EMFPA and NMFPA methods  give almost identical results (not reported in the table) in terms of shape accuracy since the advection errors for this test are negligible using any of the implemented advection methods. For this test, however, the EMFPA and NMFPA methods consumes CPU-times around a 20\% higher and only 1\% lower, respectively, than that consumed by the FMFPA method. The errors reported in Table~\ref {table:translation} are due to the reconstruction step and to the roundoff errors of the  geometric operations involved during the simulation. 
A more detailed analysis on this subject will be carried out elsewhere.

\begin{figure}[htbp]
\begin{center}
\includegraphics{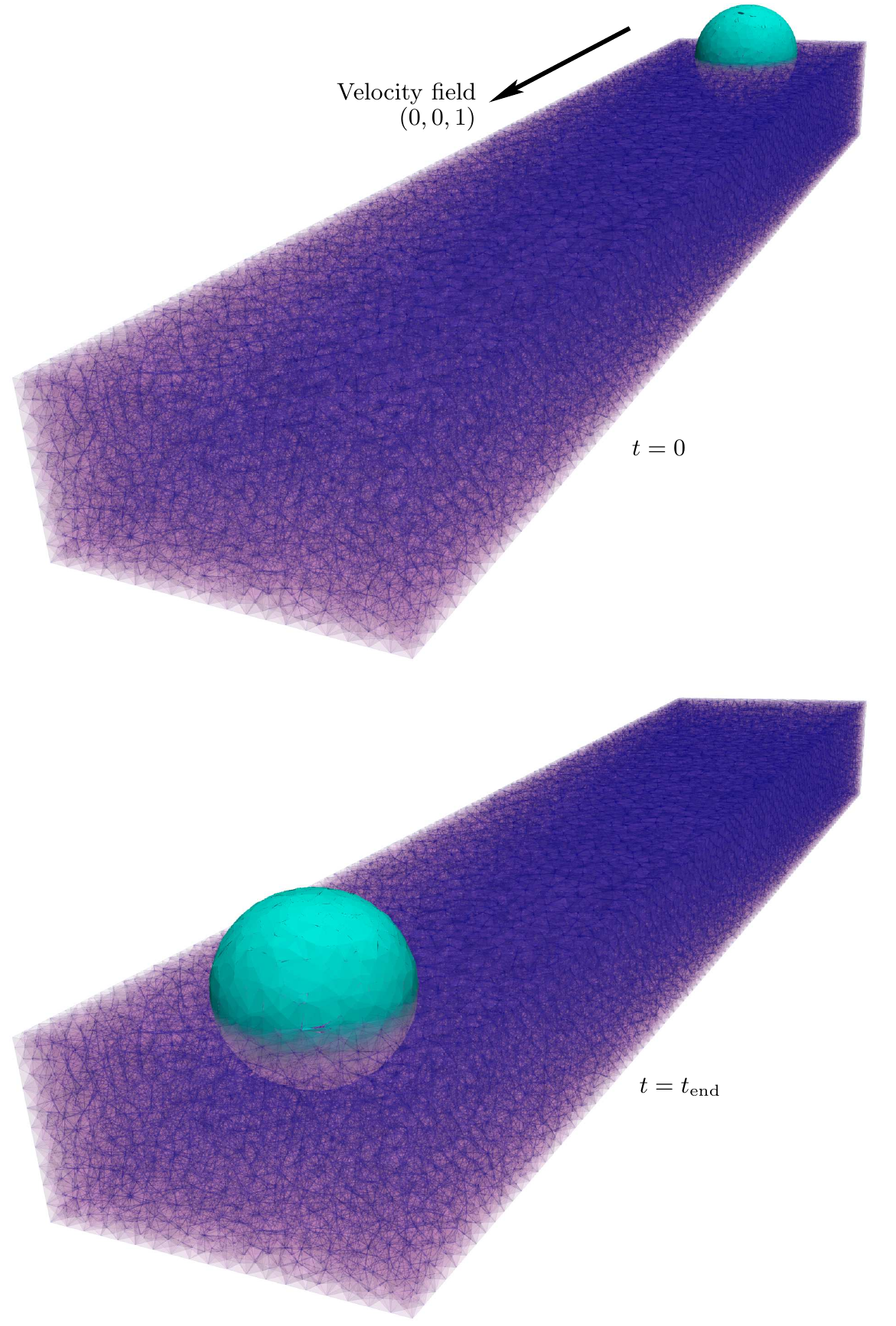}
\caption{Results for the translation case obtained using the FMFPA and CLCIR methods, the tetrahedral grid with $\mathcal{N}=41$ and $\mbox{CFL}=0.5$. The grid is clipped and made partially transparent to better see the PLIC interfaces reconstructed at $t = 0$ (top picture) and $t = t_\mathrm{end}$ (bottom picture).}
\label{plic-grid-trans-tetra41}
\end{center}
\end{figure}

\begin{figure}[htbp]
\begin{center}
\includegraphics{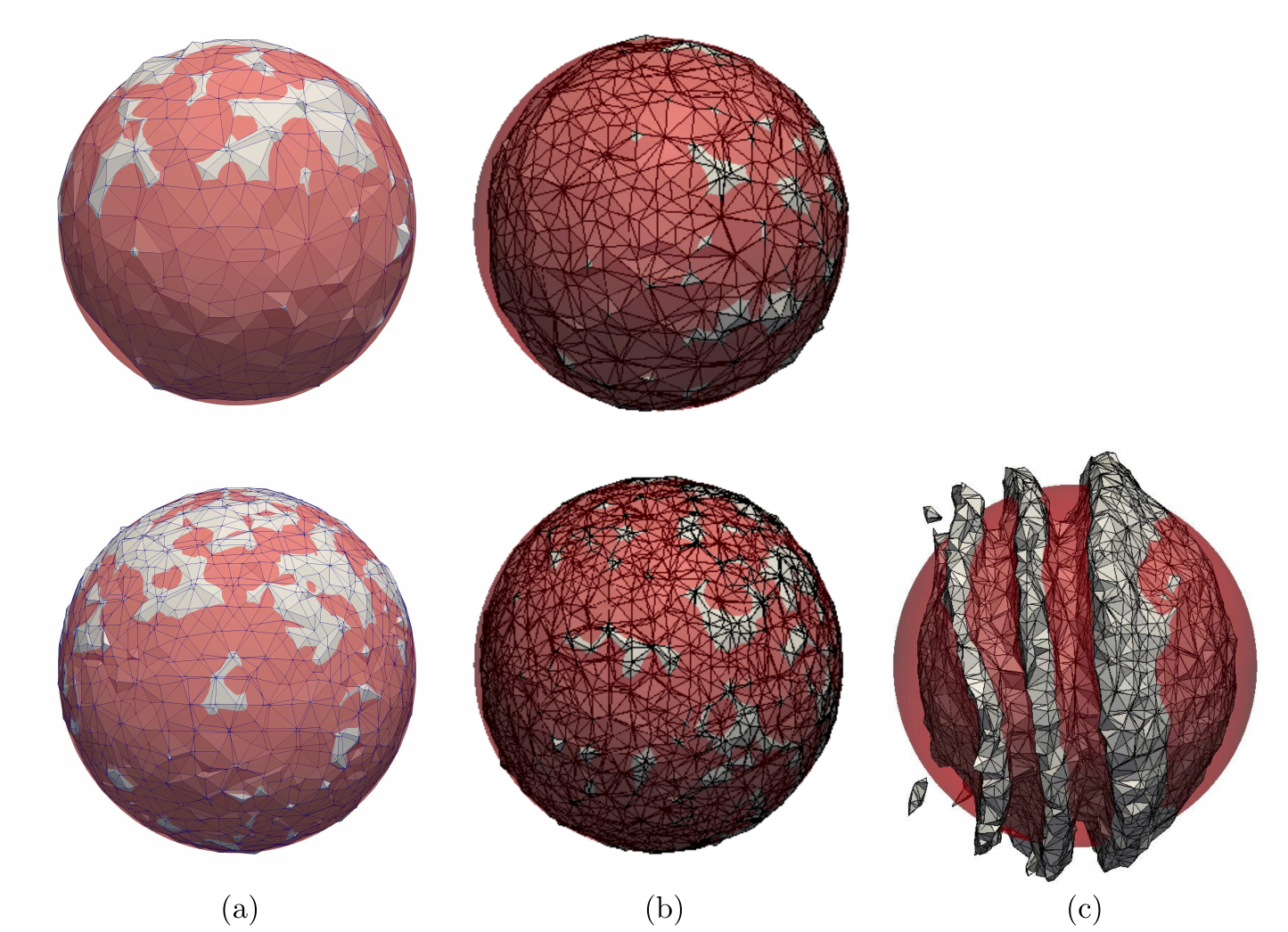}
\caption{Isosurfaces, corresponding to the 0.5 isovalue of the fluid volume fraction, extracted on the tetrahedral grids with $\mathcal{N}=41$ (top pictures) and 71 (bottom pictures) at the end of the translation test ($t=t_\mathrm{end}$). Results obtained using (a) \texttt{gVOF} (FMFPA and CLCIR),  (b) isoAdvector \cite{roenby16} and (c) MULES \cite{roenby16} methods. The exact solution is represented in transparent red color.}
\label{plic-grid-trans-tetra4171}
\end{center}
\end{figure}

Fig.~\ref{plic-grid-trans-tetra41} shows the PLIC interface at $t = 0$ (top picture) and $t = t_\mathrm{end}$ (bottom picture) obtained using the tetrahedral grid with $\mathcal{N}=41$. The grid is clipped and made partially transparent to better see the reconstructed PLIC interfaces.
A visual comparison with the isoAdvector and MULES methods is presented in Fig.~\ref{plic-grid-trans-tetra4171}, where results for the isosurfaces, corresponding to the 0.5 isovalue of the fluid volume fraction, extracted 
at the end of the translation test ($t=t_\mathrm{end}$) can be seen. It can be observed that the extracted isosurfaces obtained using the FMFPA and CLCIR methods are better adjusted to the exact surfaces represented in transparent red color than those of the isoAdvector method (the differences with the exact solution can be clearly appreciated by observing colors change and specially the gap between the extracted isosurface and the exact sphere on the left of each picture) and considerably much better than those of the MULES method, which produces a highly distorted fluid body even for this simple translation test.

\subsubsection{Rotation test}

In this test, a sphere of fluid with radius 0.15, which is initially centered at (0.5,0.75,0.5), is rotated to complete a revolution at $t_\mathrm{end}=2\pi$ around an axis parallel to the $z$-axis and centered in a domain $[0,1]\times [0,1] \times [0,1]$. The results obtained with \texttt{gVOF} (the FMFPA method combined with the CLCIR, SWIR and LSFIR methods and the NMFPA method combined with the CLCIR method) are compared in Table~\ref{table:rotation} with those \jl{obtained} by \citet{jofre14}, using an advection method similar to the EMFPA method and the second-order LVIRA \cite{pilliod97} reconstruction method, and \citet{scheufler19}, using a new version of the isoAdvector method  (isoAdvector-plicRDF) which is improved by a distance function reconstructed from PLIC interfaces. All these results were obtained using $\mathrm{CFL}=1$.
\begin{table}[htp]
\caption{Results for the rotation test using cubic grids and $\mathrm{CFL}=1$. Comparison with results from \cite{scheufler19} and \cite{jofre14}. The execution times $\widetilde{t}_\mathrm{cpu}$
reported in \cite{scheufler19} were obtained using four cores of a dual Intel Xeon 2687W v2 processor.}
\begin{center}
\scalebox{0.75}{\begin{tabular}{ccccc}
\hline
Grid size, $\mathcal{N}$ & \jl{$E_\mathrm{shape}^{L_1}(t=2\pi)$}  & \jl{$E_\mathrm{vol}^{L_1}(t=2\pi)$} & \jl{$E_\mathrm{bound}^{L_\infty}(t=2\pi)$}  & $\widetilde{t}_\mathrm{cpu}$ (s)  \\
\hline
\multicolumn{5}{l}{\texttt{gVOF} (NMFPA, CLCIR)} \\
32 & $6.74\times10^{-4}$ & $1.5\times 10^{-4}$ & $7.4\times 10^{-8}$  & 0.0055 \\
64 & $2.06\times10^{-4}$ & $6.7\times 10^{-5}$ & $2.2\times 10^{-8}$  & 0.019 \\
128 & $6.52\times10^{-5}$ & $3.0\times 10^{-5}$ & $1.5\times 10^{-9}$  & 0.090 \\
256 & $2.31\times10^{-5}$ & $1.3\times 10^{-5}$ & $1.0\times 10^{-10}$  & 0.64 \\
\multicolumn{5}{l}{\texttt{gVOF} (FMFPA, CLCIR)} \\
32 & $3.33\times10^{-4}$ & $8.7\times 10^{-18}$ & $6.9\times 10^{-19}$  & 0.0055 \\
64 & $9.31\times10^{-5}$ & $2.3\times 10^{-17}$ & $2.7\times 10^{-20}$  & 0.020 \\
128 & $2.40\times10^{-5}$ & $7.3\times 10^{-17}$ & $2.7\times 10^{-19}$  & 0.092 \\
256 & $6.34\times10^{-6}$ & $1.8\times 10^{-16}$ & $2.5\times 10^{-19}$  & 0.65 \\
\multicolumn{5}{l}{\texttt{gVOF} (FMFPA, SWIR)} \\
32 & $3.65\times10^{-4}$ & $3.8\times 10^{-17}$ & $4.0\times 10^{-19}$  & 0.013 \\
64 & $9.11\times10^{-5}$ & $0.0$ & $2.7\times 10^{-20}$  & 0.040 \\
128 & $2.38\times10^{-5}$ & $5.9\times 10^{-17}$ & $2.7\times 10^{-19}$  & 0.16 \\
256 & $6.37\times10^{-6}$ & $6.6\times 10^{-17}$ & $2.5\times 10^{-19}$  & 0.94 \\
\multicolumn{5}{l}{\texttt{gVOF} (FMFPA, LSFIR)} \\
32 & $3.62\times10^{-4}$ & $2.3\times 10^{-17}$ & $1.5\times 10^{-18}$  & 0.0074 \\
64 & $9.30\times10^{-5}$ & $3.6\times 10^{-17}$ & $2.7\times 10^{-20}$  & 0.019 \\
128 & $2.38\times10^{-5}$ & $3.8\times 10^{-17}$ & $2.4\times 10^{-19}$  & 0.092 \\
256 & $6.35\times10^{-6}$ & $9.7\times 10^{-17}$ & $1.4\times 10^{-19}$  & 0.66 \\
\multicolumn{5}{l}{isoAdvector-plicRDF \cite{scheufler19}} \\
32 & $7.50\times10^{-4}$ & $3.5\times 10^{-18}$ & $5.4\times 10^{-21}$  & 0.034 \\
64 & $1.86\times10^{-4}$ & $2.7\times 10^{-16}$ & $7.4\times 10^{-22}$  & 0.1 \\
128 & $4.77\times10^{-5}$ & $3.6\times 10^{-15}$ & $8.9\times 10^{-23}$  & 0.44 \\
256 & $1.41\times10^{-5}$ & $2.5\times 10^{-14}$ & $5.4\times 10^{-14}$  & 2.6 \\
\multicolumn{5}{l}{\citet{jofre14}} \\
32 & $5.47\times10^{-4}$ & $--$ & $--$  & $--$ \\
64 & $1.29\times10^{-4}$ & $--$ & $--$  & $--$ \\
128 & $3.46\times10^{-5}$ & $--$ & $--$  & $--$ \\
\hline
\end{tabular}}
\end{center}
\label{table:rotation}
\end{table}%
As in the above test, the EMFPA method produces an accuracy almost indentical to that of FMFPA method but with higher time consumptions.  Note that the final shape errors are reduced around 50\% with respect to those of isoAdvector-plicRDF and around 30\% with respect to those obtained by \citet{jofre14}, which represents a significant accuracy improvement. Results for the computational efficiency and volume conservation are also included in the table. The FMFPA method produces out of bounds volume errors close to  the machine precision and shows a very good computational efficiency (the advantage of the CLCIR and LSFIR methods in this \jl{respect} is clearly seen in the table).
It is also noteworthy that the NMFPA method is less accurate than the FMFPA method and not strictly conservative due to the  over/underlaps between flux polyhedra. 

\subsubsection{3D deformation test}
\label{sec:3ddef}

In this test proposed by \citet{enright2002}, a sphere of fluid with radius 0.15 and initially centered at (0.35,0.35,0.35) in a domain $[0,1]\times [0,1] \times [0,1]$, is deformed in the following velocity field:
 \begin{equation}
 \left.
 \begin{split}
u(x,y,z,t) = &2 \sin^2(\pi x) \sin(2\pi y) \sin (2\pi z) \cos (\pi t/t_\mathrm{end}), \\
v(x,y,z,t) = &- \sin(2\pi x) \sin^2(\pi y) \sin (2\pi z) \cos (\pi t/t_\mathrm{end}), \\
w(x,y,z,t) = &- \sin(2\pi x) \sin(2\pi y) \sin^2 (\pi z) \cos (\pi t/t_\mathrm{end}), \\
\end{split} \right\}
 \end{equation}
for  $t_\mathrm{end}=3$.
This test produces a high deformation requiring a very fine grid to correctly solve the thinnest fluid structures. 

\begin{figure}[htbp]
\begin{center}
\includegraphics{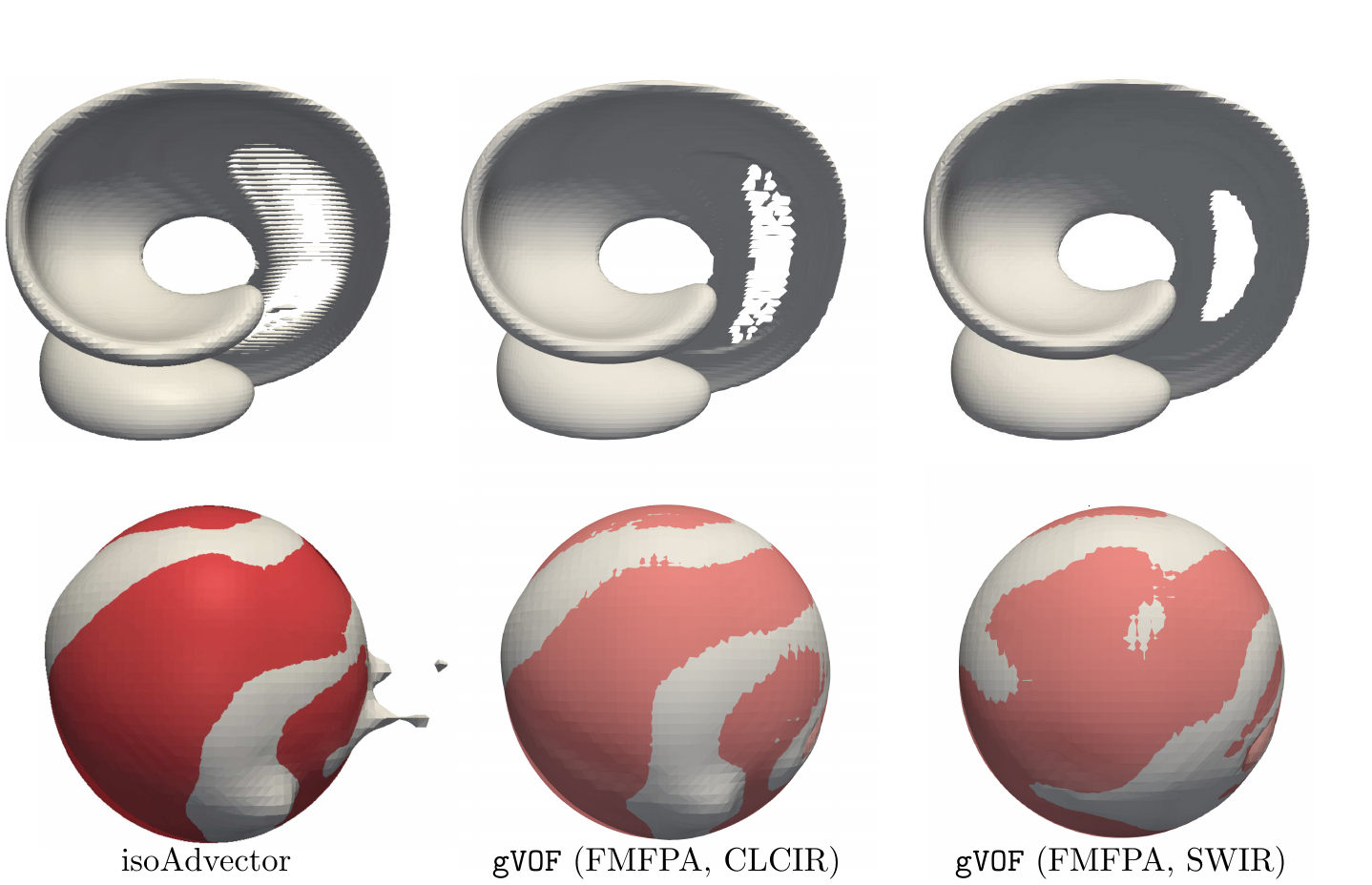}
\caption{Isosurfaces extracted on a cubic grid with $\mathcal{N}=128$ at the instant of maximum deformation (top pictures) and end instant of the test (bottom pictures). The results are compared with those presented in \cite{roenby16} using the isoAdvector method.
 }
\label{enright-cubic-vs-isoAdvector}
\end{center}
\end{figure}

A comparison with the results obtained using the isoAdvector method \cite{roenby16} is presented in Table~\ref{table:3ddef1} for cubic grids and in Table~\ref{table:3ddef1tetra} for a tetrahedral grid (when using this grid and the NMFPA methods, $\epsilon$ has been increased to $10^{-10}$ to attenuate the effect of the over/underlaps). 
\begin{table}[htp]
\caption{Results for the 3D deformation test using cubic grids and $\mathrm{CFL}=0.5$. Comparison with results from \cite{roenby16}, where the execution times $t_\mathrm{cpu}$ were obtained using  a single core of an Intel Xeon 3.1 GHz CPU (E5-2687W).}
\begin{center}
\scalebox{0.9}{\begin{tabular}{ccccc}
\hline
Grid size, $\mathcal{N}$ & \jl{$E_\mathrm{shape^*}^{L_1}(t=3)$}  & \jl{$E_\mathrm{vol}^{L_1}(t=3)$} & \jl{$E_\mathrm{bound}^{L_\infty}(t=3)$}  & $t_\mathrm{cpu}$ (s)  \\
\hline
\multicolumn{5}{l}{\texttt{gVOF} (NMFPA, CLCIR)} \\
64 & $1.52\times10^{-1}$ & $8.2\times 10^{-6}$ & $4.0\times 10^{-10}$  & 11  \\
128 & $3.13\times10^{-2}$ & $4.0\times 10^{-6}$ & $2.3\times 10^{-10}$  & 106 \\
256 & $4.40\times10^{-3}$ & $1.8\times 10^{-6}$ & $1.3\times 10^{-11}$  & 1426 \\
\multicolumn{5}{l}{\texttt{gVOF} (FMFPA, CLCIR)} \\
64 & $1.39\times10^{-1}$ & $6.8\times 10^{-17}$ & $4.0\times 10^{-19}$  & 13  \\
128 & $3.04\times10^{-2}$ & $6.6\times 10^{-17}$ & $1.4\times 10^{-19}$  & 113 \\
256 & $4.23\times10^{-3}$ & $5.5\times 10^{-15}$ & $6.8\times 10^{-20}$  & 1455 \\
\multicolumn{5}{l}{\texttt{gVOF} (EMFPA, CLCIR)} \\
64 & $1.39\times10^{-1}$ & $3.6\times 10^{-17}$ & $1.4\times 10^{-19}$  & 16 \\
128 & $3.04\times10^{-2}$ & $5.2\times 10^{-18}$ & $1.2\times 10^{-19}$  & 135 \\
256 & $4.23\times10^{-3}$ & $5.5\times 10^{-15}$ & $1.0\times 10^{-19}$  & 1617 \\
\multicolumn{5}{l}{\texttt{gVOF} (FMFPA, SWIR)} \\
64 & $7.39\times10^{-2}$ & $3.5\times 10^{-18}$ & $1.2\times 10^{-19}$  & 31 \\
128 & $2.15\times10^{-2}$ & $2.6\times 10^{-17}$ & $1.4\times 10^{-19}$  & 248 \\
256 & $4.39\times10^{-3}$ & $5.4\times 10^{-15}$ & $7.8\times 10^{-20}$  & 2672 \\
\multicolumn{5}{l}{\texttt{gVOF} (FMFPA, LSFIR)} \\
64 & $7.65\times10^{-2}$ & $4.9\times 10^{-17}$ & $1.2\times 10^{-19}$  & 12 \\
128 & $1.90\times10^{-2}$ & $5.7\times 10^{-17}$ & $8.7\times 10^{-20}$  & 112 \\
256 & $4.41\times10^{-3}$ & $5.5\times 10^{-15}$ & $7.7\times 10^{-20}$  & 1505 \\
\multicolumn{5}{l}{isoAdvector \cite{roenby16}} \\
64 & $2.2\times10^{-1}$ & $3.7\times 10^{-15}$ & $0$  & 173 \\
128 & $4.7\times10^{-2}$ & $3.7\times 10^{-14}$ & $3.0\times 10^{-13}$  & 2626 \\
256 & $1.2\times10^{-2}$ & $2.3\times 10^{-13}$ & $1.1\times 10^{-10}$  & 46706 \\
\hline
\end{tabular}}
\end{center}
\label{table:3ddef1}
\end{table}%
\begin{table}[htp]
\caption{Same results as in Table~\ref{table:3ddef1}, but using the tetrahedral grid with $\mathcal{N}=317$ from \cite{roenby16b}.}
\begin{center}
\scalebox{1}{\begin{tabular}{cccc}
\hline
 \jl{$E_\mathrm{shape^*}^{L_1}(t=3)$}  & \jl{$E_\mathrm{vol}^{L_1}(t=3)$} & \jl{$E_\mathrm{bound}^{L_\infty}(t=3)$}  & $t_\mathrm{cpu}$ (s)  \\
\hline
\multicolumn{4}{l}{\texttt{gVOF} (NMFPA, CLCIR)} \\
$4.06\times10^{-3}$ & $3.3\times 10^{-8}$ & $4.7\times 10^{-12}$  & 4217 \\
\multicolumn{4}{l}{\texttt{gVOF} (FMFPA, CLCIR)} \\
$4.01\times10^{-3}$ & $2.6\times 10^{-9}$ & $1.3\times 10^{-13}$  & 4985 \\
\multicolumn{4}{l}{\texttt{gVOF} (EMFPA, CLCIR)} \\
$3.94\times10^{-3}$ & $4.0\times 10^{-9}$ & $8.4\times 10^{-14}$  & 7196 \\
\multicolumn{4}{l}{\texttt{gVOF} (EMFPA, SWIR)} \\
$4.63\times10^{-3}$ & $9.8\times 10^{-9}$ & $5.4\times 10^{-14}$  & 39128 \\
\multicolumn{4}{l}{\texttt{gVOF} (EMFPA, LSFIR)} \\
$7.92\times10^{-3}$ & $1.22\times 10^{-8}$ & $4.9\times 10^{-12}$  & 7783 \\
\multicolumn{4}{l}{isoAdvector \cite{roenby16}} \\
$--$ & $8.9\times 10^{-5}$ & $--$  & $\simeq 259200$ \\
\hline
\end{tabular}}
\end{center}
\label{table:3ddef1tetra}
\end{table}%
From the above results, it is evident the advantage in accuracy of the \texttt{gVOF} algorithms (on average, a reduction of around 60\% in the shape errors is observed in  Table~\ref{table:3ddef1}). This improvement can be visually observed from results like those shown in Fig.~\ref{enright-cubic-vs-isoAdvector}.
When using cubic grids, the FMFPA results are slightly more accurate than the NMFPA results and almost identical to the EMFPA results. The EMFPA method increases the consumed CPU-time in around 20\% \jl{and  30\%} compared to the FMFPA and NMFPA methods, respectively.  When using the tetrahedral grid, the EMFPA method is slightly more accurate but with an increment of around 45\% and 70\% in the consumed CPU time compared to the FMFPA and NMFPA methods, respectively. 
Although the shape error value for the case with the tetrahedral grid of Table~\ref{table:3ddef1} and the isoAdvector method is not available, 
 a volume variation of 0.63\% and 
 a CPU time of around 3 days at the end of the test were reported in \cite{roenby16} using a single core of an Intel Xeon 3.10GHz CPU (E5-2687W). The very good performance and computational efficiency of \texttt{gVOF}, especially when using the CLCIR or LSFIR methods, is clearly observed in the tables.

Now, a comparison with the isoAdvector-plicRDF method \cite{scheufler19} and many of the most advanced geometric-unsplit VOF methods recently published is presented in Tables~\ref{table:3ddef2} and \ref{table:3ddef2b} \j{(the last three combined EMFPA-like advection methods with second-order reconstruction methods)}.
\begin{table}
\caption{Results for the 3D deformation test using cubic grids and $\mathrm{CFL}=0.5$. Comparison with results from (the resources used to execute the tests is included in parenthesis) \cite{scheufler19} (4 cores of a dual Intel Xeon 2687W v2). 
}\label{table:3ddef2}
\begin{center}
\scalebox{1}{\begin{tabular}{ccccc}
\hline
Grid size, $\mathcal{N}$ & \jl{$E_\mathrm{shape}^{L_1}(t=3)$}  & \jl{$E_\mathrm{vol}^{L_1}(t=3)$} & \jl{$E_\mathrm{bound}^{L_\infty}(t=3)$}  & $\widetilde{t}_\mathrm{cpu}$ (s)  \\
\hline
\multicolumn{5}{l}{\texttt{gVOF} (FMFPA, CLCIR)} \\
32 & $6.24\times10^{-3}$ & $1.0\times 10^{-16}$ & $1.6\times 10^{-19}$  & 0.0074 \\
64 & $1.97\times10^{-3}$ & $6.8\times 10^{-17}$ & $4.0\times 10^{-19}$  & 0.026 \\
128 & $4.29\times10^{-4}$ & $6.6\times 10^{-17}$ & $1.4\times 10^{-19}$  & 0.12 \\
256 & $5.98\times10^{-5}$ & $5.5\times 10^{-15}$ & $6.8\times 10^{-20}$  & 0.74 \\
\multicolumn{5}{l}{\texttt{gVOF} (FMFPA, SWIR)} \\
32 & $3.55\times10^{-3}$ & $3.0\times 10^{-16}$ & $1.7\times 10^{-19}$  & 0.019 \\
64 & $1.04\times10^{-3}$ & $3.5\times 10^{-18}$ & $1.2\times 10^{-19}$  & 0.064 \\
128 & $3.04\times10^{-4}$ & $2.6\times 10^{-17}$ & $1.4\times 10^{-19}$  & 0.25 \\
256 & $6.20\times10^{-5}$ & $5.4\times 10^{-15}$ & $7.8\times 10^{-20}$  & 1.30 \\
\multicolumn{5}{l}{\texttt{gVOF} (FMFPA, LSFIR)} \\
32 & $3.95\times10^{-3}$ & $1.4\times 10^{-16}$ & $1.9\times 10^{-19}$  & 0.0090 \\
64 & $1.08\times10^{-3}$ & $4.9\times 10^{-17}$ & $1.2\times 10^{-19}$  & 0.025 \\
128 & $2.69\times10^{-4}$ & $5.7\times 10^{-17}$ & $8.7\times 10^{-20}$  & 0.12 \\
256 & $6.24\times10^{-5}$ & $5.5\times 10^{-15}$ & $7.7\times 10^{-20}$  & 0.77 \\
\multicolumn{5}{l}{isoAdvector-plicRDF \cite{scheufler19}} \\
32 & $8.36\times10^{-3}$ & $1.1\times 10^{-16}$ & $2.1\times 10^{-14}$  & 0.058 \\
64 & $3.25\times10^{-3}$ & $9.7\times 10^{-16}$ & $6.4\times 10^{-20}$  & 0.15 \\
128 & $6.57\times10^{-4}$ & $3.7\times 10^{-15}$ & $1.1\times 10^{-15}$  & 0.52 \\
256 & $9.54\times10^{-5}$ & $2.0\times 10^{-14}$ & $2.4\times 10^{-16}$  & 2.63 \\
\hline
\end{tabular}}
\end{center}

\end{table}%
\begin{table}[htbp]
\caption{Continuation of Table~\ref{table:3ddef2}. Comparison with results from    
\cite{ivey17} (no execution time reported; the results were obtained with CFL sufficiently low to get  $E_\mathrm{vol}^{L_1}(8)<10^{-12}$), \cite{owkes14} (4 cores of a dual Intel Xeon X5670), \cite{jofre14} (no execution time reported) and \cite{maric18} (4 cores of an Intel Xeon E5-2680 v3). 
}\label{table:3ddef2b}
\begin{center}
\scalebox{0.9}{\begin{tabular}{ccccc}
\hline
Grid size, $\mathcal{N}$ & \jl{$E_\mathrm{shape}^{L_1}(t=3)$}  & \jl{$E_\mathrm{vol}^{L_1}(t=3)$} & \jl{$E_\mathrm{bound}^{L_\infty}(t=3)$}  & $\widetilde{t}_\mathrm{cpu}$ (s)  \\
\hline
\multicolumn{5}{l}{NIFPA-1 \cite{ivey17}$^*$} \\
33 & $6.8\times10^{-3}$ & $<1.4\times 10^{-14}$ & $<1.4\times 10^{-14}$  & $--$ \\
65 & $2.2\times10^{-3}$ & $<1.4\times 10^{-14}$ & $<1.4\times 10^{-14}$  & $--$ \\
129 & $4.9\times10^{-4}$ & $<1.4\times 10^{-14}$ & $<1.4\times 10^{-14}$  & $--$ \\
257 & $1.3\times10^{-4}$ & $<1.4\times 10^{-14}$ & $<1.4\times 10^{-14}$  & $--$ \\
\multicolumn{5}{l}{\citet{liovic05}} \\
32 & $7.41\times10^{-3}$ & $--$ & $--$  & $--$ \\
64 & $1.99\times10^{-3}$ & $--$ & $--$  & $--$ \\
128 & $3.09\times10^{-4}$ & $--$ & $--$  & $--$ \\
256 & $7.03\times10^{-5}$ & $--$ & $--$  & $--$ \\
\multicolumn{5}{l}{\citet{owkes14}} \\
32 & $6.98\times10^{-3}$ & $1.2\times 10^{-15}$ & $1.2\times 10^{-17}$  & 0.78 \\
64 & $2.10\times10^{-3}$ & $2.5\times 10^{-15}$ & $2.3\times 10^{-17}$  & 2.85 \\
128 & $5.63\times10^{-4}$ & $1.7\times 10^{-14}$ & $2.8\times 10^{-17}$  & 12.2 \\
256 & $1.01\times10^{-4}$ & $3.9\times 10^{-14}$ & $4.7\times 10^{-17}$  & 45.5 \\
\multicolumn{5}{l}{\citet{jofre14}} \\
32 & $6.92\times10^{-3}$ & $--$ & $--$  & $--$ \\
64 & $2.43\times10^{-3}$ & $--$ & $--$  & $--$ \\
128 & $6.37\times10^{-4}$ & $--$ & $--$  & $--$ \\
\multicolumn{5}{l}{\citet{maric18}} \\
32 & $5.86\times10^{-3}$ & $2.5\times 10^{-15}$ & $0.0$  & 0.69 \\
64 & $1.56\times10^{-3}$ & $6.0\times 10^{-15}$ & $0.0$  & 2.81 \\
128 & $3.08\times10^{-4}$ & $1.6\times 10^{-14}$ & $0.0$  & 12.0 \\
\hline
\end{tabular}}
\end{center}

\end{table}%

\begin{figure}[htbp]
\begin{center}
\includegraphics{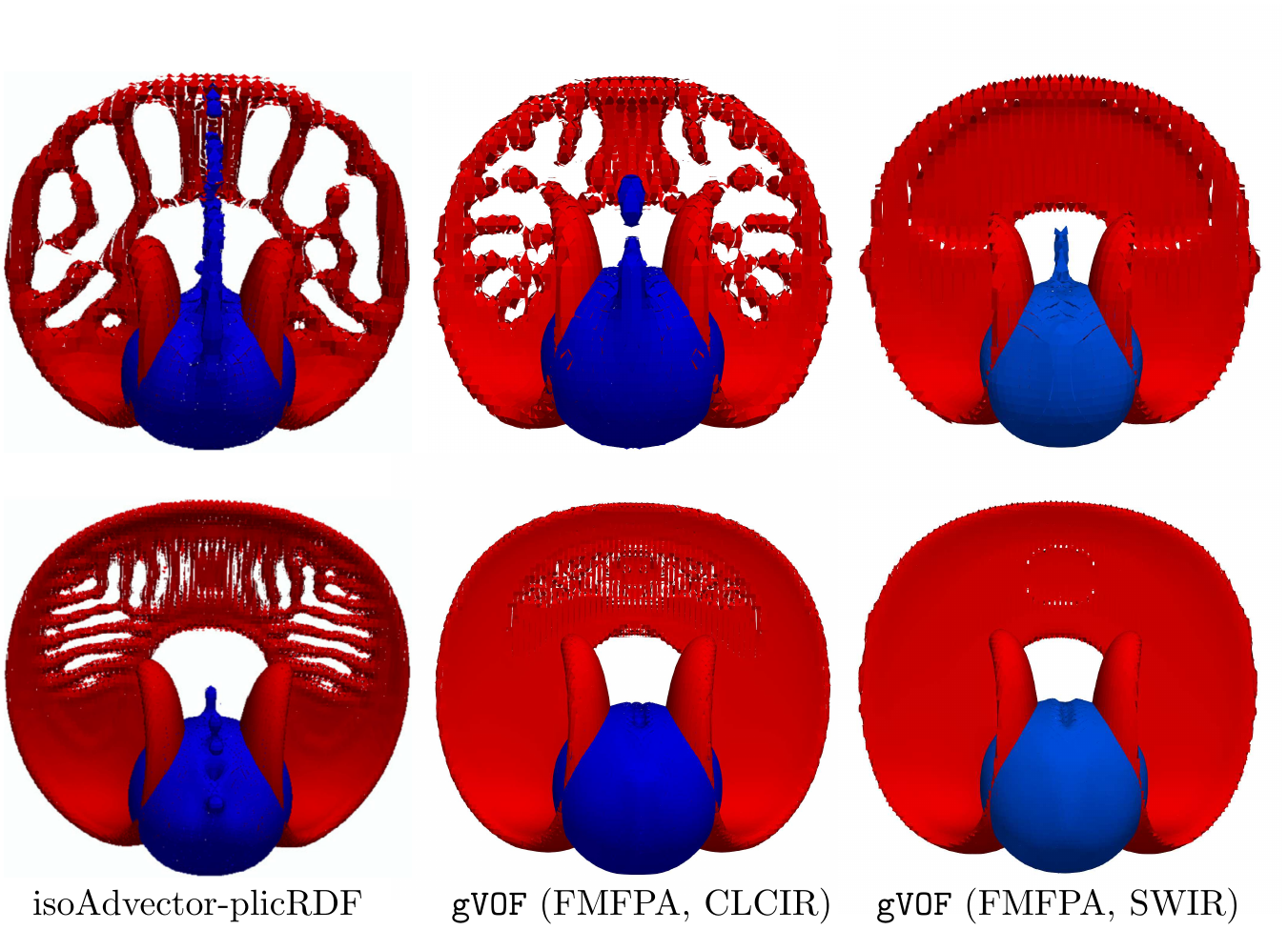}
\caption{PLIC interfaces at $t=1.5$ (in red color) and  $t=3$ (in blue color) obtained using the isoAdvector-plicRDF method \cite{scheufler19} (left pictures) and \texttt{gVOF}, combining FMFPA method with CLCIR and SWIR methods (middle and right pictures, respectively), for the 3D deformation test with two cubic grids with $\mathcal{N}=64$ (top pictures) and 128 (bottom pictures), and $\mbox{CFL}=0.5$. 
}
\label{enright-cubic-vs-isoAdvector-plicRDF}
\end{center}
\end{figure}

On average, all the methods included in the table show second-order convergence
\j{and \texttt{gVOF} provides the lowest shape errors.
  Although SWIR  and LSFIR are the most accurate \jl{methods} for $\mathcal{N}\le 128$, the CLCIR method provides acceptable shape errors with a very good computational efficiency and higher accuracy when the grid resolution is sufficiently high to correctly solve the thinnest fluid structures at the maximum fluid body deformation time.
In \cite{maric18}, where 
an unsplit advection EMFPA-like advection method is combined with a modified version of the Swartz reconstruction method proposed by \citet{dyadechko05}, zero bounding errors are reported, which is somewhat surprising given the high geometric complexity of the unsplit advection method.}
The results obtained by \citet{owkes14} and \citet{jofre14}, who also use  EMFPA-like advection methods and, respectively, the second-order ELVIRA and LVIRA reconstruction methods \cite{pilliod04}, are relatively similar and on average their shape errors are around  90\% higher than  those of \texttt{gVOF}.
A visual comparison with the results obtained with the isoAdvector-plicRDF method, which shows an accuracy close to that of \citet{owkes14} and \citet{jofre14}, can also be seen on Fig.~\ref{enright-cubic-vs-isoAdvector-plicRDF}, where the improvement achieved using \texttt{gVOF} is clearly seen. Note that SWIR method better preserves the integrity of the thin fluid structures (the LSFIR method has a similar behavior in this respect).
\j{The NIFPA-1 method proposed by \citet{ivey17} combined with the embedded height-function method proposed also by \citet{ivey15} provides accuracies comparable to those of other advanced geometric unsplit VOF methods. Although, as it was mentioned, the results of $\widetilde{t}_\mathrm{cpu}$ included in the table should only be considered as a qualitative reference, one of the reasons for the clear advantage of \texttt{gVOF} with respect to others in terms of computational efficiency, is the use of the non-convex version of VOFTools \cite{lopez19,lopez20}, which avoids the use of costly techniques to decompose the generally non-convex flux polyhedra into convex sub-polyhedra.}

\begin{figure}[htbp]
\begin{center}
\includegraphics{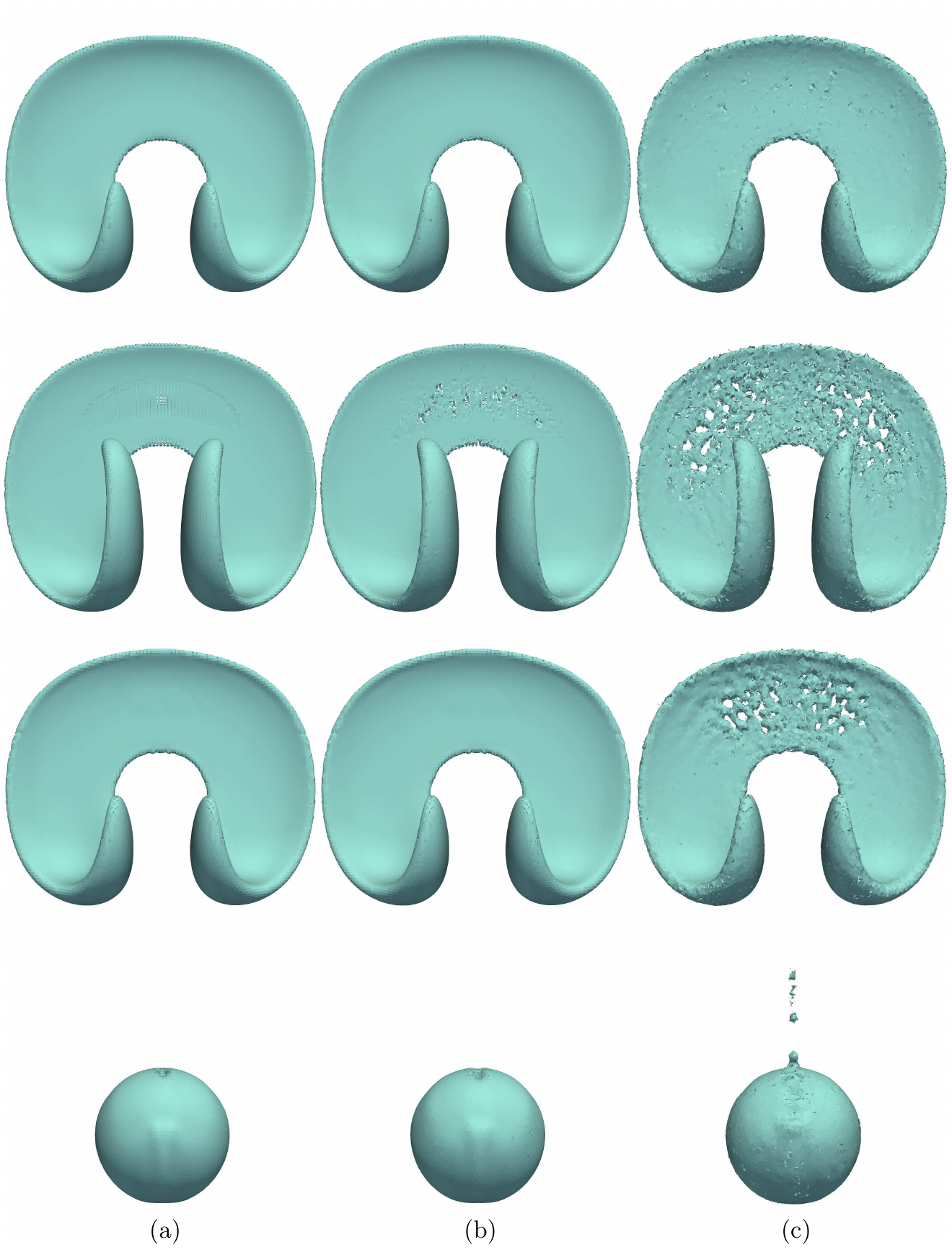}
\caption{PLIC interfaces for the 3D deformation test at  $t=1.0$, 1.5, 2.0 and 3.0 (from top to bottom) obtained using \texttt{gVOF} (EMFPA and CLCIR), $\mbox{CFL}=0.5$  and (a) cubic, (b) non-convex distorted cubic and (c) tetrahedral grids with $\mathcal{N}=160$.}
\label{enright-emfpa-clcir-160}
\end{center}
\end{figure}

Fig.~\ref{enright-emfpa-clcir-160} shows the PLIC interfaces for the 3D deformation test at different instants obtained using \texttt{gVOF} (EMFPA and CLCIR), $\mbox{CFL}=0.5$  and different grids of $\mathcal{N}=160$ with convex and non-convex cells, where the capacity of \texttt{gVOF} to solve tests of high interface deformation on 3D arbitrary grids can be clearly seen.

\subsubsection{2D deformation test}
\label{sec:2ddef}

In this test introduced by \citet{bell89}, a cylinder of fluid with radius 0.15, symmetry axis parallel to the $y$-axis and initially centered at (0.5,0,0.75) in a domain $[0,1] \times [0,1] \times [0,1]$, is deformed in the following velocity field:
\begin{equation}
\left.
\begin{split}
u(x,y,z,t) = &-2 \sin^2(\pi x) \sin(\pi z) \cos (\pi z) \cos (\pi t/t_\mathrm{end}), \\
v(x,y,z,t) = &0, \\
w(x,y,z,t) = & 2 \sin^2(\pi z) \sin(\pi x) \cos (\pi x) \cos (\pi t/t_\mathrm{end}), \\
\end{split} \right\}
\end{equation}
for $t_\mathrm{end}=8$. Note that this 2D deformation test is simulated in this work using 3D grids with a single cell 
along the $y$-axis. These quasi-2D grids will be denoted hereafter as square or triangular grids.

\begin{figure}[htbp]
\begin{center}
\includegraphics{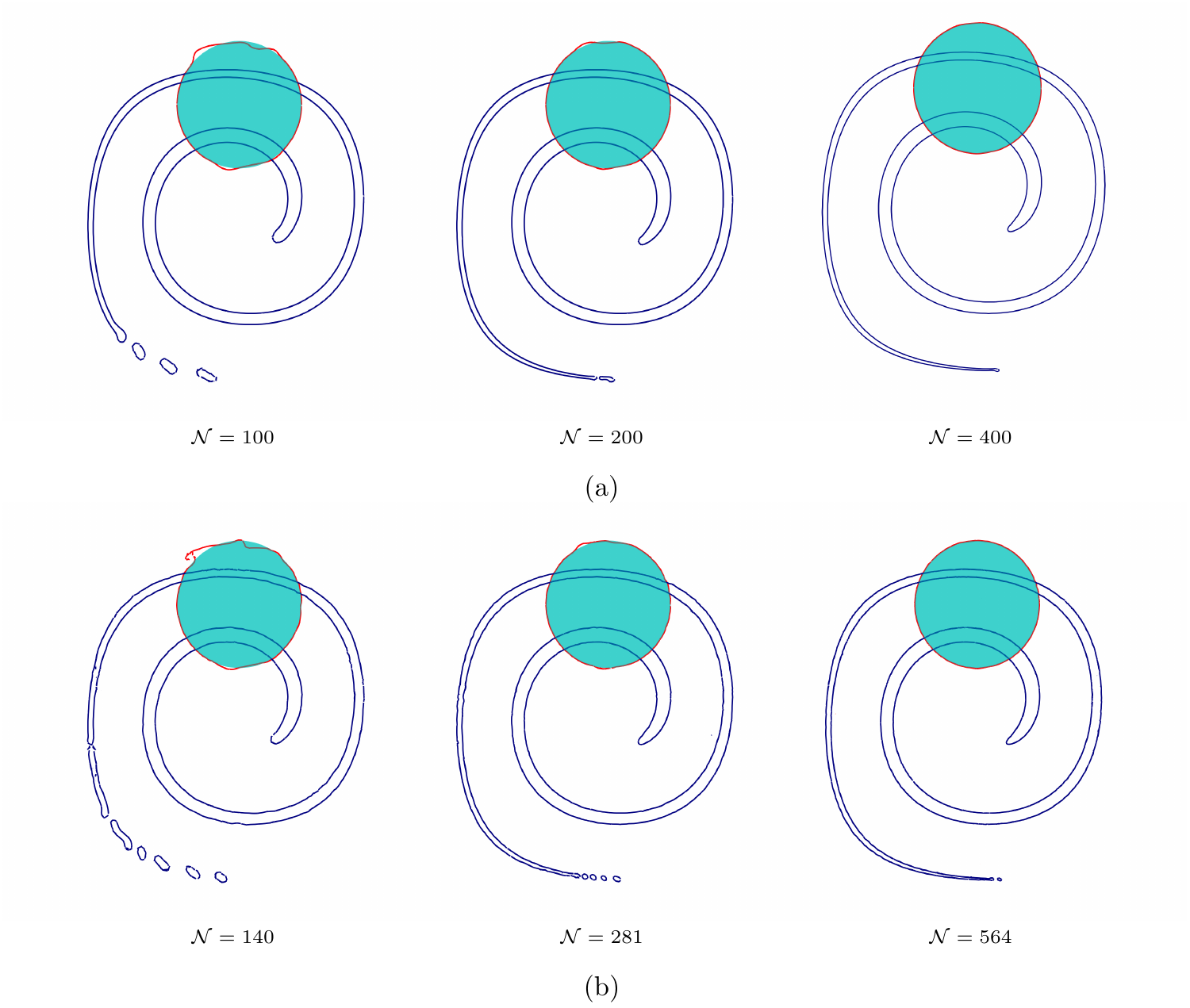}
\caption{PLIC interfaces (in blue for $t=4$ and in red for $t=8$)   obtained using FMFPA and CLCIR methods with $\mathrm{CFL}=0.5$ and different grids. (a) Square grids with $\mathcal{N}=100$, 200 and 400. (b) Triangular grids with $\mathcal{N}=140$, 281 and 564. The exact solution at the end of the test is represented in light blue color.}
\label{fig:a4}
\end{center}
\end{figure}

Table~\ref{table:2ddef1} compares the \jl{$E_\mathrm{shape^*}^{L_1}$} error values \jl{at $t=8$} obtained using \texttt{gVOF} (combining the FMFPA method with the CLCIR, SWIR and LSFIR methods) and $\mbox{CFL}=0.5$ with those obtained using the  isoAdvector and MULES  methods \cite{roenby16} on square  and unstructured grids available in \cite{roenby16b} (the results corresponding to the MULES method were obtained using $\mbox{CFL}=0.1$). 
A substantial reduction of the error values, being on average around a 60\% with respect to the isoAdvector method and higher than 90\% with respect to the MULES method, has been achieved by the \texttt{gVOF} code. 
The corresponding PLIC interfaces obtained with \texttt{gVOF} (FMFPA and CLCIR) at $t=4$ (dark blue color) and $t=8$ (red color) can be seen in Fig.~\ref{fig:a4} (the exact fluid region at $t=8$ is depicted in light blue color). Note that for 2D problems, the EMFPA and FMFPA methods produce equivalent  flux polyhedron shapes, although, as mentioned, the EMFPA method uses more faces and vertices for each flux polyhedron to construct the flux region, affecting this, obvioulsy, to the computational efficiency. As expected, the results obtained using the NMFPA method, which are omitted in this test for brevity,  are less accurate than those obtained using the EMFPA or FMFPA method. 
\begin{table}[htbp]
\caption{\jl{$E_\mathrm{shape^*}^{L_1}$} error values \jl{at $t=8$} for the 2D deformation test using  different grids and $\mbox{CFL}=0.5$. Comparison with the isoAdvector and MULES methods \cite{roenby16}. The results presented for MULES were obtained in \cite{roenby16} using $\mbox{CFL}=0.1$.}
\label{vortex-in-a-box-openfoam}
\begin{center}
  \begin{tabular}{cccccc}
    \hline
 Grid size,    & \multicolumn{3}{c}{\texttt{gVOF}, FMFPA$+$}   & isoAdvector   & MULES  \\
 \cline{2-4}
 $\mathcal{N}$ & CLCIR & SWIR & LSFIR &\cite{roenby16} & \cite{roenby16} \\
    \hline
    \multicolumn{6}{l}{\it Square grids} \\
        $100$ & $2.71\times 10^{-2}$ & $1.57\times 10^{-2}$ & $1.61\times 10^{-2}$  & $4.7\times 10^{-2}$ & -- \\
    $200$ & $7.60\times 10^{-3}$ & $3.60\times 10^{-3}$ & $3.82\times 10^{-3}$  & $1.2\times 10^{-2}$ & $7.2\times 10^{-2}$ \\
    $400$ & $1.24\times 10^{-3}$ & $1.32\times 10^{-3}$ & $1.77\times 10^{-3}$  & $2.3\times 10^{-3}$ & -- \\
    \multicolumn{6}{l}{\it Triangular grids \cite{roenby16b}} \\
   140  & $3.30\times 10^{-2}$ & $2.23\times 10^{-2}$ & $2.23\times 10^{-2}$  & $5.4\times 10^{-2}$ & -- \\
   281 & $8.77\times 10^{-3}$ & $5.47\times 10^{-3}$ & $6.15\times 10^{-3}$  & $2.0\times 10^{-2}$ & $6.6\times 10^{-1}$ \\
   564 & $2.45\times 10^{-3}$ & $8.18\times 10^{-4}$ & $1.62\times 10^{-3}$  & $9.5\times 10^{-3}$ & -- \\
    \hline
  \end{tabular}
\end{center}
\label{table:2ddef1}
\end{table}

Now, 
Table~\ref{vortex-in-a-box2} compares the $E_\mathrm{shape}^{L_1}(8)$ error values obtained using \texttt{gVOF} (combining now the EMFPA method with the CLCIR and SWIR methods) with those obtained using several of the most advanced geometric VOF methods developed to date.
 The shape error reduction achieved by \texttt{gVOF} for this test of high deformation is around  60\% and 50\%  compared to the methods implemented by \citet{owkes14} and \citet{maric18}, respectively, and around  80\% and 70\% compared to the isoAdvector-plicRDF method of \citet{scheufler19} and NIFPA-1 method of \citet{ivey17}, respectively, which represents a remarkable achievement.
\begin{table}[htb]
\caption{\jl{$E_\mathrm{shape}^{L_1}$} error values \jl{at $t=8$} for the 2D deformation test using square grids and different  $\mathrm{CFL}$ values. Comparison with recent advanced geometric-unsplit VOF methods. 
}
\label{vortex-in-a-box2}
\begin{center}
\scalebox{0.8}{\begin{tabular}{ccccccc}
  \hline
Grid size,   &  \multicolumn{2}{c}{\texttt{gVOF}, EMFPA $+$}   &  Owkes and & isoAdvector-   & Mari\'c   & NIFPA-1   \\
\cline{2-3} 
$\mathcal{N}$   &  CLCIR & SWIR & Desjardins \cite{owkes14} & plicRDF \cite{scheufler19}  &  et al. \cite{maric18} &  \cite{ivey17}$^*$ \\
    \hline
\multicolumn{7}{l}{\em $\mbox{CFL}=1$} \\
        $64$ & $3.51 \times 10^{-3}$ & $ 4.15\times 10^{-3}$ & -- & $8.70\times 10^{-3}$ & $5.74\times 10^{-3}$  & -- \\
    $128$ & $ 8.85\times 10^{-4}$ & $ 7.20\times 10^{-4}$ & -- & $ 2.27\times 10^{-3}$ & $1.45\times 10^{-3}$ & -- \\
    $256$ & $1.94 \times 10^{-4}$ & $ 2.14\times 10^{-4}$  &  -- & $ 1.61\times 10^{-3}$ & $3.77 \times 10^{-4}$ & -- \\
    $512$ & $6.37 \times 10^{-5}$ & $ 4.45 \times 10^{-5}$ &  -- & $1.93 \times 10^{-3}$ & -- & -- \\
    $1024$ & $1.59 \times 10^{-5}$ & $ 1.35\times 10^{-5}$ & --  & $ 2.75\times 10^{-3}$ & -- & -- \\
\multicolumn{7}{l}{\em $\mbox{CFL}=0.5$} \\
        $64$ & $6.35 \times 10^{-3}$ & $4.37 \times 10^{-3}$ & $ 7.75\times 10^{-3}$ & $1.26\times 10^{-2}$ & --  & $1.2 \times 10^{-2}$ \\
    $128$ & $ 1.19\times 10^{-3}$ & $ 8.18\times 10^{-4}$ & $ 1.87\times 10^{-3}$ & $ 2.61\times 10^{-3}$ & -- & $2.7 \times 10^{-3}$ \\
    $256$ & $2.07 \times 10^{-4}$ & $1.72 \times 10^{-4}$ & $ 4.04\times 10^{-4}$  & $ 5.71\times 10^{-4}$ & -- & $5.4 \times 10^{-4}$ \\
    $512$ & $5.45 \times 10^{-5}$ & $4.11 \times 10^{-5}$ & $ 8.32\times 10^{-5}$  & $1.04 \times 10^{-4}$ & -- & $1.7 \times 10^{-4}$ \\
    $1024$ & $1.32 \times 10^{-5}$ & $1.11 \times 10^{-5}$ & $ 2.35\times 10^{-5}$  & $ 3.50\times 10^{-5}$ & -- & $4.6 \times 10^{-5}$ \\
    \hline
    \multicolumn{6}{l}{\footnotesize *These results were obtained for $\mathcal{N}+1$ and a CFL sufficiently low to get  $E_\mathrm{vol}^{L_1}(8)<7\times 10^{-14}$.}
\end{tabular}}
\end{center}
\end{table}

\section{Coupling \texttt{gVOF} with an in-house CFD code}
\label{sec:drop-impact}

To show the performance of \texttt{gVOF} coupled with an in-house code that solves the flow conservation equations (previous results obtained with this code coupled with VOF and level-set methods can be found in \cite{lopez09,hernandez08,gomez05,lopez05}), the impact of a water \jl{drop} onto a free surface is solved. The code solves the
conservation equations on both sides of the interface using a projection method.
The projection step incorporates a
continuous surface tension model based on the balanced-force
algorithm proposed by \citet{francois06}, in which 
the interface curvature is computed 
using the height function technique that incorporates the improvements proposed in \cite{lopez09} and \cite{lopez10}.
The pressure Poisson equation resulting from the projection step, which is the most expensive part of the code, is solved using a preconditioned Krylov solver with the help of \texttt{LIS} (library of iterative solvers)  \cite{lis}.
A CFL number of 0.2 was used to determine the time step as
$$
\Delta t= \textrm{min} \left( \sqrt{\frac{\left[\rho_l+\rho_g \right] h^3}{4\pi \sigma}}, \frac{h}{u_\mathrm{max}}\right)\textrm{CFL},
$$
where $\rho_l$ and $\rho_g$ and the densities of the liquid and gas phases, respectively, $\sigma$ is the surface tension coefficient, $h$ is the cell size and $u_\mathrm{max}$ is the maximum absolute value of the velocity components at each instant.

\begin{figure}[htbp]
\begin{center}
\includegraphics{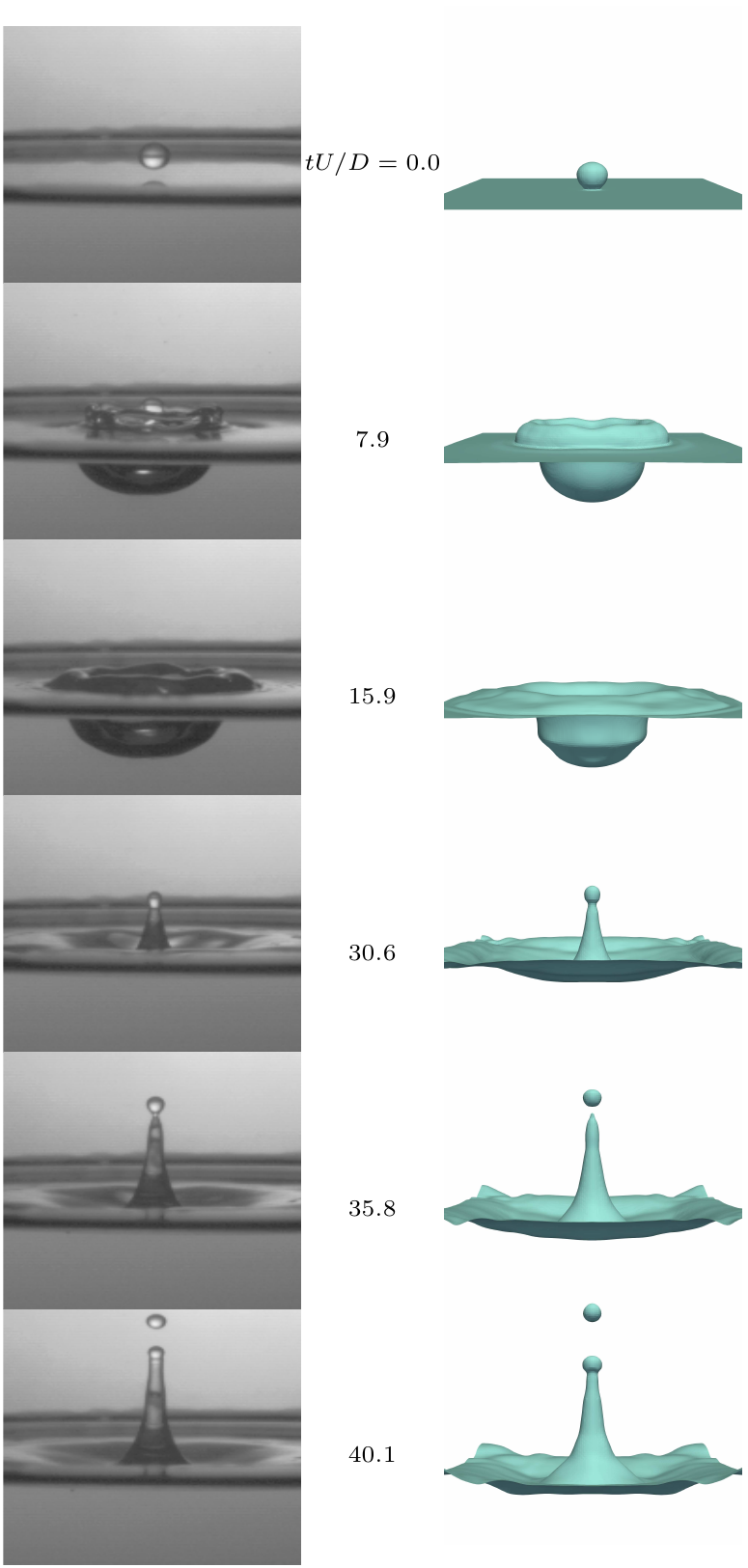}
\end{center}
\caption{Water \jl{drop} impact test. Comparison between the experimental results for the interface shape \cite{hernandez08} (left pictures) and numerical predictions for the  
 0.5-isosurfaces obtained using the EMFPA and CLCIR methods (right pictures) at different instants after the \jl{drop} had made contact with the pool surface.}
\label{drop-impact-icaso6-emfpa-clcir}
\end{figure}

\begin{figure}[htbp]
\begin{center}
\includegraphics{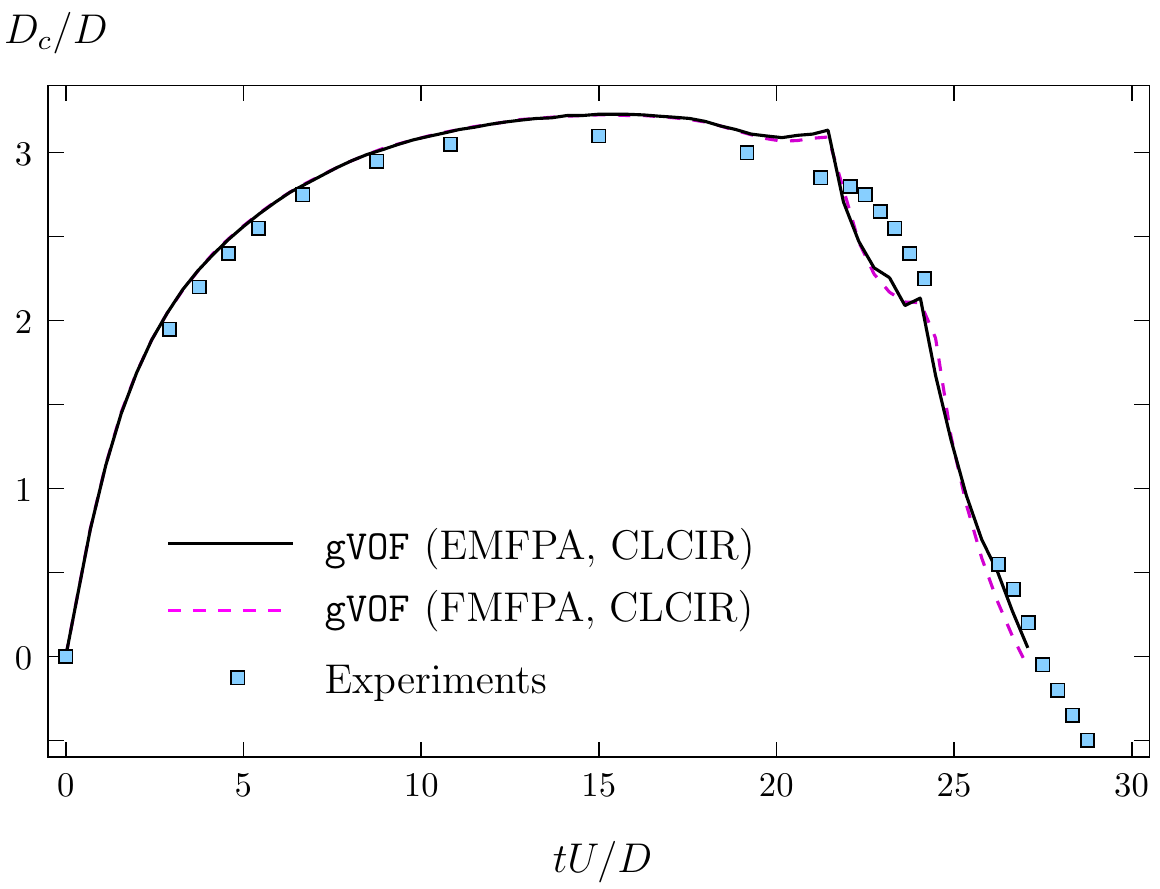}
\end{center}
\caption{Comparison between numerical predictions for the \jl{drop} impact cases of Fig.~\ref{drop-impact-icaso6-emfpa-clcir} and experimental results \cite{hernandez08} for the evolution of the free-surface depth at the symmetry axis, $D_c$, .}
\label{drop-impact-comp}
\end{figure}

A water \jl{drop} of diameter $D=2.9$ mm impacting a
deep water pool with velocity $U=2.5$  m$\,$s$^{-1}$ is
considered. The Froude and Weber numbers are  $Fr = 220$ and $We = 248$, respectively.  Due to the symmetry of the problem, only
one quarter of the physical domain was considered. The computational
domain used was $8D\times 3.5 D \times 3.5 D$. The pool depth was $4.5D$, the water \jl{drop} was initially
located at a height equal to $6.0D$ and the domain was discretized
on a grid of $175\times 70\times 70$ cells. To reach the
desired impact velocity, a fictitious gravitational force was used
to accelerate the drop. Fig.~\ref{drop-impact-icaso6-emfpa-clcir} shows results for the 0.5-isosurfaces obtained using the EMFPA and CLCIR methods at different instants after the \jl{drop} had made contact with the pool surface (right pictures). A relatively good degree of agreement can be observed with the visualization results obtained experimentally  in \cite{hernandez08} (left pictures\jjj{; note that the images of the interface are magnified when observed through water).} A quantitative comparison can be seen in Fig.~\ref{drop-impact-comp}, where numerical and experimental results of the evolution of the free-surface depth at the symmetry axis, $D_c$, are compared. A reasonable agreement can be observed for the cavity depth evolution during growth and collapse processes.
Note that the numerical predictions obtained using the EMFPA and FMFPA methods are very closed for this \jl{drop} impact test. On average, the CPU time consumed by the advection and reconstruction schemes represent around 20\% of the total.
In all simulations, the net change in total volume at the end of the test was lower than $4\times 10^{-6}$\%.

\section{Parallel performance}
\label{sec:parallel}

\begin{figure}[htbp]
\begin{center}
\includegraphics{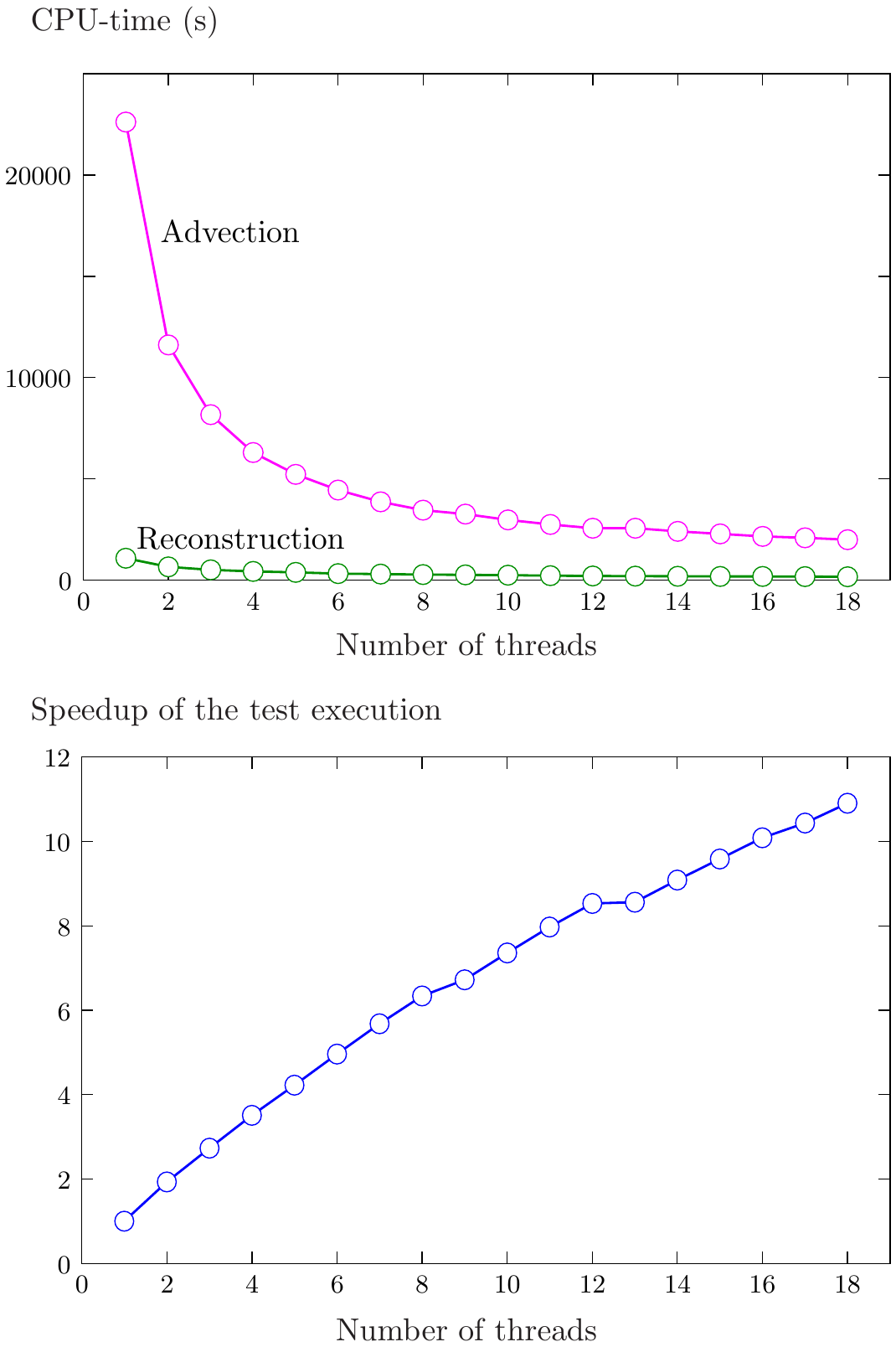}
\caption{CPU time (top picture) of the interface reconstruction and fluid advection, and execution speedup of the 3D deformation test using \texttt{gVOF} (EMFPA and CLCIR), a cubic grid with $\mathcal{N}=256$ and $\mbox{CFL}=0.5$ as a function of the number of threads.}
\label{cputime-rec-adv-thread}
\end{center}
\end{figure}

To assess the parallel performance of the \texttt{gVOF} \jjj{package}, the 3D deformation test of Section~\ref{sec:3ddef} was executed using an increasing number of threads. The same conditions as in Table~\ref{table:3ddef2} but using the EMFPA and CLCIR methods, and a cubic grid with $\mathcal{N}=256$ are considered for this analysis.
The top picture in Fig.~\ref{cputime-rec-adv-thread} shows the CPU times of the interface reconstruction and fluid advection as a function of the number of threads used during the execution of the test case. 
The corresponding execution speedup is also shown  in Fig.~\ref{cputime-rec-adv-thread} (bottom picture). Note that fluid advection spends around  twenty times the CPU time consumed by the interface reconstruction when using a single thread, while this difference tends to reduce to around ten when the number of threads increases.

\section{Conclusions}

The \texttt{gVOF} package, which is provided as open source, includes routines for the implementation of advanced geometric VOF methods on arbitrary grids that achieve accuracies and computational efficiencies which are cutting edge. 
The package uses  the \texttt{VOFTools} and \texttt{isoap} libraries to implement a VOF initialization method and several PLIC interface reconstruction and unsplit advection methods valid for arbitrary grids with convex or non-convex cells.
The routines are written in \texttt{FORTRAN} and can be used with \texttt{C} programs through interface routines included in the distributed software. To improve the computational efficiency, the \texttt{OpenMP} application programming interface is  used.
To assess the performance of the \jh{package}, interface reconstruction and advection tests are also included in the supplied software.
A thorough comparison with existing state-of-the-art
 geometric VOF methods has been carried out and has provided very favorable results.
Also, the \jh{package} has been incorporated into an existing in-house code to simulate the impact of a water \jl{drop} on a free surface. \jh{The numerical results have been compared with experimental results, and a good agreement has been found.}

\section*{Acknowledgments}

The authors gratefully acknowledge the support of the Spanish
Ministerio de Ciencia, Innovación y Universidades - Agencia Estatal de Investigación and FEDER through projects \jjj{DPI2017-87826-C2-1-P, DPI2017-87826-C2-2-P, PID2020-120100GB-C21 and PID2020-120100GB-C22.}

\appendix

\section{Nomenclature}
\label{nomenclature}

\begin{center}
\begin{tabular}{p{2.5cm}p{10cm}}
\hfill$A=$ & matrix \jl{in} Eq.~(\ref{eq:lsgir1}); area of the cell face \\
\hfill$\vec{b}=$ & vector \jl{in} Eq.~(\ref{eq:lsgir2}) \\
\hfill$C=$ & constant that determines the position of the PLIC interface \\
\hfill$D=$ & \jl{drop} diameter \\
\end{tabular}

\begin{tabular}{p{2.5cm}p{10cm}}
\hfill$D_c=$ & free-surface depth at the symmetry axis \\
\hfill$\mathcal{D}=$ & dimensions number (2 for 2D and 3 for 3D) \\
\end{tabular}

\begin{tabular}{p{2.5cm}p{10cm}}
\hfill$E_\mathrm{bound}^{L_1} =$ & average $E_\mathrm{bound}^{L_\infty} (t)$ value during the complete simulation \\
\hfill$E_\mathrm{bound}^{L_\infty} =$ & maximum $E_\mathrm{bound}^{L_\infty} (t)$ value during the complete simulation \\
\hfill$E_\mathrm{bound}^{L_\infty} (t)=$ & maximum unboundedness error of the fluid volume at a time $t$ \\
\hfill$E_\mathrm{rec}^{L_1}=$ & interface reconstruction error \\
\end{tabular}

\begin{tabular}{p{2.5cm}p{10cm}}
\hfill$E_\mathrm{shape}^{L_1}=$ & interface shape error \\
\hfill$E_\mathrm{shape^*}^{L_1}=$ & relative interface shape error \\
\hfill$E_\mathrm{vol}^{L_1}=$ & fluid volume error \\
\hfill$f=$ & auxiliary VOF function \\
\end{tabular}

\begin{tabular}{p{2.5cm}p{10cm}}
\hfill$F=$ & fluid volume fraction (discretized version of $f$) \\
\end{tabular}

\begin{tabular}{p{2.5cm}p{10cm}}
\hfill$\widetilde{F}=$ & Taylor series expanded value of $F$ \\
\hfill$F^\mathrm{e}=$ & exact fluid volume fraction of a cell \\
\end{tabular}

\begin{tabular}{p{2.5cm}p{10cm}}
\hfill$F^*=$ & fluid volume fraction interpolated at a grid node \\
\end{tabular}

\begin{tabular}{p{2.5cm}p{10cm}}
\hfill$Fr=$ & Froude number \\
\hfill$h=$ & cubic cell size \\
\end{tabular}

\begin{tabular}{p{2.5cm}p{10cm}}
\hfill$h_x,h_y,h_z=$ & sizes along the coordinate axis $x,y,z$ of the minimum-size rectangular parallelepiped that encloses a cell \\
\hfill$I=$ & number of vertices of the first face in a flux polyhedron \\
\hfill$n=$ & current time step \\
\end{tabular}

\begin{tabular}{p{2.5cm}p{10cm}}
\hfill$\vec{n}=$ & unit vector normal to the interface pointing into the fluid or normal to the cell face pointing out of the cell  \\
\hfill$\mathcal{N}=$ & grid size \\
\hfill$N_\mathrm{CELL}=$ & number of grid cells in the computational domain \\
\end{tabular}

\begin{tabular}{p{2.5cm}p{10cm}}
\hfill$N_\mathrm{FACE}=$ & number of grid faces in the computational domain \\
\hfill$N_\mathrm{STEP}=$ & number of time steps required to complete the advection test \\
\hfill$\widetilde{N}_\mathrm{CELL}=$ & number of grid cells in an equivalent unit domain \\
\hfill$t=$ & time \\
\hfill$t_0=$ & previous time of an advection test \\
\end{tabular}

\begin{tabular}{p{2.5cm}p{10cm}}
\hfill$t_\mathrm{adv}=$ & total CPU time consumed by the advection step \\
\hfill$t_\mathrm{cpu}=$ & total execution CPU time \\
\hfill$t_\mathrm{end}=$ & end time of an advection test \\
\hfill$t_\mathrm{rec}=$ & total CPU time consumed by the reconstruction step \\
\end{tabular}

\begin{tabular}{p{2.5cm}p{10cm}}
\hfill$\widetilde{t}_\mathrm{cpu}=$ & average total execution CPU time per time step \\
\end{tabular}

\begin{tabular}{p{2.5cm}p{10cm}}
\hfill$U=$ & impact velocity  \\
\end{tabular}

\begin{tabular}{p{2.5cm}p{10cm}}
\hfill$\vec{u}=$ & velocity vector  \\
\hfill${u}_{max}=$ & maximum absolute value of the velocity components  \\
\end{tabular}

\begin{tabular}{p{2.5cm}p{10cm}}
\hfill$u,v,w=$ & components of $\vec{u}$ \\
\hfill$V_d=$ & volume of the flux region  \\
\end{tabular}

\begin{tabular}{p{2.5cm}p{10cm}}
\hfill$V_{d_T}=$ & total net flux volume at the cell  \\
\hfill$V_F=$ & volume of the fluid advected through a cell face  \\
\hfill$V_{F_T}=$ & total net volume of fluid that leaves (or enters) the cell  \\
\hfill$V_\Omega=$ & volume of the grid cell $\Omega$  \\
\end{tabular}

\begin{tabular}{p{2.5cm}p{10cm}}
\hfill$w=$ & weighting factor \\
\hfill$We=$ & Weber number \\
\end{tabular}

\begin{tabular}{p{2.5cm}p{10cm}}
\hfill$\vec{x}=$ & position vector  \\
\end{tabular}

\begin{tabular}{p{2.5cm}p{10cm}}
\hfill$x,y,z=$ & Cartesian coordinates \\
\end{tabular}

\begin{tabular}{p{2.5cm}p{10cm}}
\multicolumn{2}{l}{{\em\bf Subscripts}} \\
\hfill$i=$ & face vertex index \\
\hfill$j=$ & cell face index \\
\hfill$k=$ & neighbor grid cell index \\
\end{tabular}

\begin{tabular}{p{2.5cm}p{10cm}}
\multicolumn{2}{l}{{\em\bf Superscripts}} \\
\hfill$n=$ & \jl{time step} \\
\end{tabular}

\begin{tabular}{p{2.5cm}p{10cm}}
\multicolumn{2}{l}{{\em\bf Greek characters}} \\
\hfill$\alpha=$ & facet angle in the triangulated isosurface \\
\hfill$\beta=$ & parameter in Eq.~(\ref{lsgir-beta}) \\
\hfill$\Delta t =$ & time step \\
\hfill$\epsilon=$ & fluid volume fraction tolerance \\
\hfill$\Omega=$ & grid cell\\
\end{tabular}

\begin{tabular}{p{2.5cm}p{10cm}}
\hfill$\Omega^p=$ & flux polyhedron \\
\end{tabular}

\begin{tabular}{p{2.5cm}p{10cm}}
\hfill$\rho_g=$ & gas density \\
\end{tabular}

\begin{tabular}{p{2.5cm}p{10cm}}
\hfill$\rho_l=$ & liquid density \\
\hfill$\sigma=$ & surface tension coefficient \\
\end{tabular}
\end{center}

\begin{center}
\begin{tabular}{p{2.5cm}p{10cm}}
\multicolumn{2}{l}{\em\bf Acronyms} \\
\hfill $\mathrm{CFD}=$ & computational fluid dynamics \\
\hfill $\mathrm{CFL}=$ & Courant-Friedrich-Levy number \\
\end{tabular}

\begin{tabular}{p{2.5cm}p{10cm}}
\hfill $\mathrm{CIBRAVE}=$ & coupled interpolation-bracketed analytical volume enforcement \\
\hfill $\mathrm{CLCIR}=$ & conservative level contour interface reconstruction \\
\hfill $\mathrm{ELCIR}=$ & extended level contour interface reconstruction \\
\hfill $\mathrm{EMFPA}=$ & edge-matched flux polygon/polyhedron advection \\
\hfill $\mathrm{FMFPA}=$ & face-matched flux polyhedron advection \\
\hfill $\mathrm{LLCIR}=$ & local level contour interface reconstruction \\
\hfill $\mathrm{LSFIR}=$ & least-squares fit interface reconstruction \\
\hfill $\mathrm{LSGIR}=$ & least-squares gradient interface reconstruction \\
\hfill $\mathrm{NIFPA}=$ & non-intersecting flux polyhedron advection \\
\hfill $\mathrm{NMFPA}=$ & non-matched flux polyhedron advection \\
\hfill $\mathrm{PLIC}=$ & piecewise linear interface calculation \\
\hfill $\mathrm{SLIC}=$ & simple line interface calculation \\
\hfill $\mathrm{SWIR}=$ & Swartz interface reconstruction \\
\hfill $\mathrm{VOF}=$ & volume of fluid \\
\end{tabular}
\end{center}









\end{document}